\documentclass[12pt]{article}

\usepackage{titlesec,color}
\usepackage{setspace}

\titleformat{\section}{\centering\large\bfseries}{\thesection. }{0em}{\MakeUppercase}
\titleformat{\thesection}{\flushleft}{\thesubsection }{0em}{\MakeUppercase}
\usepackage{amsmath,amssymb,amsfonts,amsthm}
\usepackage{multirow,ulem}
\usepackage[flushleft]{threeparttable}
\usepackage[authoryear]{natbib}
\setcitestyle{aysep={}} 

\usepackage{mathrsfs}
\newcommand{\blind}{1}
\newcommand{\inctable}{1}

\usepackage{latexsym}
\usepackage{amsmath}
\usepackage{multirow}

\newtheorem{theorem}{Theorem}
\newtheorem{lemma}{Lemma}

\newtheorem{example}{Example}

\newcommand{ \bzero}{\mathbf{0}}

\newcommand{\bbeta}{\boldsymbol{\beta}}           

\newcommand{\btheta}{{\boldsymbol{\theta}}}

\newcommand{\bcalI}{{\boldsymbol{\mathcal{I}}}}
\newcommand{\calI}{\mathcal{I}}
\newcommand{\bI}{{\boldsymbol{I}}}

\newcommand{\ba}{{\boldsymbol{a}}}
\newcommand{\bSigma}{{\boldsymbol{\Sigma}}}
\newcommand{\bS}{{\boldsymbol{S}}}
\newcommand{\bp}{{\boldsymbol{p}}}
\newcommand{\br}{{\boldsymbol{r}}}
\newcommand{\bbm}{{\boldsymbol{m}}}
\newcommand{\bP}{{\boldsymbol{P}}}
\newcommand{\bD}{{\boldsymbol{D}}}
\newcommand{\bV}{{\boldsymbol{V}}}

\newcommand{\en}{\end{equation}}
\newcommand{\bea}{\begin{eqnarray}}
\newcommand{\eea}{\end{eqnarray}}

\newcommand{\ea}{\end{array}}

\newcommand{\bx}{{\boldsymbol{x}}}
\newcommand{\bh}{{\boldsymbol{h}}}
\newcommand{\bX}{{\boldsymbol{X}}}
\newcommand{\bZ}{{\boldsymbol{Z}}}
\newcommand{\bH}{{\boldsymbol{H}}}
\newcommand{\bB}{{\boldsymbol{B}}}
\newcommand{\bu}{{\boldsymbol{u}}}

\newcommand{\bone}{{\boldsymbol{1}}}

\newcommand{\by}{{\boldsymbol{y}}}
\newcommand{\bean}{\begin{eqnarray*}}
\newcommand{\eean}{\end{eqnarray*}}
\newcommand{\var}{{\boldsymbol{V}}}
\def\T{{ \mathrm{\scriptscriptstyle T} }}

\begin{document}

\date{}


\if1\blind
{
\title{\bf Hypotheses Testing from Complex Survey Data Using Bootstrap Weights: A Unified Approach}
\author{Jae-kwang Kim\thanks{Jae-kwang Kim is Professor, Department of Statistics, Iowa State University, Ames, IA 50011, USA (Email: {\tt jkim@iastate.edu})}\,, J.~N.~K. Rao\thanks{J.~N.~K. Rao is Distinguished Research Professor, School of Mathematics and Statistics, Carleton University, Ottawa, ON K1S 5B6, Canada (Email: {\tt jrao@math.carleton.ca})}\, and Zhonglei Wang\thanks{Zhonglei Wang is Assistant Professor, Wang Yanan Institute for Studies in Economics (WISE) and School of Economics, Xiamen University, Xiamen, Fujian 361005, PRC (Email: {\tt wangzl@xmu.edu.cn})} }

\maketitle
} \fi

\if0\blind
{
\bigskip
\bigskip
\bigskip
\begin{center}
	{\LARGE\bf Hypotheses Testing from Complex Survey Data Using Bootstrap Weights: A Unified Approach}
\end{center}
\medskip
} \fi

\bigskip
\begin{abstract}

Standard statistical methods that do not take proper account of the complexity of survey design can lead to erroneous inferences when applied to survey data due to unequal selection probabilities, clustering, and other design features. In particular, the actual type I error rates of tests of hypotheses using standard methods can be much bigger than the nominal significance level. Methods that take account of survey design features in testing hypotheses have been proposed, including Wald tests and quasi-score tests that involve the estimated covariance matrices of parameter estimates. In this paper, we present a unified approach to hypothesis testing  that does not require computing the covariance matrices by constructing bootstrap approximations to weighted likelihood ratio statistics and weighted quasi-score statistics and  establish the asymptotic validity of the proposed bootstrap tests. In addition, we also consider hypothesis testing from categorical data and present a bootstrap procedure for testing simple goodness of fit and independence in a two-way table. In the simulation studies, the type I error rates of the proposed approach are much closer to their nominal significance level compared with the naive likelihood-ratio test and quasi-score test.  An application to data from an educational survey under a logistic regression model is also presented.
\end{abstract}

\noindent%
{\it Keywords:} Likelihood-ratio test;  Quasi-score test; Wald test; Wilks' theorem.

\doublespacing

\section{Introduction}

Testing statistical hypotheses is one of the fundamental problems in statistics. In the parametric model approach,  hypothesis testing  can be implemented using Wald test, likelihood-ratio test, or score test. In each case, a test statistic  is computed and then  compared with the $100\alpha \%$-quantile  of the reference distribution which is the limiting distribution of the test statistic under the null hypothesis, where $\alpha$ is the nominal significance level. The limiting distribution is often a chi-squared distribution \citep{shao2003mathematical}.

Statistical inference with survey data, however,  involves additional steps incorporating the sampling design features.  \citet{Korn1999} and \cite{Chambers2003} provided  comprehensive overviews of the methods for analyzing survey data from complex sampling designs. In  hypothesis testing with survey data,  the limiting distribution of the test statistic is not generally a chi-squared distribution.
Rather, it can be expressed as a weighted sum of several  independent  random variables from $\chi^2(1)$, which is a chi-square distribution with one degree of freedom, and the weights depend on unknown parameters \citep{lumley2014tests}. To handle such problems, one may consider some corrections to the test statistics to obtain a chi-square limiting distribution approximately. Such an approach usually involves computing ``design effects'' associated with the test statistics. \citet{Scott1981CHI}, \citet{rao1984chi} and \citet{rao1998quasi} used this approach to obtain quasi-score tests for survey data.

In this paper, we use a different approach of computing the limiting distribution by bootstrap approximation. \citet{beaumont2009practical} investigated a general weighted  bootstrap method for hypothesis testing  under complex sampling designs in the context of Wald tests for linear regression analysis. In this paper, we  generalize the bootstrap testing idea of \citet{beaumont2009practical} 
and present a unified approach   using  the bootstrap method to obtain the limiting distribution of test statistics   under complex sampling designs, including  Poisson sampling, probability proportional to size sampling with replacement and stratified multi-stage cluster sampling with  clusters selected with replacement. To do this, we first establish the bootstrap central limit theorem under regularity conditions.
The sampling design is allowed to be informative in the sense that the sampling mechanism depends on the response variables being sampled \citep{Pfeffermann1993}.
Once the bootstrap central limit theorem is established, the proposed bootstrap method can be applied to  Wald test, likelihood-ratio test, and score test.

The proposed bootstrap method is also applied for testing  simple goodness of fit and  independence in a two-way table for categorical survey data. 
Earlier work \citep{Scott1981CHI,rao1984chi} developed corrected test statistics based on design effects, and it is known as Rao-Scott first-order and second-order corrections. Rao-Scott methods are design-based, but they can also be applied to make inference for parameters in the super-population model if the design-based variance dominates the total variance with respect to the super-population and the sampling mechanism, where the design-based variance refers to the one with respect to the sampling mechanism conditional on the finite population. Rao-Scott corrections have been implemented in several software packages including SAS and Stata \citep{scott2007rao}. The proposed bootstrap method is attractive as it does not require computing the design effect for Rao-Scott corrections.

In Section 2, the basic setup is introduced. The proposed bootstrap method is presented in Section 3. In Section 4, the bootstrap likelihood-ratio test  using survey-weighted log-likelihood ratio is introduced. In Section 5, the proposed bootstrap method is applied to  survey-weighted quasi-score test.   In Section 6, the proposed bootstrap method is applied to the simple goodness-of-fit test for categorical survey data. In Section 7, test of independence in two-way tables is covered.
Results from two simulation studies  are presented in Section 8. An application to data from an educational survey under a logistic regression model is presented in Section 9. Concluding remarks are made in Section 10. 

\section{Basic Setup}\label{sec: basic}

Suppose that a finite population $U$ of size $N$ is randomly generated from a super-population model with density function $f( y ; \btheta_0)$ for some $\btheta_0 \in \Theta \subset
\mathbb{R}^p
$, where $\Theta$ is the parameter space. From the finite population $U$, a probability sample $A$ of size $n$ is selected,  and the corresponding sampling weight is denoted as $w_i$.  We are interested in making inference about $\btheta_0$.

The pseudo maximum likelihood estimator  of $\btheta_0$ is  given as
$$ \hat{\btheta} = \mbox{arg} \max_{\btheta \in \Theta} l_w( \btheta) , $$
where $l_w( \btheta) = N^{-1} \sum_{i \in A} w_il(\btheta;y_i) $ is the survey-weighted log-likelihood or pseudo log-likelihood with $l(\btheta;y_i)=\log f( y_i; \btheta)$. More precisely, a subscript $N$ could  be used to index the population, the sample, the sampling weights, the pseudo maximum likelihood estimator and other quantities, but it is suppressed in the sequel for simplicity.

Often, the solution $\hat{\btheta}$ can be obtained by solving the weighted score equation
\begin{equation}
\hat{\bS}_w ( \btheta) =  \frac{ \partial }{ \partial \btheta} l_w ( \btheta)  = N^{-1} \sum_{i \in A} w_i \bS( \btheta; y_i) = \bzero,
\label{1}
\end{equation}
where $\bS( \btheta; y_i)  = \partial l(\btheta;y_i)/\partial\btheta$.
Since the sampling weights are used in (\ref{1}), $\hat{\bS}_w ( \btheta)$ is design-consistent for  $ \bS(\btheta) = N^{-1} \sum_{i=1}^N \bS( \btheta; y_i)$ under regularity conditions even when the sampling design is informative. 
The following theorem establishes the limiting distribution for $\hat{\btheta}$.
\begin{theorem}\label{theo: CLT (2)}
    Under regularity conditions in Section~\ref{ss: regularity conditions}  of the Supplementary Material, we have 
	\begin{equation}
	\sqrt{n} (  \hat{\btheta} - \btheta_0 )  \longrightarrow N ( \bzero, \bSigma_\btheta )
	\label{2}
	\end{equation}
	in distribution as $n\to\infty$ with respect to the joint distribution of  the super-population model and the sampling mechanism,
	where
	$
	\bSigma_\btheta = \bcalI(\btheta_0)^{-1}  \bSigma_S(\btheta_0) \{\bcalI(\btheta_0)^{-1}\}^{\T},
	$
	$\bcalI(\btheta)= E\{  \bI( \btheta; Y)  \} $, $\bI( \btheta; y) = -  \partial^2 l(\btheta;y_i)/ \partial \btheta \partial \btheta^{\T} $, $\bSigma_S(\btheta) =\lim_{n\to\infty} [n\var\{\hat{\bS}_w(\btheta)\}]$, and $\var\{\hat{\bS}_w(\btheta)\}$ is the variance of $\hat{\bS}_w( \btheta)$ with respect to the super-population model and the sampling mechanism.
\end{theorem}
Theorem~\ref{theo: CLT (2)} is proved for general sampling designs under regularity conditions shown in Section~\ref{ss: regularity conditions}; see Section~\ref{sec: proof of theorem 1} of Supplementary Material for details of the proof. 
In this paper, we consider the case where the design-based variance $\var\{\hat{\bS}_w(\btheta)\mid U\}$ dominates $\var\{\hat{\bS}_w(\btheta)\}$, and it is often the case when the sample size $n$ is negligible compared with the population size $N$, and this assumption is the building block of the proposed bootstrap method in the next section. 

Using the second-order Taylor expansion, we  obtain
\begin{eqnarray}
l_w ( \btheta_0 ) &=& l_w ( \hat{\btheta} ) + \hat{\bS}_w^{\T}  ( \hat{\btheta} ) ( \btheta_0- \hat{\btheta}) - \frac{1}{2} ( \hat{\btheta} - \btheta_0  )^{\T} \hat{\bI}_w ( \hat{\btheta}) ( \hat{\btheta} - \btheta_0 ) + o_p (n^{-1} ) \notag \\
&=& l_w ( \hat{\btheta} )  - \frac{1}{2} ( \hat{\btheta} - \btheta_0)^{\T} \hat{\bI}_w ( \hat{\btheta}) (\hat{\btheta} - \btheta_0 ) + o_p (n^{-1} )  ,
\label{3}
\end{eqnarray}
where
$
\hat{\bI}_w (\btheta) = N^{-1} \sum_{ i \in A} w_i  \bI( \btheta; y_i).
\label{3b}
$
Define
\begin{equation}
W (\btheta_0) = - 2 n \{ l_w ( \btheta_0 ) - l_w ( \hat{\btheta})
\}
\label{4}
\end{equation}
to be the pseudo likelihood-ratio  test statistic. By  (\ref{3}), we obtain
$$ W ( \btheta_0) = n ( \hat{\btheta} - \btheta_0   )^{\T}  \hat{\bI}_w ( \hat{\btheta}) ( \hat{\btheta} - \btheta_0  )  + o_p (1) . $$
Thus, by Theorem~\ref{theo: CLT (2)},  we  have
\begin{equation}
W ( \btheta_0)   \longrightarrow \mathcal{G} = \sum_{i=1}^p c_i Z_i^2
\label{5}
\end{equation}
in distribution as $n\to\infty$, and the reference distribution  is the joint distribution of the super-population model and the sampling mechanism,
where $c_1, \ldots, c_p$ are the eigenvalues of $\bD=\bSigma_\btheta \bcalI(\btheta_0)  $, and $Z_1, \ldots, Z_p$ are $p$ independent random variables from the standard normal distribution. Result (\ref{5}) was  established by  \citet{rao1984chi} and \citet{lumley2014tests}, and it  can be regarded as  a version of the Wilks' theorem  for  survey sampling. For $p=1$, we can use $c_1^{-1} W(\btheta)$ as the test statistic with  $\chi^2(1)$ distribution as the limiting distribution under the null hypothesis, where $c_1 = \sigma^2_S(\theta_0)/I(\theta_0)$, and $\sigma^2_S(\theta)$ and $I(\theta)$ are the counterparts of $\bSigma_s(\btheta)$ and $\bI(\btheta)$, respectively. Unless the sampling design is simple random sampling and the sampling fraction is negligible, the limiting distribution does not reduce to the standard chi-squared distribution with $p$ degrees of freedom.

\section{Bootstrap calibration}\label{sec: bootstrap}


We propose using a bootstrap method to approximate the limiting distribution in (\ref{5}).
Such a bootstrap calibration is very attractive in practice since there is no need to derive the analytic form of the limiting distribution of the test statistic.  
Given the sample $A$, the proposed bootstrap method is  as follows:
\begin{enumerate}
	\renewcommand{\labelenumi}{Step \arabic{enumi}}
	\item  Generate a rescaling vector $\br^*$ of size $n=|A|$   from  a bootstrap distribution  with $E_*(\br^*)=\bone_n$, and the covariance of $\br^*$ is determined by the sampling design, where $|A|$ is the cardinality of  set $A$, $E_*(\cdot)$ is the expectation with respect to the bootstrap distribution conditional on the sample $A$, and  $\bone_n$ is a vector of 1 with length $n$. The bootstrap weights are chosen in such a way that $E_*\{\bS^*_w(\btheta)\} = \hat{\bS}_w(\btheta)$ and $\var_*\{\bS^*_w(\btheta)\} = \hat{\bV}\{\hat{\bS}_w(\btheta)\}$, where $\hat{\bS}_{w}^*(\btheta) = \sum_{i\in A} w_i^*  \bS( \btheta; y_i)$, $w_i^*=r_i^* w_i$,  $\var_* (  \cdot ) $ is the bootstrap variance conditional on the sample $A$, and $ \hat{\bV}\{\hat{\bS}_w(\btheta)\}$ 
is 
a consistent design-based variance estimator of $\hat{\bS}_w(\btheta)$.
	\item  Obtain a bootstrap replicate of $\hat{\btheta}$, denoted as $\hat{\btheta}^*$,  by solving $\hat{\bS}_{w}^*(\btheta) =0$.

 \item Repeat the first two steps for $M$ times, where $M$ is a large positive integer.
\end{enumerate}

The proposed bootstrap method can be viewed as a nontrivial extension of the exchangeably weighted bootstrap \citep{Praestgaard1993} to complex sampling designs. Different distributions may be used to generate $\br^*$ under different sampling designs, and we will present three examples to demonstrate the proposed bootstrap method later.


Our goal is to show  the following asymptotic result for $\hat{\btheta}^*$.
\begin{theorem}\label{theo: (6)}
 Under regularity conditions in Section \ref{ss: regularity conditions} of the Supplementary Material, we have
    \begin{equation}
	\sqrt{n} ( \hat{\btheta}^* - \hat{\btheta})\mid A \,{\longrightarrow }\,  N( \bzero,{\bSigma}_\btheta )
	\label{lemma1}
	\end{equation}
	in distribution
	as $n \rightarrow \infty$, and 
	the reference distribution on the left side of (\ref{lemma1}) is the bootstrap sampling distribution conditional on  the   sample $A$, where is ${\bSigma}_\btheta $ defined in Theorem~\ref{theo: CLT (2)}.
\end{theorem}
 Theorem~\ref{theo: (6)} establishes the bootstrap central limit theorem for $\hat{\btheta}^*$, and the limiting distribution coincides with the one for $\hat{\btheta}$. Thus, the proposed bootstrap method can be used to approximate the sampling distribution of $\hat{\btheta}$ which can be used to make inference for $\btheta_0$.  Theorem~\ref{theo: (6)} is proved for general sampling designs under regularity conditions shown in Section~\ref{ss: regularity conditions}, which can be verified under three complex sampling designs; see Section~\ref{sec: proof of thereom 2} of the Supplementary Material for details.

For estimating the mean of a finite population, \citet{chao1985}  first proved the bootstrap central limit theorem under simple random sampling without replacement, and  \citet{bickel1984asymptotic} proved the bootstrap central limit theorem  under stratified random sampling without replacement.  
We now present  three  sampling designs that satisfy the bootstrap central limit theorem in Theorem~\ref{theo: (6)} in the following examples: Poisson sampling, probability proportional to size sampling with replacement and stratified multi-stage  cluster sampling with  clusters selected with replacement. Poisson sampling is often applied when administrative files are available, and it can also simplify the variance estimator under some complex sampling designs; see \citet{beaumont2012generalized} for details.  Stratified multi-stage  cluster sampling  with  clusters selected with replacement is extensively used in socio-economic surveys when the first stage sampling fractions within strata are negligible, as is often the case in large-scale sample surveys. We have also presented our bootstrap method for stratified random sampling without replacement in the Supplementary Material without technical details.

\begin{example}\label{ex: 1} (Poisson sampling) 

Under Poisson sampling, the sample selection is made from   independent Bernoulli distributions with parameter $\pi_i$ for $i=1,\ldots,N$. Specifically, let $I_i$, following  Bernoulli($\pi_i$) distribution,   be the sampling indicator of  $y_i$, and $I_i=1$ if and only if $y_i$ is selected. The rescaling factor is 
$r_i^* = 1+   m_i^*- \pi_i $, where $m_i^* \sim Bernoulli (\pi_i)$.  The bootstrap replicate of $\hat{\bS}_w(\btheta)$ is $\hat{\bS}_w^*(\btheta) = N^{-1}\sum_{i \in A} r_i^*w_i \bS(\btheta;y_i)$ with $w_i=\pi_i^{-1}$, and it satisfies $E_* \{\hat{\bS}_w^*(\btheta)\} = \hat{\bS}_w(\btheta)$ and 
$V_*\{\hat{\bS}_w^*(\btheta)\} = N^{-2}\sum_{i \in A}  \pi_i^{-1} (1-\pi_i) \bS(\btheta;y_i)^{\otimes2}
$, where $\bx^{\otimes2} = \bx\bx^{\T}$ for a vector $\bx$.
In Section~\ref{ss: pois sampling} of the Supplementary Material, we show that the proposed bootstrap method satisfies the conditions for the bootstrap central limit theorem in Theorem~\ref{theo: (6)}.
\end{example}

\begin{example}\label{ex: 2} (Probability proportional to size sampling with replacement) 
 
Under probability proportional to size sampling, a sample of size $n$ is independently generated from  $U$ with replacement, and the selection probability of $y_i$ is $p_i$  with $\sum_{i=1}^N p_i = 1$. Then, we have $\hat{\bS}_w(\btheta) = (Nn)^{-1} \sum_{i=1}^n p^{-1}_{a(i)}\bS(\btheta;y_{a(i)}) $, where $a(i)$ is the index of the element selected at the $i$-th draw. A design-unbiased variance estimator is 
	$ \hat{V}\{\hat{\bS}_w(\btheta)\} = \{N^{2}n(n-1)\}^{-1} \bS_{vec}(\theta)(\bI_n - \mathcal{P}_{\bone,n})\bS{}^{\T}_{vec}(\theta)$, where $\bS_{vec}(\theta) = (p^{-1}_{a(1)}\bS(\btheta;y_{a(1)}),\ldots,p^{-1}_{a(n)}\bS(\btheta;y_{a(n)}))$ is a $p\times n$ matrix, $\bI_n$ is the $n\times n$ identity matrix,  $\mathcal{P}_{\bone,n} = \bone_n(\bone_n^{\T}\bone_n)^{-1}\bone_n^{\T}$ is the projection matrix to the linear space spanned by $\bone_n$, and $\bone_n$ is a vector of one with length $n$.
	
		Then,  the bootstrap replicate of $\hat{\bS}_w(\btheta)$ is $\hat{\bS}_w^*(\btheta) = (Nn)^{-1} \sum_{i=1}^n r_i^*p^{-1}_{a(i)} \bS(\btheta;y_{a(i)})$ with 
		the 		rescaling factor is $r_{i}^* = k_nm_{i}^*$, where $\bbm^*=(m_{1}^*,\ldots,m_{n}^*)^{\T}$ is generated using a multinomial distribution with $n-1$ trials and a success probability vector $n^{-1}\bone_n$, and  $k_n= n/(n-1)$. Using the property of the multinomial distribution, we can verify that $E_*
		\{\hat{\bS}_w^*(\btheta) \} = \hat{\bS}_w(\btheta)$ and $\var_* \{ \hat{\bS}_w^*(\btheta) \}= \hat{V}\{\hat{\bS}_w(\btheta)\}$.  Furthermore, in Section~\ref{ss: pps sampling} of the Supplementary Material, we show that the proposed bootstrap method satisfies the conditions for the bootstrap central limit theorem in Theorem~\ref{theo: (6)}.
\end{example}

\begin{example}\label{ex: 3} (Stratified multi-stage cluster  sampling  with  clusters selected with replacement) 
	
Under stratified multi-stage  cluster sampling, a popular approach is to assume that $m_h (\ge 2)$ clusters are selected with replacement from the $h$-th stratum, with the selection probabilities $p_{hi}$ satisfying $\sum_{i=1}^{M_h} p_{hi} = 1$  for $h=1,\ldots,H$ \citep{rao1988}, where $M_{h}$ is the number of clusters in the $h$-th stratum.  Within each sampled cluster, denote $\hat{\bS}_{w,hi}(\btheta)$ to be a design-unbiased estimator  of the cluster mean  under the second and subsequent stages.  For simplicity, we still use $\hat{\bS}_{w,hi}(\btheta)$ for the case when the $(hi)$-th cluster is fully observed. 
Then, the weighted score function is $\hat{\bS}_w(\btheta) = \sum_{h=1}^HW_h\hat{\bS}_{w,h}(\btheta)$, where $W_h = N_h/N$, $N=\sum_{h=1}^HN_h$, $N_h = \sum_{i=1}^{M_h}M_{hi}$, $M_{hi}$ is the size of the $(hi)$-th cluster, $\hat{\bS}_{w,h}(\btheta) = m_h^{-1}\sum_{i=1}^{m_h}\tilde{\bS}_{w,hi}(\btheta)$, $\tilde{\bS}_{w,hi}(\btheta) = d_{a(hi)}p_{a(hi)}^{-1}\hat{\bS}_{w,a(hi)}(\btheta)$, $d_{a(hi)} = M_{a(hi)}/N_h$, and $a(hi)$ is the index of the $i$-th selected cluster   in the $h$-th stratum.  

The bootstrap replicate of $\hat{\bS}_w(\btheta)$ can be obtained in the following way. For the $h$-th stratum, generate $\bbm_h^*=(m_{h1}^*,\ldots,m_{hn_h}^*)^{\T}$ by a multinomial distribution with $m_h-1$ trials and a success probability vector $m_h^{-1}(1,\ldots,1)^{\T}$ of length $m_h$.  Then, the bootstrap weighted score function is $\hat{\bS}^*_w(\btheta) = \sum_{h=1}^HW_h\hat{\bS}^*_{w,h}(\btheta)$, where $\hat{\bS}_{w,h}^*(\btheta) = m_h^{-1}\sum_{i=1}^{m_h}r_{hi}^*\tilde{\bS}_{w,hi}(\btheta)$, $r_{hi}^*=k_hm_{hi}^*$, and $k_h = m_h/(m_h-1)$; \citet{rao1992some} proposed a similar rescaling bootstrap method under stratified multi-stage cluster sampling, but they did not investigate the corresponding theoretical properties. Using a similar argument in Example~\ref{ex: 2}, we can show that $E_*
\{\hat{\bS}_w^*(\btheta) \} = \hat{\bS}_w(\btheta)$ and $\var_* \{ \hat{\bS}_w^*(\btheta) \}= \hat{V}\{\hat{\bS}_w(\btheta)\}$, where $\hat{V}\{\hat{\bS}_w(\btheta)\} = \sum_{h=1}^HW_h^2\{m_h(m_h-1)\}^{-1} \tilde{\bS}_{h,vec}(\theta)(\bI_{m_h} - \mathcal{P}_{\bone,m_h})\tilde{\bS}{}^{\T}_{h,vec}(\theta)$, and $$\tilde{\bS}_{h,vec}(\theta) = (d_{a(h1)}p^{-1}_{a(h1)}\hat{\bS}_{w,a(h1)}(\btheta),\ldots,d_{a(hm_h)}p^{-1}_{a(hm_h)}\hat{\bS}_{w,a(hm_h)}(\btheta)).$$  In Section~\ref{sec: stscl} of the Supplementary Material, we  show that the bootstrap central limit theorem in Theorem~\ref{theo: (6)}  holds with $n=\sum_{h=1}^Hm_h$. 
\end{example}

Under stratified simple random sampling without replacement \citep{bickel1984asymptotic},   a bootstrap method is proposed  in Section~\ref{ss: stratified random sampling} of the  Supplementary Material, and  it is essentially the same as the one considered by \citet{rao1988} when estimating the population mean;  see \citet{rao1988} and \citet{beaumont2012generalized} for details.

To establish a bootstrap version of the Wilks' theorem in (\ref{5}), we use
\begin{equation*}
W^* ( \hat{\btheta} ) = -2 n \{ l_w^*(\hat{\btheta}) - l_w^* ( \hat{\btheta}^* ) \}
\label{8}
\end{equation*}
as the bootstrap version of $W ( \btheta_0) $ in (\ref{4}).
The following theorem establishes the limiting distribution of $W^* ( \hat{\btheta} )$. 
\begin{theorem}\label{theo: 1}
Under regularity conditions in Section \ref{ss: regularity conditions}  of the Supplementary Material, we further obtain 
	\begin{equation*}
	W^* ( \hat{\btheta} ) \mid A   {\,\longrightarrow\, }   \mathcal{G}   \label{8b}
	\end{equation*} in distribution
	as $n \rightarrow \infty$, and 
	the reference distribution is the bootstrap sampling distribution conditional on  the   sample $A$,  where $\mathcal{G}$ is the same limiting  distribution in (\ref{5}).
\end{theorem}
The proof of Theorem~\ref{theo: 1} is given in Section~\ref{ss: theorem 1} of the Supplementary Material.
By  (\ref{5}) and Theorem~\ref{theo: 1}, we conclude that the likelihood ratio test statistic generated by the proposed bootstrap method has the same asymptotic distribution as the one in (\ref{4}). Thus, we can use the conditional distribution of $W^* ( \hat{\btheta} )$ to approximate that of $W ({\btheta}_0 )$, which validates the proposed bootstrap method for hypothesis testing problems.

\section{Likelihood-ratio test}\label{sec: lrt}
Without loss of generality, denote $\btheta$ to be the true parameter $\btheta_0$ in the following two sections.
Let $\Theta_0 \subset \Theta$ be the parameter space under the null hypothesis, so the null hypothesis can be written as $H_0: \btheta \in \Theta_0$. In this section, we consider $H_0: \btheta_{2} = \btheta_2^{(0)}$ for some known vector $\btheta_2^{(0)}$, where $\btheta=(\btheta{}^\T_{1}, \btheta{}^\T_{2}){}^\T$. Thus, we have
$ \Theta_0 = \{ \btheta  \in \Theta ; \btheta_{2} = \btheta_2^{(0)} \} .$
Let $\hat{\btheta}_{1}^{(0)} $ be the profile pseudo maximum likelihood estimator of $\btheta_{1}$ under  $H_0: \btheta_{2} = \btheta_2^{(0)}$, which can be obtained by maximizing
$ l_{w} ( \btheta_1, \btheta_2^{(0)} ) $
with respect to $\btheta_1$.

The pseudo likelihood-ratio test statistic for testing $H_0: \btheta_{2} = \btheta_2^{(0)}$ is defined as
\begin{equation}
W(\btheta_2^{(0)}) =  - 2 n \{ l_w ( \hat{\btheta}^{(0)}  ) - l_w ( \hat{\btheta})  \},
\label{lr}
\end{equation}
where $\hat{\btheta}^{(0)} = ( \hat{\btheta}_{1}^{(0)}{}^\T, \btheta_2^{(0)}{}^\T ){}^\T$. Under simple random sampling with replacement, $W(\btheta_2^{(0)}) $ in (\ref{lr}) is asymptotically distributed as  $\chi^2( q)$  with $q= p- p_0$ and $p_0= \mbox{dim} ( \Theta_0) $, and the reference distribution  is the joint distribution of the super-population model and the sampling mechanism, where $\mbox{dim}(\Theta)$ is the dimension of $\Theta$.
The following theorem, proved by \citet{lumley2014tests},  presents the limiting distribution of the pseudo likelihood-ratio test statistic in (\ref{lr}) for general sampling designs.
\begin{lemma}\label{th: 2}
	Under $H_0: \btheta_{2} = \btheta_2^{(0)}$ and  regularity conditions in Section \ref{ss: regularity conditions}  of the Supplementary Material,
	\begin{equation}
	W(\btheta_2^{(0)}) { \,\longrightarrow\, }   \mathcal{G}_1 = \sum_{i=1}^{q} c_{i} Z_i^2
	\label{8c}
	\end{equation} in distribution as $n\to\infty$,	and the reference distribution  is the joint distribution of the super-population model and the sampling mechanism,
	where   $c_{1} \ge c_2 \ge \cdots \ge c_q >0$ are the eigenvalues of $\bP =  \bSigma_{\btheta,2}   \bcalI_{22\cdot 1}(\btheta_{1},\btheta_2^{(0)})$, 
	$Z_1, \ldots, Z_q$ are $q$ independent random variables from the standard normal distribution, 
$\bSigma_{\btheta,2}$ is the asymptotic variance of $n^{1/2}\hat{\btheta}_2$ in (\ref{2}),	$\bcalI_{22 \cdot 1}(\btheta_{1},\btheta_2) = \bcalI_{22}(\btheta_{1},\btheta_2) - \bcalI_{21}(\btheta_{1},\btheta_2) \{ \bcalI_{11}(\btheta_{1},\btheta_2)\} ^{-1}\bcalI_{12}(\btheta_{1},\btheta_2)
	,$ and
	$\bcalI_{ij}(\btheta_{1},\btheta_2) =E[-\partial^2 l\{(\btheta_{1},\btheta_2);Y\}/(\partial \btheta_i \partial \btheta{}^\T_j ) ]$
	for $i,j=1,2$.
\end{lemma}
\noindent\citet{lumley2014tests} proposed to estimate the limiting distribution in (\ref{8c}) using a design-based estimator of $\bP = \bSigma_{\btheta,2} \bcalI_{22\cdot 1}(\btheta_{1},\btheta_2^{(0)}) $.

We  consider an alternative test using a  novel application of the bootstrap  method discussed in Section~\ref{sec: bootstrap}.
To do this,  a bootstrap version of the pseudo likelihood-ratio test statistic in (\ref{lr})  is obtained as
\begin{equation}
W^*(\hat{\btheta}_{2})=  - 2 n \{ l_w^* ( \hat{\btheta}_{1}^{*(0)}, \hat{\btheta}_{2}) - l_w^* ( \hat{\btheta}^*)
\},
\label{w2}
\end{equation}
where
$ \hat{\btheta}_{1}^{*(0)} = \mbox{arg} \max_{\btheta_1} l_w^* ( \btheta_1, \hat{\btheta}_{2}) $, and $\hat{\btheta}_{2}$ is the second component of $\hat{\btheta} = ( \hat{\btheta}{}^\T_{1}, \hat{\btheta}{}^\T_{2}){}^\T$. 


The following theorem establishes that  the conditional distribution of $W^*(\hat{\btheta}_{2})$, given  sample $A$, converges in distribution to $\mathcal{G}_1$ as $n\to\infty$, the same limiting distribution as $W(\btheta_2^{(0)})$ in (\ref{lr}).
\begin{theorem}\label{theo: 3}
Under regularity conditions in Section \ref{ss: regularity conditions}  of the Supplementary Material,
	\begin{equation}
W^*(\hat{\btheta}_{2})\mid A  { \,\longrightarrow\, }     \mathcal{G}_1
	\label{8d}
	\end{equation} in distribution as $n\to\infty$, and 
	the reference distribution is the bootstrap sampling distribution conditional on  the   sample $A$, where $\mathcal{G}_1$ is defined in Lemma~\ref{th: 2}.
\end{theorem}

The proof of Theorem~\ref{theo: 3} is given in Section~\ref{ss: theorem 3} of the Supplementary Material. By Theorem~\ref{theo: 3}, we can use the empirical distribution of $W^*(\hat{\btheta}_{2})$ in (\ref{w2}) to approximate the sampling distribution of  $W({\btheta}_2^{(0)})$ under $H_0: \btheta_{2} = \btheta_2^{(0)}$. Thus, the $p$-value for testing  $H_0$ using $W({\btheta}_2^{(0)})$  can be obtained by computing the proportion of $W^*(\hat{\btheta}_{2})$ greater than $W({\btheta}_2^{(0)})$.

\section{Quasi-score test}\label{sec: qst}

In this section, we consider another test without assuming a parametric super-population  model.  In the context of generalized linear regression  analysis with survey data, \citet{rao1998quasi} developed a design-based test procedure using   regression model assumptions in the super-population model.  Let $y$ be the study variable of interest and $\bx$  be the vector of auxiliary variables  used in the regression model.
Suppose that the super-population model satisfies
$
E( Y_i \mid \bx_i) = \mu ( \bx_i; \btheta)
\label{5.1}
$
 for known function $\mu( \cdot)$ with unknown parameter $\btheta$. Also, we assume a ``working'' model for the variance
$
V( Y_i \mid \bx_i) = V_0 ( \mu_i)
\label{5.2}
$ 
with known $V_0( \cdot)$ and $\mu_i =  \mu ( \bx_i; \btheta)$. Under this setup, a consistent estimator of $\btheta$ can be obtained by solving the quasi-score function of $\btheta$ given by
\begin{equation}
\hat{\bS}_w ( \btheta) = N^{-1}  \sum_{i \in A} w_i  \bu(\btheta; y_i) = \bzero,
\label{5.3}
\end{equation}
where $\bu (\btheta; y_i) = ( y_i - \mu_i) \{ V_0( \mu_i) \}^{-1} \left( \partial \mu_i / \partial \btheta \right)$. Under regularity conditions \citep{binder1983variances}, the solution $\hat{\btheta}$ of (\ref{5.3}) can be expressed as $\hat{\btheta} = \mbox{arg} \max_{\btheta \in \Theta}N^{-1}\sum_{i\in A}w_i l(\btheta;y_i)$, where 
$l(\btheta;y_i)$ is the quasi-likelihood function \citep{wedderburn1974} satisfying $\partial l(\btheta; y)/\partial \mu = \bu (\btheta; y)$. Thus,   the asymptotic theory in Section 2  follows directly. 
  That is, 
\begin{equation*}
\sqrt{n} (  \hat{\btheta} - \btheta )  {\,\longrightarrow\, }  N ( \bzero, \bSigma_\btheta )
\label{5-4}
\end{equation*}
in distribution as $n\to\infty$, and the reference distribution  is the joint distribution of the super-population model and the sampling mechanism,
where
$
\bSigma_\btheta = \bcalI(\btheta)^{-1}  \bSigma_S(\btheta) \{\bcalI(\btheta)^{-1}\}{}^{\T},
$
$ \bcalI(\btheta)= E\{ \bI( \btheta; Y) \}$, $\bI(\btheta ; y) = -\partial \bu (\btheta; y) /\partial \btheta{}^\T $, and $\bSigma_S(\btheta) =\lim_{n\to\infty} [n\var\{\hat{\bS}_w(\btheta)\}]$.

We now consider  the quasi-score test  of   $H_0: \btheta_{2} = \btheta_2^{(0)}$ under the above setup.
Given  $\btheta= (\btheta{}^\T_1, \btheta{}^\T_2){}^\T$,  we can write
\begin{equation}
\hat{\bS}_w ( \btheta) = \begin{pmatrix}
\hat{\bS}_{w1} ( \btheta) \\ \hat{\bS}_{w2} ( \btheta)
\end{pmatrix}
= N^{-1}\left( \begin{array}{ll}
\sum_{i \in A} w_i \bu_1 (\btheta ; y_i) \\
\sum_{i \in A} w_i \bu_2 (\btheta ; y_i)
\end{array}
\right)
\label{5.5}
\end{equation}
with  $\bu_j  ( \btheta ; y_i) =   ( y_i - \mu_i) \{ V_0( \mu_i) \}^{-1} \left( \partial \mu_i / \partial \btheta_j \right)$ for $j=1,2$.
Let $\hat{\btheta}_1^{(0)}$ be the solution to
$$ \hat{\bS}_{w1}  \left( \btheta_1, \btheta_2^{(0)} \right) = \bzero,  $$
where $\hat{\bS}_{w1} ( \btheta) $ is defined in (\ref{5.5}).  The quasi-score test for $H_0: \btheta_{2} = \btheta_2^{(0)}$ is
\begin{equation}
X_{QS}^2 \left( {\btheta}_2^{(0)}  \right) = \hat{\bS}_{w}^{\T}  \left( \hat{\btheta}^{(0)} \right) \left\{  \hat{\bI}_{w} \left( \hat{\btheta}^{(0)} \right) \right\}^{-1} \hat{\bS}_{w} \left( \hat{\btheta}^{(0)}\right),
\label{5.6}
\end{equation}
where $\hat{\btheta}^{(0)} = ( \hat{\btheta}^{(0)}_1{}^\T , \btheta_2^{(0)}{}^\T ){}^\T$, and $\hat{\bI}_w ( \btheta ) = N^{-1} \sum_{i \in A} w_i \bI ( \btheta; y_i) $.
Now, based on the partition $\hat{\btheta}=(\hat{\btheta}{}^\T_1, \hat{\btheta}{}^\T_2){}^\T$, we can write
$$ \hat{\bI}_w ( \btheta) = \left(
\begin{array}{ll}
\hat{\bI}_{w11} ( \btheta) & \hat{\bI}_{w12} ( \btheta) \\
\hat{\bI}_{w21} ( \btheta) & \hat{\bI}_{w22} ( \btheta)
\end{array}
\right).
$$
Since $\hat{\btheta}^{(0)} $ satisfies $\hat{\bS}_{w1} ( \hat{\btheta}^{(0)} ) =\bzero$, the test statistic in (\ref{5.6}) is algebraically equivalent to
\begin{equation}
X_{QS}^2\left({\btheta}_2^{(0)} \right)  = \hat{\bS}_{w 2 }^{\T} \left( \hat{\btheta}^{(0)}\right)  \left\{  \hat{\bI}_{w 22 \cdot 1} \left( \hat{\btheta}^{(0)} \right) \right\}^{-1} \hat{\bS}_{w 2 } \left( \hat{\btheta}^{(0)}\right),
\label{11}
\end{equation}
where
\begin{equation}
\hat{\bI}_{w 22 \cdot 1} ( \btheta) =\hat{\bI}_{w22} ( \btheta) - \hat{\bI}_{w21} (\btheta) \{ \hat{\bI}_{w11} ( \btheta) \}^{-1} \hat{\bI}_{w12} ( \btheta) .
\label{12}
\end{equation}
The matrix $ \hat{\bI}_{w 22 \cdot 1}$ in (\ref{12}) is a $q\times q$ matrix with $q=\mbox{dim} ( \Theta_0)$, while the matrix $\hat{\bI}_w ( \btheta)$ in (\ref{5.6}) is a $p \times p$ matrix with $p =\mbox{dim} (\Theta)$. Thus, if $p$ is much larger than $q$, the computation of (\ref{11}) is more efficient than the computation of (\ref{5.6}).
The following result establishes the limiting distribution of (\ref{11}), and its  proof is given in Section \ref{ss: theorem 4} of the Supplementary Material.
\begin{theorem}\label{cor: 2}
	Under $H_0: \btheta_{2} = \btheta_2^{(0)}$ and the regularity conditions in Section \ref{ss: regularity conditions}  of the Supplementary Material,
	$$nX_{QS}^2\left(\btheta_2^{(0)}\right) {\,\longrightarrow\,} \mathcal{G}_1$$ in distribution
	as $n \rightarrow \infty$,  and the reference distribution  is the joint distribution of the super-population model and the sampling mechanism, where $\mathcal{G}_1$ is defined in Lemma~\ref{th: 2}.
\end{theorem}

We now develop a bootstrap method to approximate the limiting distribution of the quasi-score test statistic in (\ref{11}).
Similarly to the likelihood-ratio method in Section~4, we first develop a bootstrap version of the quasi-score equation, that is,
\begin{equation*}
\hat{\bS}_w^* ( \btheta) = N^{-1}  \sum_{i \in A} w_i^*  \bu(\btheta; y_i) = \bzero.
\label{13}
\end{equation*}
Let $\hat{\btheta}_1^{*(0)} $ be the solution to $\bS_{w1}^* ( {\btheta}_1, \hat{\btheta}_{2}) =\bzero$.
Then, the bootstrap version of the quasi-score test statistic $X_{QS}^2 (\btheta_2^{(0)}) $ is
\begin{equation}
X_{QS}^{2*}( \hat{\btheta}_{2})  = \hat{\bS}_{w 2 }^{*\T} \left( \hat{\btheta}^{*(0)}\right) \left\{  \hat{\bI}_{w 22 \cdot 1}^* \left( \hat{\btheta}^{*(0)} \right) \right\}^{-1} \hat{\bS}_{w 2 }^* \left( \hat{\btheta}^{*(0)}\right), \label{eq: x2}
\end{equation}
where $\hat{\btheta}^{*(0)} = ( \hat{\btheta}_1^{*(0)}{}^\T ,  \hat{\btheta}_{2}{}^\T){}^\T $, and $\hat{\bI}_{w 22 \cdot 1}^*$ is computed similarly to (\ref{12}) using the bootstrap weights. Similarly to Theorem~\ref{cor: 2}, we have the following result.
\begin{theorem}\label{cor: 3}
Under  regularity conditions in Section \ref{ss: regularity conditions}  of the Supplementary Material, we have
	$$nX_{QS}^{2*}( \hat{\btheta}_{2})\mid A {\,\longrightarrow\,}  \mathcal{G}_1$$
	in distribution as $n\to\infty$, and 
	the reference distribution is the bootstrap sampling distribution conditional on  the   sample $A$, where $\mathcal{G}_1$ is defined in Lemma~\ref{th: 2}.
\end{theorem}
The proof of Theorem~\ref{cor: 3} is essentially the same as that of Theorem~\ref{cor: 2}. Given the  sample $A$, we can use $X_{QS}^{2*}( \hat{\btheta}_{2})$ to approximate the sampling distribution of $X_{QS}^2(\btheta_2^{(0)})$ in (\ref{11}). The bootstrap distribution can be used to control the size of the test  based on  $X_{QS}^2(\btheta_2^{(0)})$ in (\ref{11}).

\begin{example}
	For an illustration of the proposed bootstrap method, we consider a logistic regression model. Specifically, we assume that $Y\in\{0,1\}$ is a binary random variable, and  $\mathrm{logit}\{\mathrm{pr}(Y=1\mid \bX=\bx)\} =  (1,\bx{}^\T)\btheta$, where $\mathrm{logit}(p) = \log(p) - \log(1-p)$, and $\btheta = (\theta_{1},\btheta{}^\T_{2}){}^\T$. A sample $A$ is obtained by Poisson sampling, and the weighted score equation associated with the logistic regression model is
	\begin{equation}
	\hat{\bS}_w(\btheta) = N^{-1} \sum_{i \in A} w_i\{y_i-p_i(\btheta)\}(1,\bx_i)^{\T} = \bzero,
	\label{11b}
	\end{equation}
	where $\mathrm{logit} \{ p_i (\btheta) \} = \theta_1 + \bx^{\T}_i \btheta_2$.
	In this case, we obtain $\hat{\bI}_w(\hat{\btheta})= N^{-1}\sum_{i\in A}w_i\hat{p}_i(1-\hat{p}_i)(1,\bx{}^\T_i)(1,\bx{}^\T_i){}^\T$ and  $\hat{p}_i = p_i(\hat{\btheta})$.
	
	Suppose that we are interested in testing $H_0: \btheta_{2}=\btheta_2^{(0)}$. The profiled pseudo maximum likelihood estimator of $\btheta$ under $H_0$, denoted by $\hat{\btheta}^{(0)}$, can be obtained by solving (\ref{11b}) subject to $\btheta_{2}= \btheta_2^{(0)}$.
	The quasi-score test statistic is
	\begin{eqnarray}
	X_{QS}^2(\btheta_2^{(0)}) &=& N^{-1}\left\{\sum_{i\in A}w_i(y_i-\hat{p}_i^{(0)}) \bx_i^{\T} \right\}\left\{\sum_{i\in A}w_i\hat{p}_i^{(0)}(1-\hat{p}_i^{(0)}) (\bx_i - \bar{\bx}_w)^{\otimes 2}   \right\}^{-1}\left\{\sum_{i\in A}w_i(y_i-\hat{p}_i^{(0)}) \bx_i \right\}
	,\notag
	\end{eqnarray}
	where 
	$$ \bar{\bx}_w = \frac{\sum_{i\in A}w_i\hat{p}_i^{(0)}(1-\hat{p}_i^{(0)}) \bx_i }{
		\sum_{i\in A}w_i\hat{p}_i^{(0)}(1-\hat{p}_i^{(0)})
	} , $$ and
	$\hat{p}_i^{(0)} = p_i(\hat{\btheta}^{(0)} )$. The bootstrap replicates of $X_{QS}^2(\btheta_2^{(0)})$ can be   constructed by replacing the sampling weights $w_i$ and the profile pseudo maximum likelihood estimator $\hat{\btheta}^{(0)}$, respectively,  by the bootstrap weights $w_i^*$ and the bootstrap profile pseudo maximum likelihood estimator   that solves
	$
	\hat{\bS}_w^*(\btheta) =  N^{-1} \sum_{i \in A} w_i^* \{y_i-p_i(\btheta)\}(1,\bx{}^\T_i){}^\T = \bzero
	$
	subject to $\btheta_2=  \hat{\btheta}_{2}$.
\end{example}

\section{Simple goodness-of-fit test for categorical data} \label{sec: gof}

Suppose that a finite population $U$ is partitioned into $K$ categories with $U=U^{(1)} \cup \cdots \cup U^{(K)}$. Denote $p_k=N_k/N$ to be the population proportion of the $k$-th category, where $N_k$ is the size of $U^{(k)}$. 
Thus, we are implicitly assuming that the finite population is an IID realization of the super-population model following a multinomial distribution with $K$ categories.  

In this section, we are interested in the simple goodness-of-fit test $H_0: p_k=p_{k}^{(0)}$ for $ k=1, \ldots, K$, where $(p_{1}^{(0)}, \ldots, p_{K}^{(0)} ){}^\T$ is a pre-specified vector satisfying $\sum_{k=1}^K p_{k}^{(0)} = 1$.  
From the sample $A$, we compute $\hat{p}_k = \hat{N}_k/ \hat{N}$ as an estimator of $p_k$, where
$\hat{N}_k= \sum_{i \in A_k} w_i $ is a design-unbiased estimator of $N_k$, $A_k = A \cap U^{(k)}$,
and $\hat{N} = \sum_{k=1}^K \hat{N}_k=\sum_{i \in A} w_i$. The estimated proportions can  be obtained by {$\hat{\bp} = \mbox{arg} \max_{\bp \in \Theta}\sum_{i\in A}w_il(\bp;\bx_i)$, where $l(\bp;\bx_i) = 
\sum_{k=1}^K x_{ik} \log (p_k)$,  with $x_{ik}=1$ if the $i$-th element belongs to the $k$-th category and 0 otherwise, $\bp=(p_1,\ldots,p_K)^\T$}. 

Then, the Pearson chi-squared goodness-of-fit test statistic for $H_0$ is 
\begin{equation*}
X^2(\bp^{(0)})= n \sum_{k=1}^K \left( \hat{p}_k - p_{k}^{(0)} \right)^2/ p_{k}^{(0)},
\end{equation*}
where $\bp^{(0)} = ({p}_{1}^{(0)}, \ldots, {p}_{K-1}^{(0)} ){}^\T$. 
If we assume a multinomial distribution for  the super-population model, we can compute the likelihood-ratio test statistic as
$$ W(\bp^{(0)}) =2n \sum_{k=1}^K \hat{p}_k \log \left( \frac{\hat{p}_k }{ p_{k}^{(0)} } \right) .$$
Denoting $\hat{{\bp}} = (\hat{p}_1, \ldots, \hat{p}_{K-1} ){}^\T$, we have
$$ \sqrt{n} \left( \hat{\bp} - \bp^{(0)} \right) {\,\longrightarrow\,} N( \bzero, \bSigma_p)  $$
in distribution as $n\to\infty$ under $H_0$,  and the reference distribution  is the joint distribution of the super-population model and the sampling mechanism, where $\bSigma_p$ is the asymptotic variance of $\sqrt{n}\hat{\bp}$. Under simple random sampling   with replacement, $\bSigma_p$ is equal to $\bP_0 = \mbox{diag} (\bp^{(0)}) - \bp^{(0)}(\bp^{(0)}){}^\T $, where $\mbox{diag} (\bp^{(0)})$ is a diagonal matrix with $\bp^{(0)}$ being the vector of diagonal elements.  For other sampling designs, $\bSigma_p$ is more complicated.
Under regularity conditions,  according to \citet{Scott1981CHI}, we have
\begin{equation}
X^2(\bp^{(0)}),\,  W(\bp^{(0)}) \,\longrightarrow\, \mathcal{G}_2=\sum_{k=1}^{K-1} \lambda_kZ_k^2,
\label{b3}
\end{equation}
in distribution as $n\to\infty$ under $H_0$,  and the reference distribution  is the joint distribution of the super-population model and the sampling mechanism,
where $\lambda_1 \geq\lambda_2\geq \cdots \geq \lambda_{K-1}>0$ are the eigenvalues of the design effect matrix $\bD= \bP_0^{-1} \bSigma_p$, and  $Z_1, \ldots, Z_{K-1}$ are  $K-1$ independent random variables from the standard normal distribution.
Under simple random sampling with replacement,   the limiting distribution in (\ref{b3}) is a chi-squared distribution with $K-1$ degrees of freedom.

If  a chi-squared distribution with $K-1$ degrees of freedom is blindly used as the reference distribution for $X^2(\bp^{(0)})$, the resulting inference can be misleading. For example, under some two-stage cluster sampling design with $\lambda_i = \lambda (>1)$ for $i=1,\ldots,K-1$,  the type I error rate of using a chi-squared distribution with $K-1$ degrees of freedom is approximately equal to  $\mbox{pr} \{ X^2(\bp^{(0)}) >  \chi_{K-1}^2 ( \alpha) \} = \mathrm{pr}\{X>\lambda^{-1}\chi_{K-1}^2 ( \alpha) \}$, where $\chi_{K-1}^2 ( \alpha) $ is the $(1-\alpha)$ quantile of a chi-squared distribution with $K-1$ degrees of freedom,  $\alpha$ is a nominal significance level, and $X$ is chi-squared distributed with $K-1$ degrees of freedom. The resulting type I error rate increases with $\lambda$, and this can be arbitrarily large by increasing $\lambda$.  To overcome this problem, \citet{Scott1981CHI} proposed a first-order correction which compares $ X^2_{C}(\bp^{(0)}) =  X^2 (\bp^{(0)}) / \hat{\lambda}_{+} $ to $\chi_{K-1}^2(\alpha)$, 
where
$$ \hat{\lambda}_{+} = \frac{1}{k-1} \sum_{i=1}^k \frac{ \hat{p}_i}{ p_{i}^{(0)}} (1- \hat{p}_i) \hat{d}_i, $$
and $\hat{d}_i$ is an estimated design effect of $\hat{p}_i$, which depends on the estimated variance of $\hat{p}_i$.
The second-order Rao-Scott correction \citep{rao1984chi,ThomasRao1987}  requires the knowledge of the fully estimated covariance matrix of the estimated proportions.
Stata and other survey software use the Rao-Scott corrections as a default option.


We now apply  the proposed bootstrap  method to approximate  the limiting distribution in (\ref{b3}), without computing the estimated covariance matrix $\hat{\bSigma}_p$. To describe the proposed method,
let $\hat{{\bp}}^*$ be the estimator of the population proportion $\bp$ based on the  bootstrap weights $w_i^*$. 
The proposed bootstrap statistics of  $ X^2(\bp^{(0)})$ and $W(\bp^{(0)})$ are
\begin{eqnarray*}
	X^{2*}(\hat{\bp}) = n \sum_{i=1}^K (\hat{p}_i^* - \hat{p}_{i})^2 / \hat{p}_{i}\quad \mbox{and}\quad
	W^*(\hat{\bp}) =2n \sum_{i} \hat{p}_i^* \log \left( \hat{p}_i^* / \hat{p}_{i}\right) ,
\end{eqnarray*}
respectively. We use $\hat{p}_{i}$ in place of $p_{i}^{(0)}$ in the bootstrap test statistics.
The following theorem establishes the limiting distribution of the proposed bootstrap test statistics.
\begin{theorem}\label{theo 6}Under regularity conditions in Section \ref{ss: regularity conditions}  of the Supplementary Material,
	\begin{equation}
	X^{2*} (\hat{\bp}),\, W ^*(\hat{\bp})\mid A  { \,\longrightarrow\, }  \mathcal{G}_2
	\label{b4}
	\end{equation}in distribution as $n\to\infty$, and 
	the reference distribution is the bootstrap sampling distribution conditional on  the   sample $A$,
	where $\mathcal{G}_2$ is defined in (\ref{b3}).
	
	\label{theo: 6}
\end{theorem}

The proof of Theorem~\ref{theo: 6} is given in Section \ref{ss: theorem 6} of the Supplementary Material. By Theorem~\ref{theo: 6}, we can use the bootstrap samples to approximate the sampling distribution of the test statistics. 


\section{Test of Independence}\label{eq: toi}

We now  discuss a bootstrap test of independence in a two-way table of counts. Let $p_{ij} = N_{ij}/N$ be the  population proportion for cell $(i,j)$ with margins $p_{i+}$ and $p_{+j}$ for $i=1,\ldots,R$ and $j=1,\ldots,C$, where $R$ and $C$ are the numbers of rows and columns, and $\{ N_{ij}; i=1, \ldots, R, j=1, \ldots, C\}$ is the set of  population counts with margins $N_{i+}$ and $N_{+j}$. Let  $\hat{N}_{ij}$ be a design-unbiased estimator of $N_{ij}$ and  $\hat{p}_{ij} = \hat{N}_{ij}/ \hat{N}$. By assuming a multinomial distribution for  the super-population model, the chi-squared statistic and the likelihood-ratio 
test statistic   for testing independence $H_0: p_{ij} = p_{i+} p_{+j}$ for all $i$ and $j$ are 
\begin{eqnarray}
&\displaystyle X_{ I}^2 = n \sum_{i=1}^R \sum_{j=1}^C  \frac{ ( \hat{p}_{ij} - \hat{p}_{i+} \hat{p}_{+j} )^2 }{ \hat{p}_{i+} \hat{p}_{+j} },\label{eq: 7chi}\\ 
&\displaystyle W_{ I} = 2n \sum_{i=1}^R \sum_{j=1}^C \hat{p}_{ij} \log \left(  \frac{ \hat{p}_{ij} }{ \hat{p}_{i+} \hat{p}_{+j}  } \right) . \label{eq: 7w}
\end{eqnarray}
Since $ X_{ I}^2$ and $W_I$ are asymptotically equivalent under $H_0$, \citet{Scott1981CHI} have shown  that 
\begin{equation}
X_{I}^2,\, W_{I}  {\,\longrightarrow\, } \mathcal{G}_3=\sum_{l=1}^{d} \delta_l Z_l^2
\label{b6}
\end{equation} in distribution as $n\to\infty$   under $H_0$, and the reference distribution  is the joint distribution of the super-population model and the sampling mechanism,
where  $\delta_1 \geq \delta_2\geq \cdots \geq \delta_d>0$ are the $d$  eigenvalues of a design effect matrix discussed in  in Section \ref{ss: theorem 7} of the Supplementary Material,   $d=(R-1) (C-1)$, and  $Z_1, \ldots, Z_{d}$  are $d$ independent random samples from the standard normal distribution.

We now consider bootstrap tests of $H_0$ for two-way tables. Let $ \hat{p}_{ij}^*$ be the bootstrap cell proportion computed using the bootstrap weights, $\hat{p}_{i+}^* = \sum_{j=1}^C \hat{p}_{ij}^* $, and $ \hat{p}_{+j}^* = \sum_{i=1}^R \hat{p}_{ij}^*$. The proposed bootstrap version of  $X_{I}^2$ is  
\begin{equation}X_{I}^{2*} = n \sum_{i=1}^R \sum_{j=1}^C  \frac{ \{( \hat{p}_{ij}^* - \hat{p}_{i+}^* \hat{p}_{+j}^* )-  ( \hat{p}_{ij} - \hat{p}_{i+} \hat{p}_{+j} ) \}^2  }{( \hat{p}_{i+} \hat{p}_{+j} )}. \label{eq: 7chi*}
\end{equation}
Under $H_0$, terms in the numerator of  $X_{I}^2$ are identical to
$\{ ( \hat{p}_{ij} - \hat{p}_{i+} \hat{p}_{+j}) -  (p_{ij} - p_{i+} p_{+j} )  \}^2$. That is, the bootstrap test statistic is computed by  simply replacing $\{ \hat{p}_{ij}, \hat{p}_{i+}, \hat{p}_{+j} \}$ and $\{ p_{ij}, p_{i+}, p_{+j} \}$ with
$\{ \hat{p}_{ij}^*, \hat{p}_{i+}^*, \hat{p}_{+j}^* \}$ and   $\{
\hat{p}_{ij}, \hat{p}_{i+}, \hat{p}_{+j} \}$, respectively.

Let $\Delta_{ij}= p_{ij}/ (p_{i+} p_{+j} ) $. Since $\Delta_{ij}=1$ under $H_0$, the test statistic $W_I$ may be written as 
\begin{equation}
W_{I} = 2n \sum_{i=1}^R \sum_{j=1}^C \left[ \hat{p}_{ij} \log \left\{ \frac{\hat{p}_{ij} }{ \hat{p}_{i+} \hat{p}_{+j} \Delta_{ij}  } \right\}
- ( \hat{p}_{ij} - \hat{p}_{i+} \hat{p}_{+j} \Delta_{ij} )
\right]. \label{eq: 23a}
\end{equation} 

The bootstrap version $W_{I}^*$ is   obtained by replacing $\{ \hat{p}_{ij}, \hat{p}_{i+}, \hat{p}_{+j} \}$ with $\{ \hat{p}_{ij}^*, \hat{p}_{i+}^*, \hat{p}_{+j}^* \}$ and $\Delta_{ij}$ with $\hat{\Delta}_{ij}$ in (\ref{eq: 23a}).  That is, the proposed bootstrap version of $ W_{I}$ is 
\begin{equation} W_{I}^* = 2n \sum_i \sum_j \left[ \hat{p}_{ij}^* \log \left\{ \frac{\hat{p}_{ij}^* }{ \hat{p}_{i+}^* \hat{p}_{+j}^* \hat{\Delta}_{ij}  } \right\}
- ( \hat{p}_{ij}^* - \hat{p}_{i+}^* \hat{p}_{+j}^* \hat{\Delta}_{ij} )
\right],\label{eq: 7w*}
\end{equation}
where $\hat{\Delta}_{ij}= \hat{p}_{ij}/ ( \hat{p}_{i+} \hat{p}_{+j}) $. Note that $W_{I}^*$ is always nonnegative.

The following theorem establishes the limiting distribution of the proposed bootstrap test statistics.

\begin{theorem}\label{theo 7}
	Under regularity conditions in Section \ref{ss: regularity conditions}  of the Supplementary Material,
	\begin{equation}
	X_{I}^{2*},\, W_{I}^*  \mid A { \,\longrightarrow\, }  \mathcal{G}_3
	\label{7}
	\end{equation}
	in distribution as $n\to\infty$,  and 
	the reference distribution is the bootstrap sampling distribution conditional on  the   sample $A$, where $\mathcal{G}_3$ is defined by (\ref{b6}).

\end{theorem}

The proof of Theorem~\ref{theo 7} is given in Section \ref{ss: theorem 7} of the Supplementary Material. By Theorem~\ref{theo 7}, we can use the bootstrap distribution to approximate the sampling distribution of the test statistic $X_{I}^{2}$ or $ W_{I}$ without computing the additional terms to account for the design effects.

\section{Simulation Study}

\subsection{Probability Proportional to Size Sampling with Replacement}\label{sec: 8.1}
In this section, we test the performance of the proposed bootstrap   method under probability proportional to size sampling with replacement. A finite population of size $N$ is generated as follows. For $i=1,\ldots,N$, 
$
x_i  \sim \mathrm{U}(0,5)
$ and $
y_i\mid x_i \sim N(\theta_{1}+\theta_{2}x_i,0.2)$,
where $(\theta_{1},\theta_{2}) = (1,1)$, and $\mathrm{U}(a,b)$ is a uniform distribution on the interval $(a,b)$. To make the sampling design informative,  probability proportional to size sampling   with replacement is used to generate a sample of size $n$ with  selection probabilities $p_i\propto 1+\lvert y_i+\epsilon_i\rvert$ and $\sum_{i=1}^Np_i=1$, where $\epsilon_i \sim N(0,0.25) $.  We consider two scenarios:  $(N,n)= (1\,000,50)$ and $(N,n)=(2\,000,75)$.

We are interested in testing $H_0: \theta_{2}=\theta_2^{(0)}$ with  $\alpha=0.05$ nominal significance level and consider three different values for $\theta_2^{(0)}\in\{1,1.05,1.1\} $. The following testing methods are compared:
\begin{enumerate}
	\item Naive likelihood-ratio method with $W(\theta_2^{(0)})$ in (\ref{lr})  and  $\chi^2 (1)$ as the test statistic and the reference distribution, respectively.
	\item Naive quasi-score method with  $X_{QS}^2(\theta_2^{(0)})$ in (\ref{11})   and  $\chi^2 (1)$ as the test statistic and the reference distribution, respectively.
	\item \citet{lumley2014tests} method. The test statistic is $W(\theta_2^{(0)})/\hat{\delta}$, where $\hat{\delta} = n\hat{V}(\hat{\theta}_2)\hat{I}_{w,22\cdot1}$,  $\hat{I}_{w,22\cdot1}$ is in (\ref{12}), and $\hat{V}(\hat{\theta}_2)$ is obtained by $\hat{\bI}{}^{-1}_w ( \hat{\btheta})\hat{V}\{\hat{\bS}_w(\btheta)\}\{\hat{\bI}{}^{-1}_w ( \hat{\btheta})\}^{\T}$. The reference distribution is $F_{1,k}$, where $F_{\nu_1,\nu_2}$ is an $F$ distribution with parameters $\nu_1$ and $\nu_2$, $k$ is the degrees of freedom of the variance estimator based on the sampling design, and $k$ is obtained by subtracting the number of parameters from the effective sample size associated with the sampling design.
	\item Bootstrap likelihood-ratio method with $W(\theta_2^{(0)})$ in (\ref{lr}) being the test statistic, and the reference distribution is approximated by the empirical distribution of $W^*(\hat{\theta}_2)$ in (\ref{w2}).
	\item 	Bootstrap quasi-score method with  $X_{QS}^2(\theta_2^{(0)})$ in (\ref{11}) being the test statistic, and the reference distribution is approximated  by the empirical distribution of $X_{QS}^{2*}(\hat{\theta}_2)$ in (\ref{eq: x2}).
\end{enumerate}

For each scenario, we generate $1\,000$ Monte Carlo samples, and we consider $M=200$, $M=500$ and $M=1\,000$ iterations for both bootstrap  methods. Table \ref{tab: simulation 3} summarizes the simulation results. For both scenarios, the naive likelihood-ratio method and the quasi-score method have a significantly inflated type I error rate when the null hypothesis is true. The two naive methods are not theoretically valid because of the lack of design-effect corrections. The type I error rates of  the \citet{lumley2014tests} method  and the two proposed bootstrap testing methods are slightly larger than the nominal significance level under $H_0$, especially when the sample size is small. Besides, the power of the two bootstrap testing methods are reasonable compared to the Lumley-Scott method. The naive methods have slightly larger power but inflated type I error rates. In addition, we get similar test results regardless the number of bootstrap repetitions.  

\if1\inctable{
\begin{table}[!h]
	\centering
	\caption{Test power for the hypothesis test $H_0: \theta_2 = \theta_2^{(0)}$ based on 1\,000 Monte Carlo simulations, the true parameter is $\theta_2 = 1.00$, and the nominal significance level is 0.05. The numbers of bootstrap repetitions are set to be $M=200$, $M=500$ and $M=1\,000$.  } \label{tab: simulation 3}
	\begin{center}
    \begin{tabular}{cccccccccccc}
    \hline\hline
       \multirow{2}{*}{$(N,n)$} & \multirow{2}{*}{$\theta_2^{(0)}$} &&\multirow{2}{*}{NLR}
        &\multirow{2}{*}{NQS}&\multirow{2}{*}{LS}&\multicolumn{2}{c}{$M=200$}
        & \multicolumn{2}{c}{$M=500$}&\multicolumn{2}{c}{$M=1\,000$}\\
        &&&&&&BLR&BQS&BLR&BQS&BLR&BQS\\
        \hline
        \multirow{3}{*}{$(1\,000,50)$}&1.00&&0.09& 0.10 &0.07 &0.06& 0.06& 0.06& 0.06 &0.06& 0.07\\
        &1.05&& 0.24& 0.27& 0.21& 0.21& 0.21& 0.20& 0.20& 0.20& 0.20 \\
        &1.10&& 0.62& 0.66& 0.55& 0.55& 0.55& 0.55& 0.55& 0.55& 0.55 \\
        &&&&&&&&&&\\
        \multirow{3}{*}{$(2\,000,75)$}&1.00&&0.08& 0.09& 0.06& 0.06& 0.06& 0.06& 0.06 &0.06& 0.06\\
        &1.05&&0.32 &0.35& 0.27& 0.26& 0.26& 0.27 &0.26& 0.26& 0.26 \\
        & 1.10&&0.75 &0.76 &0.67& 0.67 &0.67& 0.67& 0.67 &0.66& 0.66 \\
        \hline
    \end{tabular}
	\end{center}
	\begin{tablenotes}
		\setlength\labelsep{0pt}
		\footnotesize
		\item NOTE: NLR, naive likelihood-ratio method; NQS, naive quasi-score method; LS, \citet{lumley2014tests} method; BLR, bootstrap likelihood-ratio method; BQS: bootstrap quasi-score method.
	\end{tablenotes}
\end{table}}\fi

\subsection{Stratified Two-Stage Cluster Sampling}
In this section, we consider a stratified two-stage cluster sampling design to test the performance of the proposed bootstrap testing methods. A finite population $U = \{(x_{hij},y_{hij}):h=1,\ldots,H; i=1,\ldots,M_h; j=1,\ldots,M_{hi}\}$  is generated based on the following steps, where $H$ is the number of strata,   $M_h$ is the number of clusters in the $h$-th stratum, and $M_{hi}$ is the size of the $(hi)$-th cluster. Let
$
a_h \sim \mathrm{Ex}(1)$ be the stratum effect, $b_{hi}\sim\mathrm{Ex}(1)$ be the cluster effect, $
M_h\mid a_h \sim 5\mathrm{Po}(a_h) +20$, $
M_{hi} \sim 5\mathrm{Po}(a_h+b_{hi}) +30$, $x_{hij}\sim \mathrm{U}(0,20)$, and $
y_{hij}\mid (a_h,b_{hi},x_{hij}) \sim N(\mu_{hij},1)
$,
where $\mathrm{Ex}(\lambda)$ is an exponential distribution with rate parameter $\lambda$, $\mathrm{Po}(\lambda)$ is a Poisson distribution with  parameter $\lambda$, $\mu_{ij} = \theta_1+ \theta_2x_{hij}+0.2a_h + 0.2b_{hi} $, and $(\theta_1,\theta_2)= (0,1)$. From the finite population, we use a stratified two-stage cluster sampling design to obtain a sample. The first-stage sampling design is probability proportional to size sampling with replacement, where the selection probability is proportional to the cluster size, and  simple random sampling is applied at the second stage.    We consider two scenarios for the number of strata: $H=20$ and $H=50$. For each scenario, $(m_1,m_2) = (5,10)$, where $m_1$ and $m_2$ are the sample sizes for the first-stage and second-stage sampling.

We test $H_0: \theta_{2}=\theta_2^{(0)}$ with $\alpha=0.05$ nominal significance level and consider two different values for $\theta_2^{(0)}$: $1$ and $1.01$.  Since the likelihood function involves intractable integral forms, we do not consider the likelihood-ratio test and only compare the bootstrap quasi-score method with the naive quasi-score method.

For each scenario, we generate $1\,000$ Monte Carlo samples, and we consider $M=200$, $M=500$ and $M=1\,000$ iterations for the bootstrap quasi-score method. Table \ref{tab: simulation two stage} summarizes the simulation results. For both scenarios, the naive  quasi-score method has a larger type I error rate under $H_0$. In contrast, the type I error rates of the bootstrap  quasi-score method are  close to the nominal significance level when the null hypothesis is true for both scenarios. Besides, the power of the proposed bootstrap testing methods is reasonable compared with the naive quasi-score method. In addition, we get similar test results regardless the number of bootstrap repetitions.  

\if1\inctable{
\begin{table}[bthp!]
	\centering
	\caption{Test power for the hypothesis test $H_0: \theta_2 = \theta_2^{(0)}$ based on 1\,000 Monte Carlo simulations, the true parameter is $\theta_2 = 1.00$, and the nominal significance level is 0.05. The numbers of bootstrap repetitions are set to be $M=200$, $M=500$ and $M=1\,000$.}\label{tab: simulation two stage}
	
	\begin{center}
		
			
			
    \begin{tabular}{ccccccc}
    \hline\hline
        $H$ &$\theta_2^{(0)}$ &&NQS&BQS\_I&BQS\_II&BQS\_III  \\
        \hline
        \multirow{2}{*}{20} & 1.00&&0.09 &0.07& 0.06& 0.06 \\
        &1.01&&0.47 &0.39 &0.39 &0.39 \\
        &&&&&&\\
         \multirow{2}{*}{50}&1.00&&0.10& 0.06& 0.06 &0.06 \\
         &1.01&&0.78 &0.69 &0.70& 0.69 \\
    \hline
    \end{tabular}
	\end{center}
	\begin{tablenotes}
		\setlength\labelsep{0pt}
		\footnotesize
		\item NOTE: NQS, naive quasi-score method; BQS\_I: bootstrap quasi-score method with $M=200$ bootstrap replications; BQS\_II: bootstrap quasi-score method with $M=500$ bootstrap replications; BQS\_III: bootstrap quasi-score method with $M=1\,000$ bootstrap replications.
	\end{tablenotes}
\end{table}}\fi

Besides, Section~\ref{sec: SS test independence} of the Supplementary Material reports results of an additional simulation study on testing independence in a two-way table. 
\section{Application}
We present an analysis of the 2011 Private Education Expenditure Survey (PEES) in South Korea using the   proposed bootstrap testing methods. The purpose of this survey is to study the relationship between private education expenditure and the academic performance of students before entering college. 

A stratified two-stage cluster sampling design  was used for the 2011 PEES, and strata consist of   16 first-tier administrative divisions, including most provinces and metropolitan cities of South Korea. For each stratum,  probability proportional to size sampling  with replacement was conducted in the first stage, and the primary sampling unit was the school. Students were randomly selected in the second stage. There are about 1\,000 sample schools and 45\,000 students involved in this survey. 


For student $i$ in the sample $A$, let $y_i$ be the academic performance assessed by the teacher, and it takes a value from 1 through 3 corresponding to low, middle and high academic performance, respectively. Associated with $y_i$, let $\bx_i$ be the vector of covariates of interest. As discussed by \citet{kim2017statistical}, we consider the following covariates: after-school education, hours taking lessons provided by the school after regular classes in a month;  father's education, 1 for college or higher and 0 otherwise;  gender, 1 for female and 0 for male;   household income per month; mother's education, 1 for college or higher and 0 otherwise; private education, hours taking private lessons in a month. 

In this section, we study the academic performance of students in middle school and high school separately, and we are interested in studying the conditional  probability of achieving  high academic performance. Specifically, consider the following logistic model,  
\begin{equation}
\mathrm{logit} \{ \mbox{pr}(Y=1\mid \bx) \} = (1,\bx{}^\T) \btheta, \label{eq: 27}
\end{equation}
where $Y=1$ if high academic performance is achieved and 0 otherwise, $\bx$ is a vector of six covariates, and $\btheta = (\theta_0,\theta_1,\ldots,\theta_6){}^\T$. We are interested in testing the null hypotheses $H_{0,i}:\theta_i = 0$ for $i=1,\ldots,6$ with $\alpha = 0.05$ nominal significance level.  Since the Wald method is widely used in practice, the naive likelihood-ratio method, naive quasi-score method, bootstrap likelihood-ratio method, bootstrap quasi-score method with 1\,000 iterations and a two-sided Wald test are compared. The  $p$-values for the two naive methods and the Wald test are obtained using reference distributions for simple random sampling. Specifically,  the reference distribution for the two naive methods is $\chi^2(1)$, and  that of the Wald test is a normal distribution with estimated variance \citep[Section 1.2.8]{fuller2009sampling} for $\hat{\btheta}$  by the sandwich formula. The $p$-values for the proposed bootstrap testing methods are obtained by bootstrap empirical distributions of the corresponding test statistics.

Results of our analysis are summarized in Table  \ref{tab: 5}. The two bootstrap testing methods and the Wald test perform approximately the same in terms of the $p$-values. The $p$-values of the  two naive  methods are approximately the same, but they differ from those of the bootstrap testing methods, especially for ``after school education'' in the middle school level and  most covariates in the high school level, as the naive methods  may not properly reflect the intra-cluster correlation in the cluster sampling. 

Based on the two bootstrap testing methods in Table \ref{tab: 5}, we have the following conclusions under $\alpha = 0.05$ nominal significance level.  Controlling other covariates, the probability of female students achieving high academic performance is significantly higher than that of male students in middle school, but the gender effect is not significant in the high school. The hours spent on private education and after-school education can increase the probability of achieving high academic performance significantly in both middle school and high school. The household income has a significantly positive influence on their child's academic performance. However, father's and mother's education level only has a significant influence during the middle school period.

\if1\inctable{
\begin{table}[ht]
	\centering
	\caption{Estimates (Est) and the $p$-values (Unit: $10^{-2}$) for testing $H_{0,i}:\theta_i=0$, where $i=1,\ldots,6$, in (\ref{eq: 27}) for the middle school and high school. }\label{tab: 5}
	\begin{center}
		\begin{tabular}{ccrrrrrrr}
			\hline\hline
			\multirow{2}{*}{School level}	& \multirow{2}{*}{Cov}&  \multirow{2}{*}{Est}&&\multicolumn{5}{c}{$p$-value ({\small Unit: $10^{-2}$})}\\
			\cline{5-9}
			&	 &  &&NLR & NQS & BLR & BQS &Wald \\ 
			
			\hline
 			  	\multirow{6}{*}{Middle School} & \multirow{1}{*}{After-school Edu} & \multirow{1}{*}{0.03} &  & 16.9 & 16.6 & 39.4 & 39.0 & 35.6 \\ 
 			  	 & \multirow{1}{*}{Father's Edu} & \multirow{1}{*}{0.42} &  & 0.0 & 0.0 & 0.0 & 0.0 & 0.0 \\ 
 			  	  & \multirow{1}{*}{Gender} & \multirow{1}{*}{0.3} &  & 0.3 & 0.3 & 0.7 & 0.6 & 0.6 \\ 
 			  	 & \multirow{1}{*}{Income} & \multirow{1}{*}{0.14} &  & 0.0 & 0.0 & 0.2 & 0.2 & 0.4 \\ 
  & \multirow{1}{*}{Mother's Edu} & \multirow{1}{*}{0.37} &  & 0.2 & 0.2 & 0.0 & 0.0 & 0.0 \\ 
 & \multirow{1}{*}{Private Edu} & \multirow{1}{*}{0.05} &  & 0.0 & 0.0 & 0.0 & 0.0 & 0.0 \\

 &  &  &  &  &  &  &  &  \\ 
 \multirow{6}{*}{High School}& \multirow{1}{*}{After-school Edu} & \multirow{1}{*}{0.03} &  & 2.0 & 2.0 & 0.3 & 0.3 & 0.1 \\ 
  & \multirow{1}{*}{Father's Edu} & \multirow{1}{*}{0.19} &  & 10.4 & 10.3 & 14.1 & 14.0 & 13.8 \\

 & \multirow{1}{*}{Gender} & \multirow{1}{*}{0.21} &  & 2.5 & 2.5 & 17.0 & 16.8 & 15.7 \\ 
 & \multirow{1}{*}{Income} & \multirow{1}{*}{0.1} &  & 0.0 & 0.0 & 0.1 & 0.1 & 0.1 \\ 
 
 & \multirow{1}{*}{Mother's Edu} & \multirow{1}{*}{0.15} &  & 20.4 & 20.4 & 41.2 & 41.1 & 40.3 \\ 
 & \multirow{1}{*}{Private Edu} & \multirow{1}{*}{0.04} &  & 0.0 & 0.0 & 3.5 & 3.0 & 3.2 \\

			\hline
		\end{tabular}
	\end{center}
	\begin{tablenotes}
		\setlength\labelsep{0pt}
		\footnotesize
		\item NOTE: Cov, covariate; Est, estimate; NLR, naive likelihood-ratio method; NQS, naive quasi-score method; BLR, bootstrap likelihood-ratio method; BQS: bootstrap quasi-score method; Wald, two-sided Wald test.
	\end{tablenotes}
\end{table}}\fi

\section{Concluding Remarks}

Many statistical agencies provide microdata files  containing survey weights and several sets of associated replication weights, in particular bootstrap weights. Standard statistical packages often permit the use of survey-weighted test statistics, and we have shown how to approximate their distributions under the null hypothesis by their bootstrap analogues computed from the bootstrap weights supplied in the data file. Using the bootstrap weights, we can easily apply hypothesis testing without using specialized software designed for complex survey data. We studied weighted likelihood-ratio tests and weighted score tests based on weighted score or estimating equations.    

We also studied the case of categorical data by developing bootstrap procedures for testing simple goodness of fit and independence in a two-way table. We plan to extend our bootstrap  method for categorical data to test hypotheses from multi-way tables of weighted counts or proportions, using a log-linear model approach proposed by \citet{rao1984chi}. 

Our theory depends on establishing bootstrap central limit theorems under the specified sampling designs, including Poisson sampling, probability proportional to size sampling with replacement and stratified multi-stage cluster sampling with  clusters selected with replacement. Under general sampling designs, the bootstrap method of \citet{Antal_Tille_2011} may satisfy the bootstrap central limit theorem and can be used for hypothesis testing similarly, which  is beyond the scope of this paper. The hypotheses we have discussed are special cases of the one considered in Section~4.2 of \citet{jiming2019robust}, so the theoretical properties of the proposed bootstrap method can  also be  investigated using the general method developed in Section~4.2 of  \cite{jiming2019robust}. 

\bibliographystyle{asa}
\bibliography{paper-ref_full_J_name}

\section*{Supplementary Material}

The supplementary material  contains regularity conditions, proofs for the theorems, and an additional simulation study for independence test. Without loss of generality, we use ``*'' to denote the quantities for the bootstrap sample, use  $y_1,\ldots,y_N$ to denote the elements of the finite population and use $Y_1,\ldots,Y_N$ to denote the corresponding random variables with respect to the super-population model. 

\renewcommand{\theequation}{S.\arabic{equation}}
\renewcommand{\thelemma}{S\arabic{lemma}}
\renewcommand{\thetheorem}{S\arabic{theorem}}
\setcounter{equation}{0}
\setcounter{theorem}{0}
\setcounter{section}{0}
\setcounter{lemma}{0}
\renewcommand{\thesection}{S\arabic{section}}

\section{Regularity conditions}\label{ss: regularity conditions} 
To discuss the asymptotic results in  Theorems~\ref{theo: CLT (2)}--\ref{theo: (6)}, we need the following regularity  conditions.
\begin{enumerate}
	\renewcommand{\labelenumi}{C\arabic{enumi}.}
	\item For $\btheta\in\Theta$,  $l(\btheta;y)$ is  concave and twice-continuously differentiable with respect to $\btheta$,  $E\{\vert l(\btheta;Y)\rvert\}<\infty$, and $E\{l(\btheta;Y)\}$ is uniquely maximized at $\btheta_0$, where  $\Theta$ is an {open and} convex set containing $\btheta_0$ as an interior point.
	\label{cond: smooth of f} 
	
	\item For $\btheta\in\Theta$, $V\{l_w(\btheta)\mid U\}\to 0$ in probability {with respect to the super-population model} as $N\to\infty$, where $V\{l_w(\btheta)\mid U\}$ is the design  variance of $l_w(\btheta)= N^{-1} \sum_{i \in A} w_i l( \btheta; y_i)$  conditional on the finite population $U$. {In addition, there exists a non-negative design variance estimator $\hat{V}\{l_w(\btheta)\mid U\}$, such that $E[\hat{V}\{l_w(\btheta)\mid U\}\mid U]=V\{l_w(\btheta)\mid U\}$ conditional on the finite population almost surely.}\label{cond: consistency l}

	\item 
	There exists a compact set $\mathcal{B}\subset\Theta$ containing $\btheta_0$ as an interior point, such that $n\var\{\hat{\bS}_w(\btheta)\mid U\}$ converges to 
	$\bSigma_S(\btheta)$   in probability {with respect to the super-population model} uniformly over $\mathcal{B}$, where $\var\{\hat{\bS}_w(\btheta)\mid U\}$ is the design  variance of $\hat{\bS}_w(\btheta)=\partial l_w( \btheta)/ \partial \btheta={N^{-1}\sum_{i\in A}w_i\bS(\btheta;y_i)}$ conditional on the finite population $U$, { $\bS( \btheta; y) = \partial l(\btheta; y)/ \partial \btheta$ is the score function of $\btheta$, }{and $\bSigma_S(\btheta)$ is positive definitive and non-stochastic with respect to the super-population model and the sampling mechanism.}  \label{cond: asymptotic con var}
	\item A central limit theorem holds for the weighted score function $\hat{\bS}_w(\btheta_0)$. Specifically,
	\begin{equation*}
		n^{1/2}\hat{\bS}_w(\btheta_0) \to N(\bzero,\bSigma_S(\btheta_0))
	\end{equation*}
	in distribution as $N\to\infty$ with respect to the super-population model and the sampling mechanism.\label{cond: clt}
	
	\item  
	$\hat{\bI}_w(\btheta)
	= N^{-1} \sum_{i \in A} w_i \bI ( \btheta; y_i)$
	converges  to 
	$\mathcal{I}(\btheta)$ in probability with respect to the super-population model and the sampling mechanism uniformly over $\mathcal{B}$ as $N\to\infty$, where $\bI ( \btheta; y) =- \partial \mathbf{S} ( \btheta; y) / \partial \btheta^\T$ is the negative Hessian matrix of $\btheta$,  $\mathcal{I}(\btheta)=E\{\bI ( \btheta; Y)\}$  with respect to the super-population model, and  $\bcalI(\btheta_0)$ is invertible. \label{cond: I}
	
	\item  For $\btheta\in\Theta$,
	{$n[
		\hat{\bV}\{\hat{\bS}_w(\btheta)\mid U\}- \var\{\hat{\bS}_w(\btheta)\mid U\}]\to0		
		$} in probability {with respect to the super-population model and the sampling mechanism} uniformly over $\mathcal{B}$,
	where $\hat{\bV}\{\hat{\bS}_w(\btheta)\mid U\}$ is a non-negative design  variance estimator of $\hat{\bS}_w(\btheta)$ conditional on the finite population $U$. 
	\label{cond: variance estimator}
	
	\item For $\btheta\in\Theta$, the bootstrap weights are non-negative and satisfy 
	$E_*\{\hat{\bS}_w^*(\btheta)\} = \hat{\bS}_w(\btheta)$ and $\bV_*\{\hat{\bS}_w^*(\btheta)\} = \hat{\bV}\{\hat{\bS}_w(\btheta)\mid U\}$ conditional on the realized sample $A$. \label{cond: boot weights}
	
	\item A central limit theorem holds for the bootstrap weighted score function $\hat{\bS}_w^*(\btheta)$ uniformly over $\mathcal{B}$. Specifically, 
	\begin{equation}
		n^{1/2}\{ \hat{\bS}_w^* ( \btheta) - \hat{\bS}_w( \btheta)
		\}  \mid A\,\longrightarrow\, N( \bzero,\bSigma_S(\btheta)  )
		\label{b1}
	\end{equation}
	in distribution  uniformly over $\mathcal{B}$, and the reference distribution is the bootstrap distribution conditional on the realized sample $A$.\label{cond: boot CLT result}
	
	\item  $\hat{\bI}_w^*(\btheta)-\hat{\bI}_w(\btheta)\to0$ in probability  uniformly over $\mathcal{B}$ 
	conditional on the sample $A$, where $\hat{\bI}_w^*(\btheta)
	= N^{-1} \sum_{i \in A} w_i^* \bI ( \btheta; y_i)$. \label{cond: boot I}
	
\end{enumerate}

In Condition~C\ref{cond: smooth of f},  the concavity of  $l(\btheta;y)$ {and the pointwise convergence of the design variance $V\{l_w(\btheta)\mid U\}$ in Condition~C\ref{cond: consistency l} are sufficient}  for the uniform convergence of $l_w(\btheta)\to E\{l(\btheta;Y)\}$  {with respect to the super-population model and the sampling mechanism} over {any compact subset of} $\Theta$ \citep[Theorem~10.8]{rockafellar1970convex},  and {such a uniform convergence of $l_w(\btheta)$}  guarantees the consistency of the point estimator $\hat{\btheta}$ as well as its bootstrap counterpart {in Lemmas~\ref{coro: consistency theta hat}--\ref{coro: consistency} below}. The continuity condition of $l(\btheta;y)$ guarantees that the weighted score function $\hat{\bS}_w(\btheta)$ is continuously differentiable with respect to $\btheta$, and such a continuity condition  applies to many classical models. \citet{Yuan1998asymptotics} and \citet{rubin-bleuer2005} also used similar continuity conditions. However, the continuity condition may fail for some cases, such as $l(\theta;y) = \lvert y-\theta\rvert$, and it is not in our scope to investigate these cases; see Theorem~5.21 of \citet{van2000asymptotic} for detailed discussion.    By the weak law of large numbers  \citep[Theorem~8.1.1]{athreya2006measure}, $E\{\vert l(\btheta;Y)\rvert\}<\infty$ implies the pointwise weak convergence of  $N^{-1}\sum_{i=1}^Nl(\btheta;y_i)$  for $\btheta\in\Theta$. If $l( \btheta ; y)=\log f( y; \btheta)$ {is a log-likelihood}, then the unique  maximizer of $E\{ l( \btheta ; Y) \}$ is equal to the true parameter when the model is identifiable and correctly specified.

{The design-unbiased variance estimator of $l_w(\btheta)$ is used to guarantee the consistency of the bootstrap estimator $\hat{\btheta}^*$, and it can be obtained under regularity conditions for general sampling designs; see \citet[Chapte~1]{fuller2009sampling} for details.} Condition~C\ref{cond: asymptotic con var} involves   convergence property of the design  variance, and it is required to get the central limit theorem of $\hat{\bS}_w(\btheta_0)$; see \citet{rubin-bleuer2005} and \citet{Laura2020} for details. By ``design  variance $\var\{\hat{\bS}_w(\btheta)\mid U\}$'', we mean the  variance of the weighted score function $\hat{\bS}_w(\btheta)$ with respect to the sampling mechanism conditional on the finite population.   Here, 
$\bSigma_S(\btheta)$ is required to be invertible over $\mathcal{B}$ to get the central limit theorem for $\bS_w^*(\hat{\btheta})$ in Condition~C\ref{cond: boot CLT result}. Specifically, we would show the bootstrap central limit theorem for $n^{1/2}\hat{\bS}_w^*(\hat{\btheta})$ by Condition~C\ref{cond: boot CLT result}. Since $\hat{\btheta}$ is random, we assume a uniform central limit theorem for $\btheta\in\mathcal{B}$ in Condition~C\ref{cond: boot CLT result}, and this is the reason why we need $\bSigma_S(\btheta)$ to be invertible for $\btheta\in\mathcal{B}$ in Condition~C\ref{cond: asymptotic con var}. Condition~C\ref{cond: clt} assumes the central limit theorem for $\hat{\bS}_w(\btheta_0)$ with respect to the  super-population model and the sampling mechanism, and it is used to derive the asymptotic distribution of $\hat{\btheta}$. In Condition~C\ref{cond: clt}, we implicitly assume that the variance of $\bS(\btheta_0) = N^{-1}\sum_{i=1}^N\bS(\btheta_0;y_i)$ is negligible  compared with the design $\var\{\hat{\bS}_w(\btheta_0)\mid U\}$ conditional on the finite population asymptotically, and such an assumption holds when the sample size is negligible compared with the finite population size, under  regularity conditions for general sampling designs. For example, the variance of $\bS(\btheta_0)$ is $\var\{\bS(\btheta_0)\} = N^{-1}\Sigma_0$ with respect to the super-population model, where  $\Sigma_0 = \var\{\bS(\btheta_0;Y)\}$ is assumed to be finite. Under simple random sampling without replacement, the design variance is $\var\{\hat{\bS}_w(\btheta_0)\mid U\}=n^{-1}(1-nN^{-1})\hat{\Sigma}_0$, where $\hat{\Sigma}_0 = (N-1)^{-1}\sum_{i=1}^N\{\bS(\btheta_0;y_i)-\bS(\btheta_0)\}^{\otimes2}$ is a consistent estimator of $\Sigma_0$ with respect to the super-population model, and $A^{\otimes2}=AA^{\T}$.  Thus, if $nN^{-1}\to0$, $\var\{\bS(\btheta_0)\}$ is negligible compared with  $\var\{\hat{\bS}_w(\btheta_0)\mid U\}$ under simple random sampling without replacement; see Sections~\ref{ss: pois sampling}--\ref{ss: pps sampling} for details under Poisson sampling and probability proportional to size sampling with replacement.   In Condition~C\ref{cond: I}, the uniform convergence of $\hat{\bI}_w(\btheta)$ is used to derive the limiting distributions of $\hat{\btheta}$  by the Slutsky's theorem \citep[Theorem~9.1.6]{athreya2006measure}.  Condition~C\ref{cond: variance estimator} guarantees a design-consistent  variance estimator, and it is satisfied under regularity conditions for  commonly used   sampling designs, including simple random sampling, probability proportional to size  sampling, and  stratified two-stage cluster sampling.  

Condition~C\ref{cond: boot weights} is a general condition for the bootstrap weights, {which are determined only by a specified sampling design. Thus, the bootstrap weights also guarantee}  $E_*\{l_w^*(\btheta)\} = l_w(\btheta)$ and $V_*\{l_w^*(\btheta)\} = \hat{V}\{l_w(\btheta)\mid U\}$ conditional on the realized sample $A$ for $\btheta\in\Theta$. We will show that the proposed bootstrap method  in Examples~\ref{ex: 1}--\ref{ex: 3} satisfies Condition~C\ref{cond: boot weights}. 
Condition~C\ref{cond: boot CLT result} is the bootstrap central limit theorem for the bootstrap weighted score function $\hat{\bS}_w^*(\btheta)$ for $\btheta\in\mathcal{B}$. Different from Condition~C\ref{cond: clt}, the uniform central limit theorem {in Condition~C\ref{cond: boot CLT result}} guarantees 
\begin{equation}
	n^{1/2}  \hat{\bS}_w^* ( \hat{\btheta})\mid A
	\,\longrightarrow\, N( \bzero,\Sigma_S(\btheta_0)) \label{eq: bootstrap CLT hat theta}
\end{equation}
in distribution with respect to the bootstrap procedure conditional on the realized sample $A$, and it is a building block
to prove Theorem~\ref{theo: (6)}. Condition~C\ref{cond: boot I} is a bootstrap version of Condition~C\ref{cond: I}, and it is used to derive the limiting distribution of $\hat{\btheta}^*$. 

\section{Proof of Theorem~1}\label{sec: proof of theorem 1}

First, we show the consistency of the point estimator $\hat{\btheta}$.
\begin{lemma}\label{coro: consistency theta hat}
	Under Conditions~C\ref{cond: smooth of f}--C\ref{cond: consistency l}, the   point estimator $\hat{\btheta}$ is consistent of $\btheta_0$ {with respect to the super-population model and the sampling mechanism}.
\end{lemma}
\begin{proof}[Proof of Lemma~\ref{coro: consistency theta hat}]
	By Condition~C\ref{cond: smooth of f} and the weak law of large numbers, we have 
	\begin{equation}
		l_N(\btheta)\to E\{l(\btheta;Y)\}\label{eq: conv lN}
	\end{equation}
	in probability {with respect to the super-population model} for $\btheta\in\Theta$, where $l_N(\btheta)=N^{-1}\sum_{i=1}^Nl(\btheta;y_i)$. 
	Since $E\{l_w(\btheta)\mid U\} = l_N(\btheta)$, it follows from Condition~C\ref{cond: consistency l} that  
	\begin{equation}
		l_w(\btheta) - l_N(\btheta) \to 0 \label{eq: conv lw}
	\end{equation}
	in probability {with respect to  the sampling mechanism} as $n\to\infty$ for $\btheta\in\Theta$.
	By (\ref{eq: conv lN})--(\ref{eq: conv lw}) and the fact that $l_w(\btheta)$ is concave in $\Theta$ according to Condition~C\ref{cond: smooth of f}, applying Theorem~10.8 of \citet{rockafellar1970convex} shows that
	\begin{equation}
		l_w(\btheta) \to E\{l(\btheta;Y)\} \label{eq: conv lw Expectation}
	\end{equation}
	in probability {with respect to the super-population model and  the sampling mechanism} uniformly over any closed and bounded subset of $\Theta$.  
	
	{Since $\Theta$ is open and contains $\btheta_0$ as an interior point by Condition~C\ref{cond: smooth of f} , there exists a positive number $\delta$ such that $\mathscr{B}=\{\btheta:\lVert\btheta-\btheta_0\rVert\leq2\delta\}\subset\Theta$.  Denote $\tilde{\theta}=\arg\max_{\btheta\in\mathscr{B}}l_w(\btheta)$, so we have 
		\begin{equation}
			l_w(\tilde{\btheta})\geq l_w({\btheta}_0).\label{eq: ltheta>=ltilde}
	\end{equation}}
	
	{If we can show that 
		\begin{equation}
			\tilde{\btheta}-\btheta_0\to0\label{eq: tilde theta}
		\end{equation}
		in probability with respect to the super-population model and the sampling mechanism, then $P(E_N)\to1$ as $N\to\infty$ with respect to the super-population model and the sampling mechanism, where $E_N$ is the event that  $\tilde{\btheta}$ lies in the ball with radius $\delta$ around $\btheta_0$. Under the event $E_N$, $\tilde{\btheta}$ is an interior point of $\mathscr{B}$, and  we need to show that 
		\begin{equation}
			l_w(\tilde{\btheta})\geq l_w(\btheta)\label{eq: l_w tilde theta goal}
		\end{equation}
		for $\btheta\in \Theta\cap\mathscr{B}^C$, where $\mathscr{B}^C$ is the complement of $\mathscr{B}$. Denote $\partial\mathscr{B}=\{\btheta:\lVert\btheta-\btheta_0\rVert =2\delta\}$ to be the boundary of $\mathscr{B}$. Then, for any $\btheta\in \Theta\cap\mathscr{B}^C$, there exists $\lambda=\lambda(\btheta,\tilde{\btheta})\in(0,1)$ such that
		$
		\lambda\btheta + (1-\lambda)\tilde{\btheta}\in\partial\mathscr{B}.
		$
		By the fact that $\tilde{\btheta}=\arg\max_{\btheta\in\mathscr{B}}$ and $\partial\mathscr{B}\subset\mathscr{B}$, we have 
		\begin{eqnarray}
			l_w(\tilde{\btheta})&\geq& l_w\{\lambda\btheta + (1-\lambda)\tilde{\btheta}\}\notag \\ 
			&\geq& \lambda l_w(\btheta) + (1-\lambda)l_w(\tilde{\btheta}),\label{eq: lw tilde theta>= theta}
		\end{eqnarray}
		where the second inequality holds since $l_w(\btheta)$ is concave by Condition~C\ref{cond: smooth of f}. By (\ref{eq: lw tilde theta>= theta}) and $\lambda>0$, we have shown (\ref{eq: l_w tilde theta goal}), so $\tilde{\btheta}$ is the maximand of $l_w(\btheta)$ over $\Theta$. That is, $\hat{\btheta}=\tilde{\btheta}$. By (\ref{eq: tilde theta}), we complete the proof of  Lemma~\ref{coro: consistency theta hat}.}
	
	{Thus, it remains to show (\ref{eq: tilde theta}).  For any $\epsilon>0$, by (\ref{eq: ltheta>=ltilde}), we have 
		\begin{equation}
			l_w(\tilde{\btheta})> l_w({\btheta}_0)-\epsilon/3.\label{eq: epsilon 3 1}
		\end{equation}
		By the uniform convergence in (\ref{eq: conv lw Expectation}) over $\mathscr{B}$, with probability approaching to 1 with respect to the super-population model and the sampling mechanism, we also have the following two results
		\begin{eqnarray}
			&E\{l(\tilde{\btheta};Y)\} > l_w(\tilde{\btheta}) - \epsilon/3,\label{eq: epsilon 3 2}\\
			&l_w(\btheta_0) > E\{l(\btheta_0;Y)\}  - \epsilon/3.\label{eq: epsilon 3 3}
		\end{eqnarray}
		Thus, by (\ref{eq: epsilon 3 1})--(\ref{eq: epsilon 3 3}), we have  
		\begin{equation}
			E\{l(\tilde{\btheta};Y)\} >l_w(\tilde{\btheta}) - \epsilon/3>l_w({\btheta}_0)-2\epsilon/3>E\{l(\btheta_0;Y)\}  - \epsilon\label{eq: lw last one}
		\end{equation}
		for any $\epsilon>0$. }
	
	{For any open set $\mathscr{O}$ containing $\btheta_0$ as an interior point, $\mathscr{O}^C\cap\mathscr{B}$ is compact. Since $E\{l(\btheta;Y)\}$ is continuous, there exists $\btheta^*\in\mathscr{O}^C\cap\mathscr{B}$ such that $E\{l(\btheta^*;Y)\} = \max_{\btheta\in\mathscr{O}^C\cap\mathscr{B}}E\{l(\btheta;Y)\}$. Since $E\{l(\btheta;Y)\}$ is uniformly maximized at $\btheta_0$ by Condition~C\ref{cond: smooth of f}, we have $E\{l(\btheta_0;Y)\}-E\{l(\btheta^*;Y)\}>0$. By choosing $\epsilon = E\{l(\btheta_0;Y)\}-E\{l(\btheta^*;Y)\}$ and by (\ref{eq: lw last one}), we conclude that $E\{l(\tilde{\btheta};Y)\}>E\{l(\btheta;Y)\}$ for $\btheta\in\mathscr{O}^C\cap\mathscr{B}$ with probability approaching 1  with respect to the super-population model and the sampling mechanism. Thus, $\tilde{\btheta}\in\mathscr{O}$ with probability approaching 1. Since $\mathscr{O}$ is arbitrary, we have proved Lemma~\ref{coro: consistency theta hat}; see Theorem~2.1 and Theorem~2.7 of \citet[Chapter~36]{engle1994handbook} for details.}
\end{proof}

\begin{proof}[Proof of Theorem~\ref{theo: CLT (2)}]
	By Lemma~\ref{coro: consistency theta hat}, we have 
	\begin{equation}
		\hat{\btheta}-\btheta_0\to0\label{eq: consitency theta hat}
	\end{equation}
	in probability  with respect to the super-population model and the sampling mechanism. Since $ \hat{\bS}_w(\btheta)$ is continuously differentiable by Condition~C\ref{cond: smooth of f} and 
	\begin{equation}
		\hat{\bS}_w(\hat{\btheta}) = 0,
	\end{equation}
	by the mean value theorem, there exists $\btheta^\dagger$ on the segment joining $\hat{\btheta}$ and $\btheta_0$, such that 
	\begin{equation}
		\hat{\bS}_w({\btheta}_0) = \frac{\partial}{\partial\btheta^\T}\hat{\bS}_w(\btheta^\dagger)(\btheta_0-\hat{\btheta}) = \hat{\bI}_w(\btheta^\dagger) (\btheta_0-\hat{\btheta}).\label{eq: S16 new}
	\end{equation}
	
	By  Condition~C\ref{cond: I} and Lemma~\ref{coro: consistency theta hat}, we have 
	\begin{equation}
		\bcalI(\btheta_0)^{-1}\hat{\bI}_w(\btheta^\dagger) \to \bI \label{eq: S17 new}
	\end{equation}
	in probability with respect to the super-population model and sampling  mechanism.
	
	Then, by (\ref{eq: S16 new}), we have 
	\begin{eqnarray}
		n^{1/2}\bcalI(\btheta_0)^{-1}\hat{\bI}_w(\btheta^\dagger) (\btheta_0-\hat{\btheta}) &=&  n^{1/2}\bcalI(\btheta_0)^{-1}\hat{\bS}_w({\btheta}_0).\label{eq: S18 new}
	\end{eqnarray}
	By Conditions~C\ref{cond: asymptotic con var}--C\ref{cond: clt}, (\ref{eq: S17 new})--(\ref{eq: S18 new}) and Slutsky's theorem, we have completed the proof of   Theorem~\ref{theo: CLT (2)}; see Theorem~4 of \citet{Yuan1998asymptotics} for details.

\end{proof}


\section{Proof of Theorem~2}\label{sec: proof of thereom 2}

First, we show the consistency of the bootstrap estimator $\hat{\btheta}^*$.
\begin{lemma}\label{coro: consistency}
	Under Conditions~C\ref{cond: smooth of f}--C\ref{cond: consistency l} and C\ref{cond: boot weights}, the bootstrap point estimator $\hat{\btheta}^*$ is consistent for $\btheta_0$.
\end{lemma}
\begin{proof}[Proof of Lemma~\ref{coro: consistency}]
	By Condition~C\ref{cond: smooth of f} and Condition~C\ref{cond: boot weights}, $l_w^*(\btheta) = N^{-1}\sum_{i\in A}w_i^*l(\btheta;y_i)$ is concave and twice continuously differentiable. By  Condition~C\ref{cond: consistency l},   $\hat{V}\{l_w(\btheta)\mid U\}\geq0$ is design-unbiased for $V\{l_w(\btheta)\mid U\}$ conditional on the finite population. That is, $E[\hat{V}\{l_w(\btheta)\mid U\}\mid U] = V\{l_w(\btheta)\mid U\}$. By Condition~C\ref{cond: consistency l}, $\hat{V}\{l_w(\btheta)\mid U\}$ is non-negative and  $E[\hat{V}\{l_w(\btheta)\mid U\}\mid U]\to0$ in probability with respect to the super-population model. Thus,  by the Markov's inequality \citep[Theorem~3.1.1]{athreya2006measure}, we conclude that $\hat{V}\{l_w(\btheta)\mid U\}\to 0$ in probability with respect to the super-population model and the sampling mechanism. By Condition~C\ref{cond: consistency l} and Condition~C\ref{cond: boot weights}, we have $V_*\{l_w^*(\btheta)\} \to 0$ in probability with respect to the bootstrap method conditional on the realized sample $A$, and it leads to \begin{equation}
		l_w^*(\btheta)- E_*\{l_w^*(\btheta)\} =l_w^*(\btheta) - l_w(\btheta)\to0\label{eq: cons LW*}
	\end{equation} in probability conditional on the realized sample $A$. By (\ref{eq: conv lw Expectation})--(\ref{eq: cons LW*}), we have shown \begin{equation}
		l _w^*(\btheta) - E\{l(\btheta;Y)\}\to0\label{eq: conv L_W*}
	\end{equation} in probability with respect to the super-population model, the sampling mechanism as well as the bootstrap method for $\btheta\in\Theta$. Furthermore, since the bootstrap weights are non-negative by Condition~C\ref{cond: boot weights}, $l_w^*( \btheta)$ is concave,  and the convergence in (\ref{eq: conv L_W*}) is uniform over any bounded and closed subset of $\Theta$.
	Thus, by Condition~C\ref{cond: smooth of f} and  (\ref{eq: conv L_W*}),  we have proved Lemma~\ref{coro: consistency} by a similar procedure used in the proof of Lemma~\ref{coro: consistency theta hat}. 
\end{proof}

\begin{proof}[Proof of Theorem~\ref{theo: (6)}]
	By (\ref{eq: consitency theta hat}) and Lemma~\ref{coro: consistency}, we conclude that
	\begin{equation}
		\hat{\btheta}^*-\hat{\btheta}\to0\label{eq: consistent theta.hat.star}
	\end{equation} in probability with respect to the super-population model, the sampling mechanism as well as the bootstrap method as $n\to\infty$.


	By the mean value theorem, we have
	\begin{eqnarray}
		\hat{\bS}_w^*(\hat{\btheta}^*) = \hat{\bS}_w^*(\hat{\btheta}) - \hat{\bI}_w^*(\tilde{\btheta}^*)(\hat{\btheta}^* - \hat{\btheta}),\label{eq: Taylor hat S w star}
	\end{eqnarray} where  $\hat{\bI}_w^*(\btheta) = - \partial \hat{\bS}_w^*(\btheta)/\partial\btheta^\T$, and $\tilde{\btheta}^*$ lies on the segment joining $\hat{\btheta}$ and $\hat{\btheta}^*$.

	Now, if we can show that 
	\begin{equation}
		\hat{\bI}_w^*(\tilde{\btheta}^*)\to {{\bcalI}(\btheta_0)}\label{eq: conv hat I w star}
	\end{equation}
	in probability with respect to the super-population model, the sampling mechanism as well as the bootstrap method, then,  by (\ref{eq: Taylor hat S w star})--(\ref{eq: conv hat I w star}), we have 
	\begin{equation}
		\hat{\bS}_w^*(\hat{\btheta}^*) = \hat{\bS}_w^*(\hat{\btheta}) - {{\bcalI}({\btheta}_0)}(\hat{\btheta} - \hat{\btheta}^*) + o_p(\hat{\btheta} - \hat{\btheta}^*),\label{eq: S24 new}
	\end{equation}
	where the $o_p$-term is with respect to the super-population model, the sampling mechanism as well as the bootstrap method.
	Since $\hat{\bS}_w^*(\hat{\btheta}^*) = 0$, by (\ref{eq: consistent theta.hat.star}) and (\ref{eq: S24 new}), we have 
	\begin{equation}
		\hat{\btheta} - \hat{\btheta}^* =    {{\bcalI}^{-1} ({\btheta}_0)}\hat{\bS}_w^*(\hat{\btheta})+o_p(1),\label{eq: hat theta - hat theta star}
	\end{equation}
	where the $o_p$-term is with respect to the super-population model, the sampling mechanism as well as the bootstrap method.
	Thus, by Conditions~C\ref{cond: clt}--C\ref{cond: boot CLT result}, (\ref{eq: hat theta - hat theta star}) and the fact that $\hat{\bS}_w(\hat{\btheta}) =0$, we have proved Theorem~\ref{theo: (6)}.

	It remains to show (\ref{eq: conv hat I w star}). Since $p$ is   fixed, it is sufficient to show  (\ref{eq: conv hat I w star}) for $p=1$. For $p=1$, we have 
	\begin{equation}
		\lvert\hat{I}_w^*(\tilde{\theta}^*)- \calI(\theta_0)\rvert\leq \lvert\hat{I}_w^*(\tilde{\theta}^*)- \hat{I}_w(\tilde{\theta}^*)\rvert + \lvert\hat{I}_w(\tilde{\theta}^*)- \calI(\tilde{\theta}^*)\rvert+ \lvert\calI(\tilde{\theta}^*)- \calI(\theta_0)\rvert.\label{eq: I convergence}
	\end{equation}
	By (\ref{eq: consitency theta hat}), (\ref{eq: consistent theta.hat.star}) and  the fact that $\theta_0$ is an interior point of $\mathcal{B}$ from Condition~C\ref{cond: asymptotic con var}, we can show that the probability for $\tilde{\theta}^*\in\mathcal{B}$ converges to 1 with respect to the super-population model, the sampling mechanism as well as the bootstrap method. Thus, by Condition~C\ref{cond: boot I}, we have  
	\begin{equation}
		\hat{I}_w^*(\tilde{\theta}^*)- \hat{I}_w(\tilde{\theta}^*)\to0
	\end{equation}
	in probability with respect tot the bootstrap method conditional on the realized sample $A$. Similarly, by Condition~C\ref{cond: I}, we can also claim  that 
	\begin{equation}
		\hat{I}_w(\tilde{\theta}^*)- \calI(\tilde{\theta}^*)\to 0
	\end{equation}
	in probability with respect to the super-population model and the sampling mechanism. By (\ref{eq: consitency theta hat}), (\ref{eq: consistent theta.hat.star}) and  the fact that $\tilde{\theta}^*$ lies on the segment joining $\hat{\theta}$ and $\hat{\theta}^*$, we conclude that $\tilde{\theta}^* - \theta_0\to0$ in probability  with respect to the super-population model, the sampling mechanism as well as the bootstrap method. Thus, by  Condition~C\ref{cond: smooth of f}, we can show that $\calI(\theta)$ is continuous on $\mathcal{B}$, so
	\begin{equation}
		\calI(\tilde{\theta}^*)- \calI(\theta_0) \to0\label{eq: I last}
	\end{equation}
	in probability  with respect to the super-population model, the sampling mechanism as well as the bootstrap method. By (\ref{eq: I convergence})--(\ref{eq: I last}), we have validated (\ref{eq: conv hat I w star}).
	
\end{proof}

\section{Poisson sampling}\label{ss: pois sampling}
To validate the proposed bootstrap method {under Poisson sampling}, it is sufficient to verify Conditions~C\ref{cond: smooth of f}--C\ref{cond: boot I} in Section~\ref{ss: regularity conditions}.  We now assume the following specific conditions.
\begin{enumerate}
	\renewcommand{\labelenumi}{A\arabic{enumi}.}
	\setcounter{enumi}{0}
	\item For $\btheta\in\Theta$, the function $l(\btheta;y)$ is  concave and twice-continuously differentiable with respect to $\btheta$,  $E\{\lvert l(\btheta;Y)\rvert^2\}<\infty$, and $E\{l(\btheta;Y)\}$ is uniquely maximized at $\btheta_0$, where  $\Theta$ is an open and convex set containing $\btheta_0$ as an interior point.
	
	\label{eq: consis Poisson}
	\item The pair $(\pi_i,y_i)$ is independent of $(\pi_j,y_j)$ for $i\neq j$.\label{cond: inpdt Poisson}
	\item  There exist two constants $0<C_1<C_2<\infty$ such that $C_1<Nn_0^{-1}\pi_i<C_2$ for $i=1,\ldots,N$ almost surely with respect to the super-population model,
	where $n_0=\sum_{i=1}^N\pi_i$ is the expected sample size satisfying $n_0=o_p(N)$ and $n_0\to\infty$ as $N\to\infty$ almost surely with respect to the super-population. \label{cond: Poi pi}
	\item There exists a compact set  $\mathcal{B}\subset\Theta$  containing $\btheta_0$ as an interior point, such that $\sup_{\btheta\in\mathcal{B}}E\lVert\bS(\btheta;Y)\rVert^{4}<\infty$, where $\lVert\bx\rVert$ is the Euclidean norm of a vector $\bx$. \label{cond: six moments}
	\item Given every $\ba\in\mathbb{R}^p$ satisfying $\lVert\ba\rVert=1$, $V\{S_{\ba,i}(\btheta_0)\}>0$ and there exists a positive  definitive matrix $\bSigma_{\ba}$ such that 
	\begin{eqnarray}
		&n_0N^{-2}\sum_{i=1}^N\pi_i^{-1}(1,S_{\ba,i}(\btheta_0))(1,S_{\ba,i}^\T(\btheta_0))^{\T}\to\bSigma_{\ba}\notag
	\end{eqnarray}
	in probability with respect to the super-population model,
	where $V\{S_{\ba,i}(\btheta_0)\}$ is the variance of $S_{\ba,i}(\btheta_0)$ with respect to the super-population model, and $S_{\ba,i}(\btheta) = \ba^{\T}\bS(\btheta,y_i)$.
	\label{cond: limit fuller}
	\item  For $\btheta\in\mathcal{B}$, there exists a  positive definitive and  non-stochastic matrix $\bSigma_S(\btheta)$ with respect to the super-population model and sampling mechanism, such that $n_0\var\{\hat{\bS}_w(\btheta)\mid U\}\to \bSigma_S(\btheta)$ in probability with respect to the super-population model uniformly over $\mathcal{B}$, where  $\var\{\hat{\bS}_w(\btheta)\mid U\} = N^{-2}\sum_{i=1}^N\pi_i^{-1}(1-\pi_i)\bS(\btheta;y_i)^{\otimes2}$ is the design variance of $\hat{\bS}_w(\btheta)$, and $\bx^{\otimes 2} = \bx\bx^\T$ for a vector $\bx$. \label{cond: Poi S6}
	\item  	$\sup_{\btheta\in\mathcal{B}}E\left\{\lVert\bI(\btheta;Y)\rVert^2\right\}<\infty$,  and $\bcalI(\btheta_0)$ is invertible, where $\bcalI(\btheta) = E\left\{\bI(\btheta;Y)\right\}$. \label{cond: z poisson}
	
\end{enumerate}

Condition~A\ref{eq: consis Poisson} is almost the same as Condition~C\ref{cond: smooth of f}, and the difference is the second moment condition   $E\{\lvert l(\btheta;Y)\rvert^2\}<\infty$, which is used to get Condition~C\ref{cond: consistency l} under Poisson sampling. By the Cram\'{e}r-Wold device \citep[Theorem~10.4.5]{athreya2006measure}, Conditions~A\ref{cond: inpdt Poisson}--A\ref{cond: limit fuller} are sufficient for the central limit theorem of $\hat{\bS}_w(\btheta_0)$ with respect to the super-population and the sampling mechanism under Poisson sampling; see Lemma~\ref{lemma: C4 Poisson} below for details. 
In addition to the pairwise independence among $\{Y_i:i=1,\ldots,N\}$,   Condition~A\ref{cond: inpdt Poisson} requires   independence among the sequence $\{\pi_i:i=1,\ldots,N\}$ as well, and it  holds trivially if $\{\pi_i:i=1,\ldots,N\}$ are non-stochastic with respect to the super-population model. Condition~A\ref{cond: Poi pi} is a regularity condition about the expected sample size and the inclusion probabilities, and it is used to show the consistency of the variance estimator as well as others. The condition $n_0=o_p(N)$   ensures that the design  variance dominates the total variance of the weighted score function with respect to the super-population model and the sampling mechanism; see Lemma~\ref{lemma: 5} below for details.  Condition~A\ref{cond: six moments} guarantees the central limit theorem for the weighted score function as well as the consistency of the design  variance estimator; see (1.3.37) of \citet{fuller2009sampling} for details. Condition~A\ref{cond: limit fuller} contains technical convergence condition at the true value $\btheta_0$; see Theorem~1.3.3 and Theorem~1.3.5 of \citet{fuller2009sampling} for details. Condition~A\ref{cond: Poi S6} guarantees the convergence of the design   variance of the weighted score function. Condition~A\ref{cond: z poisson} is necessary to show that $\hat{\bI}_w(\btheta)$ converges in probability uniformly to $\bcalI(\btheta)$ over $\mathcal{B}$, and such a condition is also used to validate the central limit theorems for $\hat{\btheta}$ and $\hat{\btheta}^*$ by \citet{Yuan1998asymptotics} and \citet{rubin-bleuer2005} as well.

Next, we validate the conditions in Section~\ref{ss: regularity conditions}  one by one.

\begin{lemma}\label{coro: cond 2}
	By Conditions~A\ref{eq: consis Poisson}--A\ref{cond: Poi pi}, Condition~C\ref{cond: consistency l} holds under Poisson sampling.
\end{lemma}
\begin{proof}[Proof of Lemma~\ref{coro: cond 2}]
	First, under Poisson sampling, consider 
	\begin{eqnarray}
		V\{l_w(\btheta)\mid U\} &=& N^{-2}\sum_{i=1}^N\frac{1-\pi_i}{\pi_i}l(\btheta;y_i)^2\notag \\ 
		&\leq&(C_1n_0)^{-1}N^{-1}\sum_{i=1}^Nl(\btheta;y_i)^2\notag \\ 
		&=& O_p(n_0^{-1}),\label{eq: var l}
	\end{eqnarray}
	where the second inequality is based on Condition~A\ref{cond: Poi pi}, and the last equality holds by Conditions~A\ref{eq: consis Poisson}--A\ref{cond: inpdt Poisson} and the weak law of large numbers. Thus, $V\{l_w(\btheta)\mid U\}\to 0$ in probability with respect to the super-population model for $\btheta\in\Theta$ under Poisson sampling.
	
	Let $\hat{V}\{l_w(\btheta)\mid U\} = N^{-2}\sum_{i\in A}\pi_i^{-2}(1-\pi_i)l(\btheta;y_i)^2$. Then, $\hat{V}\{l_w(\btheta)\mid U\}\geq 0$, and we have 
	\begin{eqnarray}
		E[\hat{V}\{l_w(\btheta)\mid U\}\mid U]
		&=& E\left\{N^{-2}\sum_{i=1}^N\frac{I_i(1-\pi_i)}{\pi_i^2}l(\btheta;y_i)^2\mid U\right\}\notag \\ 
		&=&N^{-2}\sum_{i=1}^N\frac{1-\pi_i}{\pi_i}l(\btheta;y_i)^2\notag \\ 
		&=& V\{l_w(\btheta)\mid U\}.\label{eq: S.31.nwe}
	\end{eqnarray}
	
	By (\ref{eq: var l})--(\ref{eq: S.31.nwe}), we have completed the proof of  Lemma~\ref{coro: cond 2}.
\end{proof}
\begin{lemma}\label{lemma: 5}
	Under Poisson sampling and Conditions~A\ref{cond: inpdt Poisson}--A\ref{cond: six moments}, we have 
	$$
	\sup_{\btheta\in\mathcal{B}}\hat{V}\{\ba^\T\hat{\bS}_w(\btheta)\mid U\} \asymp n_0^{-1}
	$$
	in probability with respect to the super-population model and the sampling mechanism, where $\ba\in\mathbb{R}^p$  satisfying $\lVert \ba\rVert =1$,   $\hat{V}\{\ba^\T\hat{\bS}_w(\btheta)\mid U\} = N^{-2}\sum_{i\in A}\pi_i^{-2}(1-\pi_i)\{\ba^\T\bS(\btheta;y_i)\}^2$, and $a_n\asymp b_n$ is equivalent to $a_n=O(b_n)$ and $b_n=O(a_n)$.
\end{lemma}
\begin{proof}[Proof of Lemma~\ref{lemma: 5}]
	First, consider the expectation of $\hat{V}\{\ba^\T\hat{\bS}_w(\btheta)\mid U\}$ {with respect to the super-population model and the sampling mechanism}:
	\begin{equation}
		E[	\hat{V}\{\ba^\T\hat{\bS}_w(\btheta)\mid U\}] = E(E[	\hat{V}\{\ba^\T\hat{\bS}_w(\btheta)\mid U\}\mid U]).\label{eq: exp decom}
	\end{equation}
	Under Poisson sampling,  
	\begin{eqnarray}
		E[	\hat{V}\{\ba^\T\hat{\bS}_w(\btheta)\mid U\}\mid U]&=&E\left[ \frac{1}{N^2}\sum_{i\in A}\frac{(1-\pi_i)}{\pi_i^2}\{\ba^\T\bS(\btheta;y_i)\}^2\mid U\right]\notag \\ 
		&=& \frac{1}{N^2}\sum_{i=1}^N\frac{1-\pi_i}{\pi_i}\{\ba^\T\bS(\btheta;y_i)\}^2,\label{eq: exp}
	\end{eqnarray}
	and 
	\begin{eqnarray}
		E\left[\frac{1}{N^2}\sum_{i=1}^N\frac{1-\pi_i}{\pi_i}\{\ba^\T\bS(\btheta;Y_i)\}^2\right] &\asymp& n_0^{-1},\label{eq: exp 2}
	\end{eqnarray}
	where capital letter $Y_i$ is used in (\ref{eq: exp 2}) to highlight that the expectation  is taken with respect to the super-population model, and the asymptotic order of (\ref{eq: exp 2}) holds uniformly over $\mathcal{B}$  by Conditions~A\ref{cond: Poi pi}--A\ref{cond: six moments}. Thus, by  (\ref{eq: exp decom})--(\ref{eq: exp 2}), we have shown
	\begin{equation}
		E[\hat{V}\{\ba^\T\hat{\bS}_w(\btheta)\mid U\}]\asymp n_0^{-1}\label{eq: exp f}
	\end{equation}
	uniformly over $\mathcal{B}$. 
	
	Next, consider the variance of $\hat{V}\{\ba^\T\hat{\bS}_w(\btheta)\mid U\}$ {with respect to the super-population model and the sampling mechanism}:
	\begin{equation}
		V[\hat{V}\{\ba^\T\hat{\bS}_w(\btheta)\mid U\}] = E(V[\hat{V}\{\ba^\T\hat{\bS}_w(\btheta)\mid U\}\mid U])+ V(E[\hat{V}\{\ba^\T\hat{\bS}_w(\btheta)\mid U\}\mid U]).\label{eq: var decop}
	\end{equation}
	Under Poisson sampling, we have 
	\begin{eqnarray}
		E(V[\hat{V}\{\ba^\T\hat{\bS}_w(\btheta)\mid U\}\mid U])&=& E\left[ \frac{1}{N^4}\sum_{i=1 }^N\pi_i(1-\pi_i)\frac{(1-\pi_i)^2}{\pi_i^4}\{\ba^\T\bS(\btheta;y_i)\}^4\mid U\right]\notag \\ 
		&=&\frac{1}{N^4}\sum_{i=1}^NE\left[\frac{(1-\pi_i)^3}{\pi_i^3}\{\ba^\T\bS(\btheta;y_i)\}^4\right]\notag\\
		&\asymp& \frac{1}{n_0^3N}\sum_{i=1}^NE[\{\ba^\T\bS(\btheta;y_i)\}^4]\notag \\ 
		&\asymp& \frac{1}{n_0^3} \label{eq: var decop 1}
	\end{eqnarray}
	uniformly over $\mathcal{B}$ by Conditions~A\ref{cond: inpdt Poisson}--A\ref{cond: six moments}.
	Consider 
	\begin{eqnarray}
		V(E[\hat{V}\{\ba^\T\hat{\bS}_w(\btheta)\mid U\}\mid U])&=&V\left[\frac{1}{N^2}\sum_{i=1}^N\frac{1-\pi_i}{\pi_i}\{\ba^\T\bS(\btheta;Y_i)\}^2\right]\notag \\ &=&N^{-4}\sum_{i=1}^NV\left[\frac{1-\pi_i}{\pi_i}\{\ba^\T\bS(\btheta;Y_i)\}^2\right]\notag \\ 
		&\leq&N^{-4}\sum_{i=1}^NE\left[\frac{(1-\pi_i)^2}{\pi_i^2}\{\ba^\T\bS(\btheta;Y_i)\}^4\right]\notag \\ 
		&\leq&(C_1n_0)^{-2}N^{-2}\sum_{i=1}^NE\left[\{\ba^\T\bS(\btheta;Y_i)\}^4\right]\notag \\
		&\asymp& n_0^{-2}N^{-1}\label{eq: var decop 2}
	\end{eqnarray}
	uniformly over $\mathcal{B}$, where the first equality of (\ref{eq: var decop 2}) holds by Condition~A\ref{cond: inpdt Poisson}, the third inequality holds by Condition~A\ref{cond: Poi pi}, and the last asymptotic order holds by Condition~A\ref{cond: six moments}. 
	
	Since $n_0=o(N)$ from Condition~A\ref{cond: Poi pi}, by (\ref{eq: var decop})--(\ref{eq: var decop 2}), we have shown 
	\begin{equation}
		V[\hat{V}\{\ba^\T\hat{\bS}_w(\btheta)\mid U\}] \asymp n_0^{-3}\label{eq: var f}
	\end{equation}
	with respect to the super-population model and the sampling mechanism uniformly over $\mathcal{B}$.
	By establishing (\ref{eq: exp f}) and (\ref{eq: var f}), we have proved Lemma~\ref{lemma: 5}.
\end{proof}
By (\ref{eq: exp}) and (\ref{eq: var decop 1}), we conclude that $	n[\hat{\bV}\{\hat{\bS}_w(\btheta)\mid U\}-\var\{\hat{\bS}_w(\btheta)\mid U\}]\to0$ in probability with respect to the super-population model and the sampling mechanism uniformly over $\mathcal{B}$, so we have verified Condition~C\ref{cond: variance estimator}. 
Next, we consider the central limit theorem in Condition~C\ref{cond: clt}.
\begin{lemma}\label{lemma: C4 Poisson}
	Under Poisson sampling and Conditions~A\ref{cond: inpdt Poisson}--A\ref{cond: Poi S6}, Condition~C\ref{cond: clt} holds.
\end{lemma}
\begin{proof}[Proof of Lemma~\ref{lemma: C4 Poisson}]
	Given every $\ba\in\mathbb{R}^p$ satisfying $\lVert\ba\rVert=1$,  by Conditions~A\ref{cond: inpdt Poisson}--A\ref{cond: limit fuller}, Theorem~1.3.5 of \citet{fuller2009sampling} shows
	\begin{equation}
		\hat{V}\{\ba^\T\hat{\bS}_w(\btheta_0)\mid U\}^{-1/2}\ba^\T\{\hat{\bS}_w(\btheta_0) - \bS(\btheta_0)\}\mid U\to N(0,1)\label{eq: clt 1}
	\end{equation} in distribution with respect to the sampling mechanism conditional on finite population in probability, where  $\hat{V}\{\ba^\T\hat{\bS}_w(\btheta_0)\mid U\}$ is discussed in Lemma~\ref{lemma: 5},
	and  $\bS(\btheta_0) = N^{-1}\sum_{i=1}^N\bS(\btheta_0;y_i)$.
	
	Since the finite population $\{y_1,\ldots,y_N\}$ is a random sample generated from a super-population model satisfying Conditions~A\ref{cond: six moments}--A\ref{cond: limit fuller}, by the Chebychev's inequality \citep[Corollary~3.1.3]{athreya2006measure}, we have 
	\begin{equation}
		\ba^\T\bS(\btheta_0) = \ba^{\T}[\bS(\btheta_0) - E\{\bS(\btheta_0) \}] = o_p(n_0^{-1/2})\label{eq: clt 2}
	\end{equation}
	with respect to the super-population model,
	where the second equality holds since $E\{\bS(\btheta_0)\} = 0$, and the last equality holds by Condition~A\ref{cond: Poi pi}.
	
	Thus, by Condition~A\ref{cond: Poi S6}, Lemma~\ref{lemma: 5} and (\ref{eq: clt 1})--(\ref{eq: clt 2}), Theorem~5.1 of \citet{rubin-bleuer2005} can be used to show 
	\begin{equation}
		\hat{V}\{\ba^\T\hat{\bS}_w(\btheta_0)\mid U\}^{-1/2}\ba^\T\hat{\bS}_w(\btheta_0)  \to N(0,1)\label{eq: clt 3}
	\end{equation}
	in distribution with respect to the super-population model and the sampling mechanism, which validates Condition~C\ref{cond: clt} by the Cram\'{e}r-Wold device.
	
\end{proof}

The following lemma shows the uniform convergence of $\hat{\bI}_w(\btheta)$ to $\bcalI(\btheta)$ in probability with respect to the super-population model and the sampling mechanism over $\btheta\in\mathcal{B}$ in Condition~C\ref{cond: I}.
\begin{lemma}\label{lemma: C5 Poisson}
	Under Poisson sampling, Conditions~A\ref{cond: inpdt Poisson}--A\ref{cond: Poi pi} and Condition~A\ref{cond: z poisson}, Condition~C\ref{cond: I} holds.
\end{lemma}
\begin{proof}[Proof of Lemma~\ref{lemma: C5 Poisson}]
	Consider
	\begin{eqnarray}
		&E\{\hat{\bI}_w(\btheta)\mid U\} =N^{-1}\sum_{i=1}^N \bI(\btheta;y_i),\label{eq 43}\\
		&E\left\{N^{-1}\sum_{i=1}^N \bI(\btheta;y_i)\right\}=  \bcalI(\btheta).\label{eq: 431}
	\end{eqnarray} 	
	Thus, by (\ref{eq 43})--(\ref{eq: 431}), we have 
	\begin{equation}
		E\{\hat{\bI}_w(\btheta)\} = \bcalI(\btheta)\label{eq: exp bcal I}
	\end{equation}
	with respect to the super-population model and the sampling mechanism.
	For $\ba\in\mathbb{R}^p$ satisfying $\lVert \ba\rVert =1$, consider the variance of $\ba^{\T}\hat{\bI}_w(\btheta)\ba$ with respect to the super-population model and the sampling mechanism:
	\begin{equation}
		V\{\ba^{\T}\hat{\bI}_w(\btheta)\ba\} = E[V\{\ba^{\T}\hat{\bI}_w(\btheta)\ba\mid U\}] + V[E\{\ba^{\T}\hat{\bI}_w(\btheta)\ba\mid U\}].
	\end{equation}
	First, consider
	\begin{eqnarray}
		V\{\ba^{\T}\hat{\bI}_w(\btheta)\ba\mid U\} &=& \frac{1}{N^2}\sum_{i=1}^N\frac{1-\pi_i}{\pi_i}Z^2_{\ba,i}(\btheta)\notag \\ 
		&\leq& \frac{1}{C_1n_0}\frac{1}{N}\sum_{i=1}^NZ^2_{\ba,i}(\btheta),\label{eq: 44}
	\end{eqnarray}
	where $Z_{\ba,i}(\btheta) = \ba^{\T}\bI(\btheta;y_i)\ba$, the second inequality holds by Conditions~A\ref{cond: inpdt Poisson}--A\ref{cond: Poi pi}.
	Thus, we have 
	\begin{equation}
		E[V\{\ba^{\T}\hat{\bI}_w(\btheta)\ba\mid U\}] = O(n_0^{-1})
	\end{equation}
	in probability with respect to the super-population model uniformly over $\mathcal{B}$ by Condition~A\ref{cond: z poisson}.

	Next, consider 
	\begin{eqnarray}
		V\left\{N^{-1}\sum_{i=1}^N \ba^\T\bI(\btheta;y_i)\ba\right\}  &=& N^{-1}V\{\ba^\T \bI(\btheta;Y)\ba\}\notag \\ 
		&\leq& N^{-1}E\{\lvert\ba^\T \bI(\btheta;Y)\ba\rvert^2\}\notag \\ 
		&=& O(N^{-1})\notag \\ 
		&=&o(n_0^{-1})\label{eq: var decom I}
	\end{eqnarray}
	in probability with respect to the super-population model uniformly for $\btheta\in\mathcal{B}$ by Condition~A\ref{cond: z poisson},
	where thee last equality holds by Condition~A\ref{cond: Poi pi}. 
	
	By  (\ref{eq: exp bcal I})--(\ref{eq: var decom I}) and the Markov's inequality, we can show that\begin{equation}
		\hat{\bI}_w(\btheta)\to \bcalI(\btheta)\label{eq: unif I}
	\end{equation} in probability with respect to the super-population model and the sampling mechanism uniformly for $\btheta\in\mathcal{B}$, which completes the proof of Lemma~\ref{lemma: C5 Poisson}.
\end{proof}
\begin{lemma}\label{lemma: Poisson Condition 7}
	The proposed bootstrap satisfies Condition~C\ref{cond: boot weights}.
\end{lemma}
\begin{proof}[Proof of Lemma~\ref{lemma: Poisson Condition 7}] Recall that, {from Example~\ref{ex: 1}}, the bootstrap replicate of $\hat{\bS}_w(\btheta)$ is $\hat{\bS}_w^*(\btheta) = N^{-1}\sum_{i \in A} r_i^*w_i \bS(\btheta;y_i)$ with $w_i=\pi_i^{-1}$, where $r_i^* = 1+   m_i^*- \pi_i \geq0$, and $m_i^* \sim Bernoulli (\pi_i)$. Then, $E_*(r_i^*)=1$ and $V_*(r_i^*) =\pi_i(1-\pi_i)$. Thus, we have, conditional on the realized sample $A$,
	\begin{eqnarray}
		E_*\{\hat{\bS}_w^*(\btheta)\} &=&N^{-1}\sum_{i \in A} E_*(r_i^*)w_i \bS(\btheta;y_i)\notag \\
		&=& \hat{\bS}_w(\btheta),\label{eq: Poisson Condition 7 E*}\\
		\bV_*\{\hat{\bS}_w^*(\btheta)\} &=&N^{-2}\sum_{i \in A} \frac{\pi_i(1-\pi_i)}{\pi_i^2} \bS(\btheta;y_i)^{\otimes2}.\label{eq: Poisson Condition 7 V*}
		\notag \\
		&=&\hat{\bV}\{\hat{\bS}_w(\btheta)\}
	\end{eqnarray}
	By  (\ref{eq: Poisson Condition 7 E*})--(\ref{eq: Poisson Condition 7 V*}), we have completed the proof for  Lemma~\ref{lemma: Poisson Condition 7}.
\end{proof}
The following result validates Condition~C\ref{cond: boot CLT result} under Poisson sampling.
\begin{lemma}\label{lemma: C8 Poisson}
	Under Poisson sampling and Conditions~A\ref{cond: inpdt Poisson}--A\ref{cond: six moments}, Condition~C\ref{cond: boot CLT result} holds.
\end{lemma}
\begin{proof}[Proof of Lemma~\ref{lemma: C8 Poisson}]
	Recall that  $\hat{\bS}_w^* ( \btheta) =\sum_{i\in A}w_i^*\bS(\btheta;y_i)$ is the bootstrap weighted score function, 
	where $w_i^* = w_i(1+   m_i^*- \pi_i)$ and $m_i^* \sim Bernoulli (\pi_i)$ under Poisson sampling. We have shown that $E_*\{\hat{\bS}_w^* ( \btheta) \}=\hat{\bS}_w(\btheta)$ and $\var_*\{\hat{\bS}_w^* ( \btheta) \} = \hat{\var}\{\bS_w(\btheta)\mid U\}$ in Lemma~\ref{lemma: Poisson Condition 7}.
	
	First, conditional on the realized sample $A$, consider  
	\begin{eqnarray}
		\frac{1}{N^3}\sum_{i\in A}\frac{\lvert\ba^\T \bS(\btheta;y_i)\rvert^3}{\pi_i^3}E_*(\lvert m_i^*-\pi_i\lvert^3) &=& \frac{O(1)}{N^3}\sum_{i\in A}\frac{\lvert\ba^\T \bS(\btheta;y_i)\rvert^3}{\pi_i^3},\label{eq: third moments}
	\end{eqnarray}
	where the equality holds since $E_*(\lvert m_i^*-\pi_i\lvert^3)<1$ uniformly for $i\in A$.
	In order to investigate the stochastic order of (\ref{eq: third moments}) for $\btheta\in\mathcal{B}$, consider 
	\begin{eqnarray}
		E\left\{\frac{1}{N^3}\sum_{i\in A}\frac{\lvert\ba^\T \bS(\btheta;y_i)\rvert^3}{\pi_i^3}\mid U\right\}&=&  \frac{1}{N^3}\sum_{i=1}^N\frac{\lvert\ba^\T \bS(\btheta;y_i)\rvert^3}{\pi_i^2}\notag\\ 
		&\asymp& \frac{1}{n_0^2N}\sum_{i=1}^N\lvert\{\ba^\T \bS(\btheta;y_i)\}\rvert^3\notag \\ 
		&=&O(n_0^{-2})\label{eq: 54}\notag
	\end{eqnarray} in probability with respect to the super-population model uniformly over $\mathcal{B}$ by Conditions~A\ref{cond: Poi pi}--A\ref{cond: six moments} and the H\"{o}lder's inequality \citep[Theorem~3.1.11]{athreya2006measure}, so we conclude  that the order of (\ref{eq: third moments}) is $O_p(n^{-2})$ by the Markov's inequality.
	By Lemma~\ref{lemma: 5} and (\ref{eq: third moments}), we can apply Lyapounov's central limit theorem \citep[Corollary 11.1.4]{athreya2006measure} and the Cram\'{e}r-Wold device  to get Condition~C\ref{cond: boot CLT result} proved by noting the fact that $\{r_i^*:i\in A\}$ are generated independently.
\end{proof}

It remains to verify Condition~C\ref{cond: boot I} by the following lemma. 
\begin{lemma}\label{lemma: C9 Poisson}
	Under Poisson sampling,  Conditions~A\ref{cond: inpdt Poisson}--A\ref{cond: Poi pi} and Condition~A\ref{cond: z poisson}, Condition~C\ref{cond: boot I} holds.
\end{lemma}
\begin{proof}[Proof of Lemma~\ref{lemma: C9 Poisson}]
	For $\ba\in\mathbb{R}^p$ with $\lVert\ba\rVert=1$, we consider $\ba^{\T}\hat{\bI}_w^*(\btheta)\ba$, and we have  
	\begin{eqnarray}
		E_*\{\ba^{\T}\hat{\bI}_w^*(\btheta)\ba\} &=& E_*\left\{\frac{1}{N}\sum_{i\in A}\frac{1+m_i^*-\pi_i}{\pi_i}Z_{\ba,i}(\btheta)\right\}\notag \\ 
		&=&\ba^{\T}\hat{\bI}_w(\btheta)\ba,\label{eq: 57a}
	\end{eqnarray} 
	where $Z_{\ba,i}(\btheta) = \ba^{\T}\bI(\btheta;y_i)\ba$.
	Next, consider 
	\begin{eqnarray}
		V_*\{\ba^{\T}\hat{\bI}_w^*(\btheta)\ba\}=\frac{1}{N^2}\sum_{i\in A}\frac{1-\pi_i}{\pi_i}Z^2_{\ba,i}(\btheta).\notag
	\end{eqnarray}
	To investigate the asymptotic order of $V_*\{\ba^{\T}\hat{\bI}_w^*(\btheta)\ba\}$, consider
	\begin{eqnarray}
		E\left\{\frac{1}{N^2}\sum_{i\in A}\frac{1-\pi_i}{\pi_i}Z^2_{\ba,i}(\btheta)\right\}&=&E\left\{\frac{1}{N^2}\sum_{i=1}^N(1-\pi_i)Z^2_{\ba,i}(\btheta)\right\}\notag \\ 
		&\leq& N^{-1}E\left\{Z^2_{\ba,i}(\btheta)\right\}\notag \\
		&=& O(N^{-1})\notag \\ 
		&=& o(1)
	\end{eqnarray}
	with respect to the super-population model and the sampling mechanism,
	where the third equality holds uniformly for $\btheta\in\mathcal{B}$ by Condition~A\ref{cond: z poisson}, and the last equality holds by Condition~A\ref{cond: Poi pi}. Thus, by Markov's inequality, we can show that \begin{equation}
		V_*\{\ba^{\T}\hat{\bI}_w^*(\btheta)\ba\}=o_p(1)\notag
	\end{equation} 	holds with respect to the super-population and the sampling mechanism uniformly for $\btheta\in\mathcal{B}$. 
	By the uniform convergence of $\hat{\bI}_w(\btheta)$ in Lemma~\ref{lemma: C5 Poisson} and the Markov's inequality, it follows that
	\begin{equation}
		\hat{\bI}_w^*(\btheta)\to\bcalI(\btheta)\label{eq: unif I star}
	\end{equation}
	in probability with respect to the super-population, the sampling mechanism and the bootstrap procedure uniformly for $\btheta\in\mathcal{B}$.
\end{proof}

By Conditions~A\ref{eq: consis Poisson}--A\ref{cond: boot I} and Lemmas~\ref{coro: cond 2}--\ref{lemma: C9 Poisson}, we have verified the general conditions in Section~\ref{ss: regularity conditions}. Thus, Theorems~\ref{theo: CLT (2)}--\ref{theo: (6)} hold for the proposed bootstrap method under Poisson sampling.

\section{Probability proportional to size sampling with replacement}\label{ss: pps sampling}

To validate the proposed bootstrap method {under probability proportional to size sampling with replacement}, it is sufficient to verify Conditions~C\ref{cond: smooth of f}--C\ref{cond: boot I} in Section~\ref{ss: regularity conditions}.  We now assume the following specific conditions.
\begin{enumerate}
	\renewcommand{\labelenumi}{B\arabic{enumi}.}
	\setcounter{enumi}{0}
	\item For $\btheta\in\Theta$, the function $l(\btheta;y)$ is  concave and twice-continuously differentiable with respect to $\btheta$,  $E\{\lvert l(\btheta;Y)\rvert^2\}<\infty$, and $E\{l(\btheta;Y)\}$ is uniquely maximized at $\btheta_0$, where  $\Theta$ is a convex set containing $\btheta_0$ as an interior point.
	
	\label{eq: consis PPS}
	\item There exist two constants $0<C_3<C_4<\infty$ such that 
	$C_3<Np_i<C_4$ for $i=1,\ldots,N$ almost surely with respect to the super-population model, the sample size satisfies $n=o(N)$, and $n\to\infty$ as $N\to\infty$.\label{cond: selection prob}
	\item  \label{cond: Uniform convergence PPS}There exists a compact set $\mathcal{B}\subset\Theta$ containing $\btheta_0$ as an interior point such that 
	\begin{eqnarray}
		&\sup_{\btheta\in\mathcal{B}}E(\lVert \bS(\btheta;Y)\rVert^4)=O(1),\label{eq: S} \\
		&\sup_{\btheta\in\mathcal{B}}E\left\{\lVert\bI(\btheta;Y)\rVert^2\right\}=O(1).\label{eq: I}
	\end{eqnarray}  Besides, $\bV\{\bS(\btheta_0;Y)\}$ is positive definitive, and  $\bcalI(\btheta_0)$ is invertible, where  $\bcalI(\btheta)=E\{\bI(\btheta;Y)\}$. \label{cond: second moment}
	\item $n\var\{\hat{\bS}_w(\btheta)\mid U\}$ converges to a   positive-definitive matrix $\bSigma_S(\btheta)$   in probability with respect to the super-population model uniformly over $\mathcal{B}$, where $\var\{\hat{\bS}_w(\btheta)\mid U\}$ is the design  variance of $\hat{\bS}_w(\btheta)$ conditional on the finite population $U$, and $\bSigma_S(\btheta)$ is non-stochastic with respect to the super-population model and the sampling mechanism.  \label{cond: var PPS}
	
\end{enumerate}
Condition~B\ref{eq: consis PPS} is almost the same as Condition~C\ref{cond: smooth of f}, and the difference is the second moment condition   $E\{\lvert l(\btheta;Y)\rvert^2\}<\infty$, which is used to get Condition~C\ref{cond: consistency l} under probability proportional to size sampling with replacement.   Condition~B\ref{cond: selection prob} is    used  to show the asymptotic properties of the proposed bootstrap methods under probability proportional to size sampling with replacement. In Condition~B\ref{cond: second moment}, (\ref{eq: S}) guarantees  the consistency of the  design  and bootstrap variance estimators, and it is also used to show the central limit theorems for the weighted score functions; (\ref{eq: I}) is used to show the consistency of the negative Hessian  matrix.  Condition~B\ref{cond: var PPS} is analogous to Condition~C\ref{cond: asymptotic con var}, and it is required to get the central limit theorem for the weighted score functions.

Next, we validate the conditions in Section~\ref{ss: regularity conditions}  one by one.
\begin{lemma}\label{lemma: C2 PPS}
	Under probability proportional to size sampling with replacement and Conditions~B\ref{eq: consis PPS}--B\ref{cond: selection prob}, Condition~C\ref{cond: consistency l} holds.
\end{lemma}
\begin{proof}[Proof of Lemma~\ref{lemma: C2 PPS}]
	Recall that $l_w(\btheta) = (Nn)^{-1} \sum_{i=1}^n p^{-1}_{a(i)}l(\btheta;y_{a(i)})$ under probability proportional to size sampling with replacement, where  $a(i)$ denotes the index of the element selected at the $i$-th draw. Then, 
	\begin{eqnarray}
		V\{l_w(\btheta)\mid U\} 
		&=& \frac{n}{N^2n^2}\left[\sum_{i=1}^Np_i\frac{l(\btheta;y_i)^2}{p_i^2}-\left\{\sum_{i=1}^Np_i\frac{l(\btheta;y_i)}{p_i}\right\}^2\right]\notag \\ 
		&\leq&\frac{1}{n}\left[\frac{1}{C_3N}\sum_{i=1}^N l(\btheta;y_i)^2-\left\{\frac{1}{N}\sum_{i=1}^N l(\btheta;y_i)\right\}^2\right]\notag \\ 
		&=&O_p(n^{-1})\label{eq: l var pps}
	\end{eqnarray}
	with respect to the super-population model,
	where the second inequality  holds by Condition~B\ref{cond: selection prob}, and the last equality holds by Condition~B\ref{eq: consis PPS} and the weak law of large number.
	
	Under  probability proportional to size sampling with replacement, a design-unbiased variance estimator of $l_w(\btheta)$ is \begin{eqnarray}
		\hat{V}\{l_w(\btheta)\mid U\} &=&\frac{1}{n^2N^2}\frac{n}{n-1}\left[\sum_{i=1}^n p^{-2}_{a(i)}l(\btheta;y_{a(i)})^2 - n\left\{\frac{1}{n}\sum_{i=1}^n p^{-1}_{a(i)}l(\btheta;y_{a(i)})\right\}^2\right],\label{eq: S60.new} 
	\end{eqnarray}
	and it is non-negative. 
	By (\ref{eq: l var pps})--(\ref{eq: S60.new}), we have completed the proof for Lemma~\ref{lemma: C2 PPS}.
\end{proof}

\begin{lemma}\label{lemma: C4 PPS}
	Under probability proportional to size sampling with replacement and  Conditions~B\ref{eq: consis PPS}--B\ref{cond: var PPS}, Condition~C\ref{cond: clt} holds.
\end{lemma}
\begin{proof}[Proof of Lemma~\ref{lemma: C4 PPS}]
	We use Theorem~5.1 of \citet{rubin-bleuer2005} to show Condition~C\ref{cond: clt}, and it is sufficient to show the following two results:
	\begin{equation}
		\sqrt{n}\bS(\btheta_0)\to 0\label{eq: C4 PPS part 1}
	\end{equation}
	in probability with respect to the super-population model and 
	\begin{equation}
		\sqrt{n}\{\hat{\bS}_w(\btheta_0) - \bS(\btheta_0)\}\mid U\to N(0,\Sigma_S(\btheta_0))\label{eq: C4 PPS part 2}
	\end{equation}
	in distribution with respect to the sampling mechanism conditional on the finite population in probability with respect to the super-population model. 
	
	First, we show (\ref{eq: C4 PPS part 1}) by considering
	\begin{eqnarray}
		\sqrt{n}\bS(\btheta_0)&=& \frac{\sqrt{n}}{N}\sum_{i=1}^N\bS(\btheta_0;y_i)\notag \\ 
		&=& \frac{\sqrt{n}}{N}\sum_{i=1}^N[\bS(\btheta_0;y_i)-E\{\bS(\btheta_0;y_i)\}]\notag \\ 
		&=&O_p(n^{1/2}N^{-1/2})\notag \\ 
		&=&o_p(1)\label{eq: C4 PPS part 1 proof}
	\end{eqnarray}
	with respect to the super-population model,
	where the second equality holds by Condition~B\ref{eq: consis PPS}, and the third asymptotic order holds by the weak law of large number and  (\ref{eq: S}) of Condition~B\ref{cond: second moment}, and the last equality holds by Condition~B\ref{cond: selection prob}. Thus, by (\ref{eq: C4 PPS part 1 proof}), we have validated (\ref{eq: C4 PPS part 1}).
	
	Next, we show (\ref{eq: C4 PPS part 2}) by the Lindeberg-Feller central limit theorem \citep[Proposition~2.27]{van2000asymptotic}. For simplicity, denote $\tilde{\bS}_i = n^{-1/2}N^{-1}p_{a(i)}^{-1}\bS(\btheta_0;y_{a(i)})$. Thus, we have $\sqrt{n}\hat{\bS}_w(\btheta_0)=\sum_{i=1}^n\tilde{\bS}_i$ and
	\begin{equation}
		\sum_{i=1}^n\bV(\tilde{\bS}_i) = \bV\left(\sum_{i=1}^n\tilde{\bS}_i\right) =n\bV\{\hat{\bS}_w(\btheta_0)\}\to \bSigma_S(\btheta_0)\label{eq: C4 PPS part 2 proof 1}
	\end{equation} in probability with respect to the super-population model by Condition~B\ref{cond: var PPS}, where the first equality holds by the fact that $\{\tilde{\bS}_i:i=1,\ldots,n\}$ are independent and identically distributed under probability proportional to size sampling with replacement. 
	
	For $\epsilon>0$, consider 
	\begin{eqnarray}
		&&\sum_{i=1}^nE\{\lVert \tilde{\bS}_i\rVert^2I(\lvert \tilde{\bS}_i\rVert^2>\epsilon)\mid U\} \notag \\ 
		&=&E\left[\lVert N^{-1}p_{a(1)}^{-1}\bS(\btheta_0;y_{a(1)})\rVert^2I\{\lVert N^{-1}p_{a(1)}^{-1}\bS(\btheta_0;y_{a(1)})\rVert>\sqrt{n}\epsilon\}\mid U\right]\notag \\ 
		&=& \frac{1}{N^2}\sum_{i=1}^Np_i\frac{\lVert \bS(\btheta_0;y_{i})\rVert^2}{p_i^2}I\{\lVert N^{-1}p_{i}^{-1}\bS(\btheta_0;y_{i})\rVert>\sqrt{n}\epsilon\}\notag \\ 
		&\leq& \frac{1}{C_3N}\sum_{i=1}^N\lVert \bS(\btheta_0;y_{i})\rVert^2I\{\lVert \bS(\btheta_0;y_{i})\rVert>C_3\sqrt{n}\epsilon\}\notag \\ 
		&=&o_p(1)\label{eq: C4 PPS part 2 proof 2}
	\end{eqnarray}
	with respect to the super-population model,
	where the third inequality holds by Condition~B\ref{cond: selection prob}, and the last inequality holds by (\ref{eq: S}) of Condition~B\ref{cond: second moment}, dominant convergence theorem and the weak law of large numbers. Thus, by (\ref{eq: C4 PPS part 2 proof 1})--(\ref{eq: C4 PPS part 2 proof 2}), we have shown (\ref{eq: C4 PPS part 2}), which concludes the proof of Lemma~\ref{lemma: C4 PPS}.
	
\end{proof}

\begin{lemma}\label{lemma: Condition C5  PPS}
	Under probability proportional to size sampling with replacement and  Conditions~B\ref{cond: selection prob}--B\ref{cond: second moment}, Condition~C\ref{cond: I} holds.
\end{lemma}
\begin{proof}[Proof of Lemma~\ref{lemma: Condition C5  PPS}]
	Under probability proportional to size sampling with replacement, we have $\hat{\bcalI}_w(\btheta) = (Nn)^{-1}\sum_{i=1}^np^{-1}_{a(i)}\bI(\btheta;y_{a(i)})$. 
	
	Consider
	\begin{eqnarray}
		E\{\hat{\bcalI}_w(\btheta)\} 
		&=& E\left[E\left\{\hat{\bcalI}_w(\btheta)\mid U\right\}\right]\notag \\ 
		&=& E\left[\frac{1}{N}\sum_{i=1}^Np_i\frac{\bI(\btheta;y_i)}{p_i}\right]\notag \\ 
		&=&\bcalI(\btheta) \label{eq: E I_w theta PPS}
	\end{eqnarray} 
	for $\btheta\in\Theta$, where the expectation in (\ref{eq: E I_w theta PPS}) is with respect to the super-population model and the sampling mechanism.
	
	Next, for every $\ba\in\mathbb{R}^p$ satisfying $\lVert \ba\rVert=1$, we consider the variance of $\ba^\T\hat{\bcalI}_w(\btheta)\ba$ as we have done in the proof of Lemma~\ref{lemma: C5 Poisson}, and we have 
	\begin{equation}
		V\left\{\ba^{\T}\hat{\bI}_w(\btheta)\ba\right\} = E\left[V\left\{\ba^{\T}\hat{\bI}_w(\btheta)\ba\mid U\right\}\right] + V\left[ E\left\{\ba^{\T}\hat{\bI}_w(\btheta)\ba\mid U\right\}\right].\label{eq V decompose PPS}
	\end{equation}
	First, consider
	\begin{eqnarray}
		V\left\{\ba^{\T}\hat{\bI}_w(\btheta)\ba\mid U\right\} 
		&=& \frac{1}{nN^2}\left[\sum_{i=1}^Np_i\frac{Z^2_{\ba,i}(\btheta)}{p_i^2} - \left\{\sum_{i=1}^Np_i\frac{Z_{\ba,i}(\btheta)}{p_i}\right\}^2\right]\notag \\
		&\leq& \frac{1}{C_3n}\frac{1}{N}\sum_{i=1}^NZ^2_{\ba,i}(\btheta)-\frac{1}{n}\left\{\frac{1}{N}\sum_{i=1}^NZ_{\ba,i}(\btheta)\right\}^2\notag \\ 
		&=& O_p(n^{-1})\notag
	\end{eqnarray}
	with respect to the super-population model
	uniformly over $\mathcal{B}$, where $Z_{\ba,i}(\btheta) = \ba^{\T}\bI(\btheta;y_i)\ba$, the second inequality holds by Condition~B\ref{cond: selection prob}, and the last asymptotic order holds by (\ref{eq: I}) of Condition~B\ref{cond: second moment}.
	Thus, we have 
	\begin{equation}
		E\left[V\left\{\ba^{\T}\hat{\bI}_w(\btheta)\ba\mid U\right\}\right] = O(n^{-1})\label{eq: EV PPS 2}
	\end{equation}
	uniformly over $\mathcal{B}$ by (\ref{eq: I}) in Condition~B\ref{cond: second moment}. By a similar argument to (\ref{eq: var decom I}), we can show 
	\begin{equation}
		V\left[E\left\{\ba^{\T}\hat{\bI}_w(\btheta)\ba\mid U\right\}\right] = o(n^{-1})\label{eq: VE PPS 2}
	\end{equation}
	by Condition~B\ref{cond: selection prob} and  (\ref{eq: I}) in Condition~B\ref{cond: second moment}. Thus, by (\ref{eq: E I_w theta PPS})--(\ref{eq: VE PPS 2}), we have completed the proof of Lemma~\ref{lemma: Condition C5  PPS}.
\end{proof}

\begin{lemma}\label{lemma: Condition C6  PPS}
	Under probability proportional to size sampling with replacement and  Conditions~B\ref{cond: selection prob}--B\ref{cond: second moment}, Condition~C\ref{cond: variance estimator} holds.
\end{lemma}
\begin{proof}[Proof of Lemma~\ref{lemma: Condition C6  PPS}]
	For every $\ba\in\mathbb{R}^p$ satisfying $\lVert \ba\rVert=1$,  denote $Z_{n,i} = p_{a(i)}^{-1}\ba^\T\bS(\btheta_0;y_{a(i)})$ for $i=1,\ldots,n$.   Under probability proportional to size sampling with replacement,  the design variance of $\ba^\T\hat{\bS}_w(\btheta)$ is
	\begin{eqnarray}
		{V}\{\ba^\T\hat{\bS}_w(\btheta)\mid U\}   &=& \frac{1}{n}\left[\frac{1}{N^2}\sum_{i=1}^Np_i\frac{\{\ba^\T\bS(\btheta_0;Y_i)\}^2}{p_i^2} -\left\{\frac{1}{N} \sum_{i=1}^N\ba^\T\bS(\btheta_0;Y_i)\right\}^2 \right]\label{eq: conditional variance PPS},
	\end{eqnarray}
	and a design  variance estimator of $\ba^\T\hat{\bS}_w(\btheta)$ is 
	\begin{equation}
		\hat{V}\{\ba^\T\hat{\bS}_w(\btheta)\mid U\}     
		=\frac{1}{n^2N^2}\frac{n}{n-1}\left\{\sum_{i=1}^n Z_{n,i}^2 - n\left(\frac{1}{n}\sum_{i=1}^n Z_{n,i}\right)^2\right\}.\label{eq: estimated variance PPS}
	\end{equation}
	By (\ref{eq: conditional variance PPS})--(\ref{eq: estimated variance PPS}), to show Condition~C\ref{cond: variance estimator}, it is sufficient to show
	\begin{eqnarray}
		&\displaystyle\frac{1}{nN^2}\frac{1}{n-1} \sum_{i=1}^n Z_{n,i}^2
		-\frac{1}{nN^2}\sum_{i=1}^Np_i\frac{\{\ba^\T\bS(\btheta_0;Y_i)\}^2}{p_i^2}
		=o_p(n^{-1}),\label{eq: var hat 1 PPS} \\
		&\displaystyle\frac{1}{nN}\sum_{i=1}^n Z_{n,i} - 
		\frac{1}{N} \sum_{i=1}^N\ba^\T\bS(\btheta_0;Y_i)
		=o_p(1)\label{eq: var hat 2 PPS}
	\end{eqnarray}
	with respect to the super-population model and the sampling mechanism uniformly over $\mathcal{B}$.

	To show (\ref{eq: var hat 1 PPS}), consider 
	\begin{eqnarray}
		E\left(\frac{1}{nN^2}\frac{1}{n-1} \sum_{i=1}^n Z_{n,i}^2\mid U\right) 
		&=&\frac{1}{N^2(n-1)}E( Z_{n,1}^2\mid U)  \notag \\
		&=&\frac{1}{N^2(n-1)}\sum_{i=1}^Np_i\frac{\{\ba^{\T}\bS(\btheta_0;y_i)\}^2}{p_i^2}\notag \\ 
		&=& \frac{1}{nN^2}\sum_{i=1}^Np_i\frac{\{\ba^\T\bS(\btheta_0;y_i)\}^2}{p_i^2} + o_p(n^{-1})\notag \\ 
		&=& o_p(n^{-1})\label{eq: var hat 1 PPS expectation part}
	\end{eqnarray}
	with respect to the super-population uniformly over $\mathcal{B}$,
	where the last two equalities  hold by Conditions~B\ref{cond: selection prob}--B\ref{cond: second moment} and the weak law of large number. Next, consider 
	\begin{eqnarray}
		V\left(\frac{1}{nN^2(n-1)} \sum_{i=1}^n Z_{n,i}^2\mid U\right) 
		&=& \frac{1}{nN^4(n-1)^2} V(Z_{n,1}^2\mid U)\notag \\ 
		&\leq& \frac{1}{nN^4(n-1)^2}E(Z_{n,1}^4\mid U)\notag \\ 
		&=&\frac{1}{nN^4(n-1)^2}\sum_{i=1}^N     
		p_i\frac{\{\ba^{\T}\bS(\btheta_0;Y_i)\}^4}{p_i^4}\notag \\ 
		&=&O_p(n^{-3})\label{eq: var hat 1 PPS var part}
	\end{eqnarray}
	in probability with respect to the super-population  uniformly over $\mathcal{B}$, where the last asymptotic order holds by Conditions~B\ref{cond: selection prob}--B\ref{cond: second moment} and the weak law of large number. Thus, by (\ref{eq: var hat 1 PPS expectation part})--(\ref{eq: var hat 1 PPS var part}), we have validated (\ref{eq: var hat 1 PPS}), and  we can also validate (\ref{eq: var hat 2 PPS}) in a similar manner. Thus, we have completed the proof for Lemma~\ref{lemma: Condition C6  PPS}.
\end{proof}

\begin{lemma}\label{lemma: PPS Condition 7}
	The proposed bootstrap satisfies Condition~C\ref{cond: boot weights}.
\end{lemma}
\begin{proof}[Proof of Lemma~\ref{lemma: PPS Condition 7}]
	Recall that,from Example~\ref{ex: 2}, the bootstrap weighted score function is $\hat{\bS}_w^*(\btheta) =(Nn)^{-1}k_n\bS_{vec}(\btheta)\bbm^*$, where $k_n= n(n-1)^{-1}$, $\bS_{vec}(\theta) = (p^{-1}_{a(1)}\bS(\btheta;y_{a(1)}),\ldots,p^{-1}_{a(n)}\bS(\btheta;y_{a(n)}))$ is a $p\times n$ matrix, $\bbm^*=(m_{1}^*,\ldots,m_{n}^*)^{\T}$ is generated using a multinomial distribution with $n-1$ trials and a success probability vector $n^{-1}\bone_n$, and $\bone_n$ is a vector of one with length $n$. 
	
	Since $E_*(\bbm^*) =k_n^{-1}\bone_n$, we have 
	\begin{equation}
		E_*\{\hat{\bS}_w^*(\btheta)\} = (Nn)^{-1}k_n\bS_{vec}(\btheta)k_n^{-1}\bone_n = \hat{\bS}_w(\btheta).\label{eq: Condition 7 E* PPS}
	\end{equation}
	By the basic property of multinomial distribution, we have $V_*(\bbm^*)=(n-1)n^{-1}(\bI_n - \mathcal{P}_{\bone,n})$, where $\bI_n$ is the $n\times n$ identity matrix,  and $\mathcal{P}_{\bone,n} = \bone_n(\bone_n^{\T}\bone_n)^{-1}\bone_n^{\T}$ is the projection matrix to the linear space spanned by $\bone_n$. Thus, we have 
	\begin{eqnarray}
		V_*\{\hat{\bS}_w^*(\btheta)\} &=& \frac{k_n^2}{(Nn)^{2}}\frac{n-1}{n}\bS_{vec}(\btheta)(\bI_n - \mathcal{P}_{\bone,n})\bS_{vec}^{\T}(\btheta)\notag \\ 
		&=& \frac{1}{N^{2}n(n-1)} \bS_{vec}(\theta)(\bI_n - \mathcal{P}_{\bone,n})\bS{}^{\T}_{vec}(\theta)\notag\\ 
		&=&\hat{V}\{\hat{\bS}_w(\btheta)\}.\label{eq: condition 7 V^* PPS}
	\end{eqnarray}
	By (\ref{eq: Condition 7 E* PPS})--(\ref{eq: condition 7 V^* PPS}), we have completed the proof for Lemma~\ref{lemma: PPS Condition 7}.
\end{proof}
The next lemma concerns the bootstrap central limit theorem in Condition~C\ref{cond: boot CLT result}.

\begin{lemma}\label{lemma: Condition C8  PPS}
	Under probability proportional to size sampling with replacement and  Conditions~B\ref{cond: selection prob}--B\ref{cond: second moment}, Condition~C\ref{cond: boot CLT result} holds.
\end{lemma}
\begin{proof}[Proof of Lemma~\ref{lemma: Condition C8  PPS}]
	By Condition~C\ref{cond: asymptotic con var} and Conditions~C\ref{cond: variance estimator}--C\ref{cond: boot weights}, we have 
	\begin{eqnarray}
		n\var_*\{\hat{\bS}^*_w(\btheta)\} \to \bSigma(\btheta) \label{eq: Condition 8 PPS variance}
	\end{eqnarray}
	in probability with respect to the super-population model, the sampling mechanism and the bootstrap procedure uniformly over $\mathcal{B}$. 
	
	By (\ref{eq: Condition 8 PPS variance}) and the Lindeberg-Feller central limit theorem, it remains to show
	\begin{equation}
		E_*\left[k_n^2\left\lVert \frac{\bS(\btheta;y_i^*)}{p_i^*N}  \right\rVert^2I\left\{ k_n\left\lVert \frac{\bS(\btheta;y_i^*)}{p_i^*N}  \right\rVert>\sqrt{n}\epsilon   \right\}\right]\to0 \label{eq: S.24}
	\end{equation}
	in probability with respect to the super-population model, the sampling mechanism and the bootstrap procedure uniformly over $\mathcal{B}$ for any $\epsilon>0$ , where $p_i^*$ is the selection probability for $\bS(\btheta;y_i^*)$, and   $k_n = n(n-1)^{-1}$. Consider 
	\begin{eqnarray}
		&&E_*\left[k_n^2\left\lVert \frac{\bS(\btheta;y_i^*)}{p_i^*N}  
		\right\rVert^2I\left\{ k_n\left\lVert \frac{\bS(\btheta;y_i^*)}{p_i^*N}  \right\rVert>\sqrt{n}\epsilon   \right\}\right]\notag \\ 
		&=& \frac{1}{n}\sum_{i=1}^nk_n^2\left\lVert \frac{\bS(\btheta;y_{a(i)})}{p_{a(i)}N}  
		\right\rVert^2I\left\{ k_n\left\lVert \frac{\bS(\btheta;y_{a(i)})}{p_{a(i)}N}  \right\rVert>\sqrt{n}\epsilon   \right\}.\label{eq: S.25}
	\end{eqnarray}

	Furthermore, we have 
	\begin{eqnarray}
		&& E\left[\frac{1}{n}\sum_{i=1}^nk_n^2\left\lVert \frac{\bS(\btheta;y_{a(i)})}{p_{a(i)}N}  
		\right\rVert^2I\left\{ k_n\left\lVert \frac{\bS(\btheta;y_{a(i)})}{p_{a(i)}N}  \right\rVert>\sqrt{n}\epsilon   \right\}\mid U\right] \notag\\
		&=& E\left[k_n^2\left\lVert \frac{\bS(\btheta;y_{a(1)})}{p_{a(1)}N}  
		\right\rVert^2I\left\{ k_n\left\lVert \frac{\bS(\btheta;y_{a(1)})}{p_{a(1)}N}  \right\rVert>\sqrt{n}\epsilon   \right\}\mid U\right] \notag \\
		&=& \sum_{i=1}^Np_i k_n^2\left\lVert \frac{\bS(\btheta;y_{i})}{p_{i}N}  
		\right\rVert^2I\left\{ k_n\left\lVert \frac{\bS(\btheta;y_{i})}{p_{i}N}  \right\rVert>\sqrt{n}\epsilon   \right\}\notag \\ 
		&\leq&\frac{4}{C_3N}\sum_{i=1}^N\lVert \bS(\btheta;y_{i})\rVert^2I\left\{\lVert \bS(\btheta;y_{i}) \rVert>C_3\sqrt{n}\epsilon/2   \right\},\notag \\ 
		&=&o_p(1)\label{eq: E phi part PPS}
	\end{eqnarray}
	uniformly with respect to the super-population model over $\mathcal{B}$, where the third inequality holds by Condition~B\ref{cond: selection prob} and $1/2\leq k_n^{-1}<k_n\leq2$, and the last asymptotic order holds by Condition~B\ref{cond: second moment}, the weak law of large number and the dominant convergence theorem. Thus, we have validated (\ref{eq: S.24}) by (\ref{eq: S.25})--(\ref{eq: E phi part PPS}), and it completes the proof of Lemma~\ref{lemma: Condition C8  PPS}.
\end{proof}

\begin{lemma}\label{lemma: Condition 7 PPS}
	Under probability proportional to size sampling with replacement and  Conditions~B\ref{cond: selection prob}--B\ref{cond: second moment}, Condition~C\ref{cond: boot I}   holds.  
\end{lemma}
\begin{proof}[Proof of Lemma~\ref{lemma: Condition 7 PPS}]
	For $\ba\in\mathbb{R}^p$ with $\lVert\ba\rVert=1$, we consider $$\ba^{\T}\hat{\bI}_w^*(\btheta)\ba=(Nn)^{-1} \sum_{i=1}^n r_i^*p^{-1}_{a(i)}\ba^{\T} \bI(\btheta;y_{a(i)})\ba$$ with 
	the rescaling factor is $r_{i}^* = k_nm_{i}^*$, where $\bbm^*=(m_{1}^*,\ldots,m_{n}^*)^{\T}$ is generated using a multinomial distribution with $n-1$ trials and a success probability vector $n^{-1}(1,\ldots,1)^{\T}$ of length $n$, and  $k_n= n/(n-1)$.  By a similar argument  to Lemma~\ref{lemma: PPS Condition 7}, we have 
	\begin{eqnarray}
		E_*\{\ba^{\T}\hat{\bI}_w^*(\btheta)\ba\} 
		&=&\ba^{\T}\hat{\bI}_w(\btheta)\ba\label{eq: 74a}\\
		\var_*\{\ba^{\T}\hat{\bI}_w^*(\btheta)\ba\}
		&=&\frac{1}{n(n-1)}\left[\sum_{i=1}^nZ^2_{\ba,i}(\btheta) - n\left\{\frac{1}{n}\sum_{i=1}^nZ_{\ba,i}(\btheta)\right\}^2\right],
	\end{eqnarray} 
	where  $Z_{\ba,i}(\btheta) =(Nn)^{-1} p^{-1}_{a(i)}\ba^{\T}\bI(\btheta;y_{a(i)})\ba$.
	
	We can use a similar argument to (\ref{eq: var hat 1 PPS})--(\ref{eq: var hat 1 PPS var part}) to show \begin{equation}
		\var_*\{\ba^{\T}\hat{\bI}_w^*(\btheta)\ba\}=o_p(1)\label{eq: var I pps star}
	\end{equation} with respect to the super-population model and the sampling mechanism  uniformly for $\btheta\in\mathcal{B}$ by (\ref{eq: I}) in Condition~B\ref{cond: second moment}. By (\ref{eq: 74a})--(\ref{eq: var I pps star}), we have completed the proof of Lemma~\ref{lemma: Condition 7 PPS}.
\end{proof}

\section{Stratified {Multi}-Stage Cluster Sampling}\label{sec: stscl}
\citet{krewski1981inference} and \citet[Chapter~6]{shao2012jackknife} discussed design  limiting properties for stratified {multi}-stage cluster sampling with clusters sampled with replacement conditional on the finite population, and \citet{rubin-bleuer2005} considered a super-population model for a stratified two-stage cluster sampling design.  In this paper,  we mainly adopt the asymptotic frames of \citet{krewski1981inference} and \citet{rubin-bleuer2005}.

We first consider a super-population model proposed by \citet{rubin-bleuer2005} as Example~\ref{ex: 3} under stratified multi-stage cluster sampling. Assume that the cluster size measures $M_{hi}$ are generated from a super-population model for $h=1,\ldots,H$ and $i=1,\ldots,M_h$, where $M_{hi}$ is the size of the $(hi)$th cluster, containing elements in the second and sub-sequential stages, and $M_h$ is the number of the clusters with respect to the first sampling design.  Conditional on the generated cluster sizes, we generate the finite population $U$ from a super-population model. We implicitly assume that $H\to\infty$ and the number of stages are fixed for the sampling design. Besides,  the population quantities are implicitly indexed by $H$. The following specific conditions are assumed for stratified multi-stage cluster sampling; see Example~\ref{ex: 3} of the paper for details about the notations.

\begin{enumerate}
	\renewcommand{\labelenumi}{D\arabic{enumi}.}
	\setcounter{enumi}{0}
	
	\item Conditions~C\ref{cond: smooth of f}--C\ref{cond: consistency l} hold.\label{cond: previous cond S1}
	\item $\max_{1\leq h\leq H}m_h = O(1)$ and $m_h\geq2$ for $h=1,2,\ldots$.\label{cond: strclusam clusize}
	\item $\max_{1\leq h\leq H}W_h = O_p(H^{-1})$ with respect to the super-population model.\label{cond: strclusam WEI}
	\item There exists a $\delta>0$, such that $\sum_{h=1}^HW_hE\{\lVert\tilde{\bS}_{w,h1}(\btheta)- \bS_h(\btheta)\rVert^{2+\delta}\mid U\}=O_p(1)$ with respect to the super-population model uniformly for $\btheta\in\mathcal{B}$, where $\mathcal{B}$ is a compact set containing $\btheta_0$ as an interior point,  $\tilde{\bS}_{w,h1}(\btheta)= d_{a(h1)}p_{a(h1)}^{-1}\hat{\bS}_{w,a(h1)}(\btheta)$, $\bS_h(\btheta)=N_h^{-1}\sum_{i=1}^{M_{h}}M_{hi}\bS_{hi}(\btheta)$ and $\bS_{hi}(\btheta)$ are the mean of the weighted score functions of the $h$-th stratum and $(hi)$-th cluster, respectively.\label{cond: Lyaponouv cond}
	\item  $n\sum_{h=1}^HW_h^2\var\{\tilde{\bS}_{w,a(h1)}(\btheta)\mid U\}/m_h\to\bSigma_S(\btheta)$ in probability with respect to the super-population model uniformly over $\mathcal{B}$, where $\bSigma_S(\btheta)$ is non-stochastic and positive definitive for $\btheta\in\mathcal{B}$, and $n=\sum_{h=1}^Hm_h$ is the number of sampled clusters.
	\label{cond: conv var}
	\item $\sup_{\btheta\in\mathcal{B}}\var\{\bS(\btheta)\}=o(n^{-1})$ with respect to the super-population model,   where  $\bS(\btheta)=E\{\hat{\bS}_w(\btheta)\mid U\}$. \label{cond: popu var}
	\item There exist two  constants $0<C_5<C_6<\infty$, such that 
	$C_5<M_hp_{hi}<C_6$ almost surely with respect to the super-population model for $h=1,\ldots,H$ and $i = 1,\ldots,M_h$.\label{cond: selection prob strclus}
	\item 
	$\max_{1\leq h\leq H}\sup_{\btheta\in\mathcal{B}}E\{\lVert\hat{\bI}_{w,a(h1)}(\btheta)\rVert^2\mid U \}=O_p(1)$ with respect to the super-population model,
	where $\hat{\bI}_{w, hi}(\btheta) = \partial \hat{\bS}_{w,hi}(\btheta)/\partial\btheta^{\T}$. \label{eq: I strclus}
	\item $ {\bI}(\btheta) \to\bcalI(\btheta)$  in probability with respect to the super-population model and the sampling mechanism uniformly over $\mathcal{B}$, where ${\bI}(\btheta) = E\{\hat{\bI}_w(\btheta)\mid U\}$, and $\bcalI(\btheta_0)$ is invertible.\label{cond: Conv}
\end{enumerate}
Although we may have more than two sampling stages,  we mainly focus on the clusters (primary sampling units, PSU's) in the first stage of sampling. 
Condition~D\ref{cond: strclusam clusize} shows that we focus on the stratified multi-stage cluster sampling where the number of selected PSUs is bounded within each stratum, but we assume that the number of strata diverges; see \citet{krewski1981inference} for details. Condition~D\ref{cond: strclusam WEI} rules out the existence of extremely large or small strata. Condition~D\ref{cond: Lyaponouv cond} is a typical Lyapounov-type condition, and it is used to validate the central limit theorem in Condition~C\ref{cond: clt}. Condition~D\ref{cond: conv var} shows the convergence rate of the variance of the estimation equation. Condition~D\ref{cond: popu var} shows that $\var\{\bS(\btheta)\}$ is negligible compared with the design  variance of  $\hat{\bS}_w(\btheta)$ conditional on the finite population. Condition~D\ref{cond: selection prob strclus} is similar to Condition~B\ref{cond: selection prob}, and it is used to regulate the selection probability of each PSU. Conditions~D\ref{eq: I strclus}--D\ref{cond: Conv} are used to show the convergence property of the negative Hessian matrix. 

\begin{lemma}\label{lemma: C4 STRATIFIED}
	Under stratified multi-stage cluster sampling and Condition~D\ref{cond: previous cond S1}--D\ref{cond: popu var}, Condition~C\ref{cond: clt} holds.
\end{lemma}
\begin{proof}[Proof of Lemma~\ref{lemma: C4 STRATIFIED}]
	By Conditions~D\ref{cond: strclusam clusize}--D\ref{cond: Lyaponouv cond}, \citet{krewski1981inference} established the following central limit theorem:
	\begin{equation}
		n^{1/2}\{\hat{\bS}_w(\btheta_0)-\bS(\btheta_0)\}\mid U \to N(\bzero,\bSigma_S(\btheta_0))\label{eq: central limit theorem U STRATIFIED}
	\end{equation}
	in distribution with respect to the sampling mechanism in probability regarding the super-population model,
	where  $\bS(\btheta)=E\{\hat{\bS}_w(\btheta)\mid U\}$.
	
	Besides, by Condition~D\ref{cond: popu var}, we have 
	\begin{equation}
		\bS(\btheta_0) = o_p(n^{-1})\label{eq: S(theta) STRATIFIED}
	\end{equation}
	with respect to the super-population model,
	since $E\{\bS(\btheta_0)\}=0$ by Condition~C\ref{cond: smooth of f}.
	
	By Condition~D\ref{cond: conv var} and (\ref{eq: central limit theorem U STRATIFIED})--(\ref{eq: S(theta) STRATIFIED}), Theorem~5.1 of \citet{rubin-bleuer2005} validates Lemma~\ref{lemma: C4 STRATIFIED}.
\end{proof}

\begin{lemma}\label{lemma: C5 STRATIFIED}
	Under stratified multi-stage cluster sampling, Conditions~D\ref{cond: strclusam clusize}--D\ref{cond: strclusam WEI} and Conditions~D\ref{eq: I strclus}--D\ref{cond: Conv}, Condition~C\ref{cond: I} holds.
\end{lemma}

\begin{proof}[Proof of Lemma~\ref{lemma: C5 STRATIFIED}]
	By Condition~D\ref{cond: Conv}, it is sufficient to show
	\begin{equation}
		\hat{\bI}_w(\btheta) - \bI(\btheta)\to0\label{eq: Uniform I w STRATIFIED}
	\end{equation}
	in probability with respect to the super-population and the sampling mechanism uniformly over $\mathcal{B}$. Since 
	\begin{equation}
		{\bI}(\btheta) = E\{\hat{\bI}_w(\btheta)\mid U\},
	\end{equation} it remains to show
	\begin{equation}
		V\{\hat{\bI}_w(\btheta)\mid U\}\to 0\label{eq: Var Ihat STRATIFIED}
	\end{equation}
	in probability with respect to the super-population model uniformly over  $\mathcal{B}$.
	
	Consider 
	\begin{eqnarray}
		V\{\hat{\bI}_w(\btheta)\mid U\} &=& \sum_{h=1}^HW_h^2m_h^{-1}V\{\hat{\bI}_{w,a(h1)}(\btheta)\mid U\} \notag \\ 
		&=&O(H^{-1})\sum_{h=1}^HW_hm_h^{-1}V\{\hat{\bI}_{w,a(h1)}(\btheta)\mid U\}\notag \\ 
		&=& O_p(H^{-1})\label{eq: Var Ihat STRATIFIED proof}
	\end{eqnarray}
	with respect to the super-population model uniformly over $\mathcal{B}$,
	where the second equality holds by Condition~D\ref{cond: strclusam WEI}, and the last equality holds by Conditions~D\ref{cond: strclusam clusize}--D\ref{cond: strclusam WEI} and Condition~D\ref{eq: I strclus}. Thus, we have shown (\ref{eq: Var Ihat STRATIFIED}) by (\ref{eq: Var Ihat STRATIFIED proof}) and proved Lemma~\ref{lemma: C5 STRATIFIED}.
\end{proof}
Condition~C\ref{cond: variance estimator} can be validated in a similar manner as Theorem~3.2 of \citet{krewski1981inference} based on Conditions~D\ref{cond: strclusam clusize}--D\ref{cond: Lyaponouv cond}, so we omit the proof here.
We can use a similar argument in Lemma~\ref{lemma: PPS Condition 7} to show that the proposed bootstrap method for stratified multi-stage sampling satisfies Condition~C\ref{cond: boot weights}.
\begin{lemma}\label{lemma: Condition C7 Stratified}
	The proposed bootstrap satisfies Condition~C\ref{cond: boot weights}.
\end{lemma}
\begin{proof}[Proof of Lemma~\ref{lemma: Condition C7 Stratified}]
	Recall that 
	$\hat{\bS}_w(\btheta) = \sum_{h=1}^HW_h\hat{\bS}_{w,h}(\btheta)$, where $W_h = N_h/N$, $N=\sum_{h=1}^HN_h$, $N_h = \sum_{i=1}^{M_h}M_{hi}$, $M_{hi}$ is the size of the $(hi)$-th cluster, $\hat{\bS}_{w,h}(\btheta) = m_h^{-1}\sum_{i=1}^{m_h}\tilde{\bS}_{w,hi}(\btheta)$, $\tilde{\bS}_{w,hi}(\btheta) = d_{a(hi)}p_{a(hi)}^{-1}\hat{\bS}_{w,a(hi)}(\btheta)$, $d_{a(hi)} = M_{a(hi)}/N_h$, and $a(hi)$ is the index of the $i$-th selected cluster   in the $h$-th stratum. {From Example~\ref{ex: 3}}, the bootstrap   replicate of $\hat{\bS}_w(\btheta)$ is $\hat{\bS}^*_w(\btheta) = \sum_{h=1}^HW_h\hat{\bS}^*_{w,h}(\btheta)$, where $\hat{\bS}_{w,h}^*(\btheta) = m_h^{-1}\sum_{i=1}^{m_h}r_{hi}^*\tilde{\bS}_{w,hi}(\btheta)$, $r_{hi}^*=k_hm_{hi}^*$,  $k_h = m_h/(m_h-1)$, and $\bbm_h^*=(m_{h1}^*,\ldots,m_{hn_h}^*)^{\T}$ is generated by a multinomial distribution with $m_h-1$ trials and a success probability vector $m_h^{-1}(1,\ldots,1)^{\T}$ of length $m_h$. Since $E_*(r^*_{hi}) = 1$ for $h=1,\ldots,H$ and $i=1,\ldots,m_h$, we have 
	\begin{equation}
		E_*\{\hat{\bS}^*_w(\btheta)\} = \hat{\bS}_w(\btheta).\label{eq: E^* STRATIFIED}
	\end{equation}
	
	By a similar argument  to (\ref{eq: condition 7 V^* PPS}), we can show that 
	\begin{eqnarray}
		\bV_*\{\hat{\bS}^*_w(\btheta)\} &=&\sum_{h=1}^HW_h^2\{m_h(m_h-1)\}^{-1} \tilde{\bS}_{h,vec}(\theta)(\bI_{m_h} - \mathcal{P}_{\bone,m_h})\tilde{\bS}{}^{\T}_{h,vec}(\theta)\notag \\ 
		&=&\hat{\bV}\{\hat{\bS}^*_w(\btheta)\},\label{eq: V^* STRATIFIED1}
	\end{eqnarray}
	where $\tilde{\bS}_{h,vec}(\theta) = (d_{a(h1)}p^{-1}_{a(h1)}\hat{\bS}_{w,a(h1)}(\btheta),\ldots,d_{a(hm_h)}p^{-1}_{a(hm_h)}\hat{\bS}_{w,a(hm_h)}(\btheta))$, and $\mathcal{P}_{\bone,m_h}$ is defined similarly as $\mathcal{P}_{\bone,n}$ in (\ref{eq: condition 7 V^* PPS}).
	
	By (\ref{eq: E^* STRATIFIED})--(\ref{eq: V^* STRATIFIED1}), we have completed the proof of Lemma~\ref{lemma: Condition C7 Stratified}.
\end{proof}
\begin{lemma}\label{lemma: Condition C8  STRATIFIED}
	Under stratified multi-stage cluster sampling and  Conditions~D\ref{cond: strclusam clusize}--D\ref{cond: conv var}, Condition~C\ref{cond: boot CLT result} holds.
\end{lemma}
\begin{proof}[Proof of Lemma~\ref{lemma: Condition C8  STRATIFIED}]
	We prove Lemma~\ref{lemma: Condition C8  STRATIFIED} by validating the Lyapounov condition for $\hat{\bS}^*_w(\btheta)$ with respect to the super-population model, the sampling mechanism and the bootstrap procedure. Notice that $\hat{\bS}^*_{w,h}(\btheta) = m_h^{-1}k_h\sum_{i=1}^{m_h-1}\tilde{\bS}_{w,hi}^*(\btheta)$, where $\{\tilde{\bS}_{w,h1}^*(\btheta),\ldots,\tilde{\bS}_{w,h(m_h-1)}^*(\btheta)\}$ is a random sample of size $m_h-1$ from $\{\tilde{\bS}_{w,h1}(\btheta),\ldots,\tilde{\bS}_{w,hm_h}(\btheta)\}$. Thus, we have $E_*\{\tilde{\bS}_{w,h1}^*(\btheta)\} = \hat{\bS}_{w,h}(\btheta)$, where $ \hat{\bS}_{w,h}(\btheta) = m_h^{-1}\sum_{i=1}^{m_h}\tilde{\bS}_{w,hi}(\btheta)$. For $\delta>0$ in Condition~D\ref{cond: Lyaponouv cond},  consider 
	\begin{eqnarray}
		&&\sum_{h=1}^HW_h^{2+\delta}E_*\{\lVert\tilde{\bS}_{w,h1}^*(\btheta) - \hat{\bS}_{w,h}(\btheta)\rVert^{2+\delta}\}\notag \\ &=&\sum_{h=1}^HW_h^{2+\delta}\frac{1}{m_h}\sum_{i=1}^{m_h}\lVert\tilde{\bS}_{w,hi}(\btheta) -\hat{\bS}_{w,h}(\btheta)\rVert^{2+\delta} \notag \\ 
		&\leq& C(\delta)\left\{\sum_{h=1}^HW_h^{2+\delta}\frac{1}{m_h}\sum_{i=1}^{m_h}\lVert\tilde{\bS}_{w,hi}(\btheta) -{\bS}_{h}(\btheta)\rVert^{2+\delta}+\sum_{h=1}^HW_h^{2+\delta}\lVert\hat{\bS}_{w,h}(\btheta) -{\bS}_{h}(\btheta)\rVert^{2+\delta}\right\}\notag \\\label{eq: s31}
	\end{eqnarray}
	by the  triangular inequality and the H\"{o}lder's inequality, where $C(\delta)$ is a constant determined by $\delta$.
	
	Consider the first term of (\ref{eq: s31}), 
	\begin{eqnarray}
		E\left\{\sum_{h=1}^HW_h^{2+\delta}\frac{1}{m_h}\sum_{i=1}^{m_h}\lVert\tilde{\bS}_{w,hi}(\btheta) -{\bS}_{h}(\btheta)\rVert^{2+\delta}\mid U \right\} 
		&=&\sum_{h=1}^HW_h^{2+\delta}E\{\lVert\tilde{\bS}_{w,h1}(\btheta)- \bS_h(\btheta)\rVert^{2+\delta}\mid U\}\notag \\ 
		&=& O_p(H^{-1-\delta})\label{eq: part 1 STRATIFIED}
	\end{eqnarray}
	with respect to the super-population model and the sampling mechanism,
	where the first equality holds by the fact that we use probability proportional to size sampling with replacement in the first stage sampling, and the second equality holds by Conditions~D\ref{cond: strclusam WEI}--D\ref{cond: Lyaponouv cond}.
	
	Before considering the second term of (\ref{eq: s31}), we have  
	\begin{eqnarray}
		\lVert\hat{\bS}_{w,h}(\btheta) -{\bS}_{h}(\btheta)\rVert^{2+\delta} &=& \left\lVert\frac{1}{m_h}\sum_{i=1}^{m_h}\{\tilde{\bS}_{w,hi}(\btheta) - \bS_h(\btheta)\}\right\rVert\notag \\ 
		&\leq& \left\{\frac{1}{m_h}\sum_{i=1}^{m_h}\left\lVert\tilde{\bS}_{w,hi}(\btheta) - \bS_h(\btheta)\right\rVert\right\}^{2+\delta}\notag \\ 
		&\leq&\frac{1}{m_h}\sum_{i=1}^{m_h}\left\lVert\tilde{\bS}_{w,hi}(\btheta) - \bS_h(\btheta)\right\rVert^{2+\delta},\notag
	\end{eqnarray}
	where the second inequality holds by the triangular inequality for norms, and the third inequality holds by the H\"{o}lder's inequality. Thus, we can use a similar argument to (\ref{eq: part 1 STRATIFIED}) to show that 
	\begin{equation}
		E\left\{\sum_{h=1}^HW_h^{2+\delta}\lVert\hat{\bS}_{w,h}(\btheta) -{\bS}_{h}(\btheta)\rVert^{2+\delta}\mid U\right\}=O_p(H^{-1-\delta})\label{eq: part 2 STRATIFIED}
	\end{equation}
	with respect to the super-population model and the sampling mechanism.
	Thus, by (\ref{eq: s31})--(\ref{eq: part 2 STRATIFIED}) and the Markov's inequality, we can show that 
	\begin{equation}
		\sum_{h=1}^HW_h^{2+\delta}E_*\{\lVert\tilde{\bS}_{h1}^*(\btheta) - \hat{\bS}_{w,h}(\btheta)\rVert^{2+\delta}\} = O_p(H^{-1-\delta})
	\end{equation}
	with respect to the super-population model and the sampling mechanism.
	On the other hand, by Lemma~\ref{lemma: Condition C7 Stratified}, Condition~D\ref{cond: strclusam clusize} and Condition~D\ref{cond: conv var}, we have shown that 
	\begin{equation}
		\bV_*\{\hat{\bS}^*_w(\btheta)\} \asymp H^{-1}\label{eq: Var* STRATIFIED}
	\end{equation}
	in probability with respect to the super-population model and the sampling mechanism.
	Thus, for any $\ba\in\mathbb{R}^p$ with $\lVert\ba\rVert = 1$, we have 
	\begin{equation}
		[V_*\{\ba^\T\hat{\bS}^*_w(\btheta)\}]^{-(2+\delta)/2}\sum_{h=1}^HW_h^{2+\delta}E_*\{\lVert\tilde{\bS}_{h1}^*(\btheta) - \hat{\bS}_{w,h}(\btheta)\rVert^{2+\delta}\} = O_p(H^{-\delta/2})=o_p(1)\label{eq: Lyapounov STRATIFIED}
	\end{equation}
	with respect to the super-population model and the sampling mechanism,
	where the last equality holds by the fact that $H\to\infty$. Thus, by (\ref{eq: Lyapounov STRATIFIED}) and the Cram\'{e}r-Wold device, we have completed the proof of Lemma~\ref{lemma: Condition C8  STRATIFIED}.
\end{proof}

\begin{lemma}\label{lemma: Condition 7 STRATIFIED}
	Under stratified multi-stage sampling and  Conditions~D\ref{cond: strclusam clusize}--D\ref{cond: strclusam WEI} and Condition~D\ref{eq: I strclus}, Condition~C\ref{cond: boot I}   holds.  
\end{lemma}
\begin{proof}[Proof of Lemma~\ref{lemma: Condition 7 STRATIFIED}]
	Recall that 
	$\hat{\bI}_w(\btheta) = \sum_{h=1}^HW_h\hat{\bI}_{w,h}(\btheta)$, where  $\hat{\bI}_{w,h}(\btheta) = m_h^{-1}\sum_{i=1}^{m_h}\tilde{\bI}_{w,hi}(\btheta)$, $\tilde{\bI}_{w,hi}(\btheta) = d_{a(hi)}p_{a(hi)}^{-1}\hat{\bI}_{w,a(hi)}(\btheta)$.  The bootstrap   replicate of $\hat{\bI}_w(\btheta)$ is $\hat{\bI}^*_w(\btheta) = \sum_{h=1}^HW_h\hat{\bI}^*_{w,h}(\btheta)$, where $\hat{\bI}_{w,h}^*(\btheta) = m_h^{-1}\sum_{i=1}^{m_h}r_{hi}^*\tilde{\bI}_{w,hi}(\btheta)$. 
	
	First, we can use a similar argument to (\ref{eq: E^* STRATIFIED}) to show that 
	\begin{equation}
		E_*\{\hat{\bI}^*_w(\btheta)\} = \hat{\bI}_w(\btheta).\label{eq: E^* I STRATIFIED}
	\end{equation}

	For $\ba\in\mathbb{R}^p$ with $\lVert\ba\rVert=1$, denote $Z_{\ba,hi}(\btheta)=\ba^\T \tilde{\bI}_{w,hi}(\btheta)\ba$. Then, by a similar argument  to (\ref{eq: condition 7 V^* PPS}), we can show that  
	\begin{eqnarray}
		\bV_*\{\ba^\T\hat{\bI}^*_w(\btheta)\ba\} &=&\sum_{h=1}^HW_h^2\{m_h(m_h-1)\}^{-1} \tilde{\bZ}_{h,vec}(\theta)(\bI_{m_h} - \mathcal{P}_{\bone,m_h})\tilde{\bZ}{}^{\T}_{h,vec}(\theta)\notag \\ 
		&\leq&\sum_{h=1}^HW_h^2\{m_h(m_h-1)\}^{-1}\sum_{i=1}^{m_h}Z^2_{\ba,hi}(\btheta),\label{eq: V^* STRATIFIED}
	\end{eqnarray}
	where $\tilde{\bZ}_{h,vec}(\theta) = (Z_{\ba,h1}(\btheta),\ldots,Z_{\ba,hm_h}(\btheta))$, and the second inequality holds since $\mathcal{P}_{\bone,m_h}$ is non-negative definitive.
	
	Consider 
	\begin{eqnarray}
		E\left[\sum_{h=1}^HW_h^2\{m_h(m_h-1)\}^{-1}\sum_{i=1}^{m_h}Z^2_{\ba,hi}(\btheta)\mid U\right] &\leq& \sum_{h=1}^HW_h^2E\{Z_{\ba,h1}^2(\btheta)\mid U\}\notag \\ 
		&=& O_p(H^{-1})\label{eq: V^* STRATIFIED 2}
	\end{eqnarray}
	with respect to the super-population model uniformly over $\mathcal{B}$, where the first inequality holds by Condition~D\ref{cond: strclusam clusize}, and the second inequality holds by Condition~D\ref{cond: strclusam WEI} and Condition~D\ref{eq: I strclus}. 
	
	By (\ref{eq: V^* STRATIFIED})--(\ref{eq: V^* STRATIFIED 2}) to show 
	\begin{equation}
		\var_*\{\ba^{\T}\hat{\bI}_w^*(\btheta)\ba\}=o_p(1)\notag\label{eq: V^* STRATIFIED 3}
	\end{equation} with respect to the super-population model and the sampling mechanism uniformly for $\btheta\in\mathcal{B}$ by Markov's inequality. Thus, by (\ref{eq: E^* I STRATIFIED}) and (\ref{eq: V^* STRATIFIED 3}), we have completed the proof of Lemma~\ref{lemma: Condition 7 STRATIFIED}.
\end{proof}

\section{Stratified Simple Random Sampling}\label{ss: stratified random sampling}

In this section, we propose a bootstrap method for   stratified random sampling without replacement, which is   commonly used   and has been investigated by \citet{bickel1984asymptotic}. Specifically, denote $N_h$ and $n_h$ to be the stratum population size and sample size of the $h$th stratum, respectively.  Let $N=\sum_{h}^HN_h$ and $n=\sum_{h=1}^Hn_h$ be the population size  and sample size, where $H$ is the number of strata. Simple random sampling without replacement is applied within each stratum, and we assume that the first $n_h$ elements are sampled within the $h$th stratum, for simplicity.  Let $H$, $N_h$ and $n_h$ depend on an index $\nu$ such that $n(\nu)=n_1(\nu)+\cdots+n_H(\nu)\to \infty$ as $\nu\to\infty$, and we omit the index $\nu$ without loss of generality; see \citet{bickel1984asymptotic} for details. 
\setcounter{example}{4}
\begin{example}\label{ex: 5}
	Under stratified simple random sampling without replacement, the pseudo maximum likelihood estimator $\hat{\btheta}$ is obtained by solving 
	\begin{equation*}
		\hat{\bS}_w(\btheta) = \sum_{h=1}^HW_hN_h^{-1}\sum_{i=1}^{n_h}w_{hi}\bS(\btheta;y_{hi})=\sum_{h=1}^HW_hn_h^{-1}\sum_{i=1}^{n_h}\bS(\btheta;y_{hi})=\bzero,
	\end{equation*}
	where   $W_h=N_hN^{-1}$, $w_{hi} =N_hn_h^{-1}$.
	Then,  we can show that 
	\begin{equation}
		\hat{\var}\{\hat{\bS}_w(\btheta)\mid U\} = \sum_{h=1}^HW_h^2n_h^{-1}(1-n_hN_h^{-1})(n_h-1)^{-1}\sum_{i=1}^{n_h}\{\bS(\btheta;y_{hi}) - \bar{\bS}_h(\btheta)\}^{\otimes2},\notag
	\end{equation}
	where $\bar{\bS}_h(\btheta) = n_h^{-1}\sum_{i=1}^{n_h}\bS(\btheta;y_{hi})$. 
	
	The rescaling factor of the bootstrap method is $r_{hi}^* = 1 + \{m_{hi}^*-(n_h-1)n_h^{-1}\}k_h$, where $\bbm_h^*=(m_{h1}^*,\ldots,m_{hn_h}^*)^{\T}$ is generated using a multinomial distribution with $n_h-1$ trials and a success probability vector $n_h^{-1}(1,\ldots,1)^{\T}$ of length $n_h$, and  $k_h^2 = n_h^2(1-n_hN_h^{-1})(n_h-1)^{-2}$.

\end{example}
The proposed bootstrap method is essentially the same as the one considered in Section~4 of \citet{rao1988} when estimating the population mean.

\section{Proof of Theorem~3}\label{ss: theorem 1} 

By  (\ref{eq: consitency theta hat}) and Lemma~\ref{coro: consistency},  we have  $\hat{\btheta} - \hat{\btheta}^*\to0$ in probability conditional on the sample $A$. {Thus, }by the second-order Taylor expansion, we have
\begin{equation}
	l_w^* ( \hat{\btheta} )= l_w^* ( \hat{\btheta}^{*}  ) + \hat{\bS}_w^* ( \hat{\btheta}^* )^{\T} ( \hat{\btheta}- \hat{\btheta}^* ) -  \frac{1}{2}( \hat{\btheta}- \hat{\btheta}^* )^{\T} \hat{\bI}_{w}^* ( \hat{\btheta}^*)  ( \hat{\btheta}- \hat{\btheta}^*)  + o_p (n^{-1}).
	\label{b-1}
\end{equation}
Furthermore, we have 
\begin{equation}
	\bzero = \hat{\bS}^*_w(\hat{\btheta}^*)= \hat{\bS}^*_w(\hat{\btheta}) + \hat{\bI}_{w} (\hat{\btheta})(\hat{\btheta}^* -\hat{\btheta}) + o_p(1).\label{eq: 7a}
\end{equation} By (\ref{eq: 7a}), we have \begin{eqnarray}
	-2 n \left\{ l_w^* ( \hat{\btheta}) - l_w^* ( \hat{\btheta}^* ) \right\}
	&=& n (\hat{\btheta}^* -\hat{\btheta} )^{\T} \hat{\bI}_{w} ( \hat{\btheta}^*)  ( \hat{\btheta}^* -\hat{\btheta} )  + o_p ( 1)\notag  \\
	&=& n \bS_w^* (\hat{\btheta})^{\T} \{ \hat{\bI}_{w} ( \hat{\btheta}^*)  \}^{-1}
	\bS_w^* (\hat{\btheta}) + o_p ( 1) ,  \label{b-2}  \end{eqnarray}
where the second equality follows by the continuous mapping theorem {and Condition~C\ref{cond: boot I}}. By  the central limit theorem for $ \bS_w^* (\hat{\btheta})$ conditional on the sample $A$, $ W^* (\hat{\btheta})$ converges in distribution to the weighted sum of $p$ independent $\chi^2 (1)$ variables as $n\to\infty$,  where the weights are the eigenvalues of
$\bcalI(\btheta_0)^{-1} \bSigma_S(\btheta_0)$ by Condition~C\ref{cond: asymptotic con var} and Condition~C\ref{cond: I}. Since $\bcalI(\btheta_0)^{-1} \bSigma_S(\btheta_0) = \bSigma_\btheta\bcalI(\btheta_0)^\T$ and $\bcalI(\btheta_0)$ is symmetric, Theorem \ref{theo: 1} is established.

\section{Proof of Theorem~4}\label{ss: theorem 3} 

Recall that $\hat{\btheta} = ( \hat{\btheta}{}^\T_{1}, \hat{\btheta}{}^\T_{2}){}^\T$ and  $\hat{\btheta}^{*(0)} = ( \hat{\btheta}_1^{*(0)}{}^\T , \hat{\btheta}{}^\T_{2}){}^\T$, where $ \hat{\btheta}_1^{*(0)}$ is the maximizer of $l_w^* ( \btheta_1, \hat{\btheta}_{2} )$. Then, we can obtain, similarly to (\ref{b-1}),
\begin{eqnarray*}
	l_w^* ( \hat{\btheta}  ) &=& l_w^* (\hat{\btheta}^{*(0)} ) + \hat{\bS}_{w1}^* ( \hat{\btheta}^{*(0)}  )^{\T} ( \hat{\btheta}_{1}-\hat{\btheta}_1^{*(0)}  ) -  \frac{1}{2}(\hat{\btheta}_{1}-\hat{\btheta}_1^{*(0)} )^{\T} \hat{\bI}_{w11}^* ( \hat{\btheta} )
	(\hat{\btheta}_{1}-\hat{\btheta}_1^{*(0)} )  + o_p (n^{-1}), \end{eqnarray*}
where $\hat{\bS}_{w1}^* ( {\btheta} )  = \partial  l_w^* ( \btheta) / \partial \btheta_1$ and $\hat{\bI}_{w11}^* ( \btheta) = -\partial^2 l_w^* ( \btheta) / (\partial \btheta_1 \partial \btheta_1^{\T})$. By the definition of  $\hat{\btheta}^{*(0)} $, we have $\hat{\bS}_{w1}^* ( \hat{\btheta}^{*(0)}  )=\bzero$. Thus, we have
\begin{eqnarray}
	-2 n \{ l_w^* ( \hat{\btheta} ) - l_w^* ( \hat{\btheta}^{*(0)}  ) \}
	&=& n ( \hat{\btheta}_1^{*(0)}  - \hat{\btheta}_{1} )^\T \hat{\bI}_{w11} ( \hat{\btheta})  ( \hat{\btheta}_1^{*(0)} -  \hat{\btheta}_{1} )  + o_p ( 1) \notag \\
	&=& n \bS_{w1}^* ( \hat{\btheta})^\T \{ \hat{\bI}_{w11} ( \hat{\btheta})  \}^{-1}
	\bS_{w1}^* (\hat{\btheta}) + o_p ( 1),  \label{c-1} \end{eqnarray}
where the last equality follows from
$ \hat{\btheta}_1^{*(0)} - \hat{\btheta}_{1} = \{ \hat{\bI}_{w11} (\hat{\btheta}) \}^{-1} \hat{\bS}_{w1}^* ( \hat{\btheta}) + o_p (n^{-1/2} ). $
Thus, combining (\ref{b-2}) with (\ref{c-1}), we have
\begin{eqnarray*}
	W^*(\hat{\btheta}_2) &=&
	- 2 n \{  l_w^* ( \hat{\btheta}^{*(0)}  ) - l_w^* ( \hat{\btheta}^*)  \} \\
	&=& -2 n \{ l_w^* ( \hat{\btheta}) - l_w^* ( \hat{\btheta}^* ) \} + 2n \{ l_w^* ( \hat{\btheta} ) - l_w^* ( \hat{\btheta}^{*(0)} ) \} \\
	&=& n \begin{pmatrix}
		\bS_{w1}^* ( \hat{\btheta} ) \\
		\bS_{w2}^* (\hat{\btheta})
	\end{pmatrix}^{\T}
	\left[
	\begin{array}{ll}
		\hat{\bI}_{w11} (\hat{\btheta} ) & \hat{\bI}_{w12} (\hat{\btheta} ) \\
		\hat{\bI}_{w21} ( \hat{\btheta} ) & \hat{\bI}_{w22} (\hat{\btheta} ) \\\end{array}
	\right]^{-1}\begin{pmatrix}
		\bS_{w1}^* (\hat{\btheta} ) \\
		\bS_{w2}^* (\hat{\btheta})
	\end{pmatrix}\\
	&& - n \bS_{w1}^* ( \hat{\btheta})^{\T} \{ \hat{\bI}_{w11} (\hat{\btheta})  \}^{-1}
	\bS_{w1}^* ( \hat{\btheta}) + o_p ( 1) \\
	&=& n \{ \bS_{w 2}^* (\hat{\btheta} )  - \hat{\bB}_{21}   \bS_{w1}^* ( \hat{\btheta}) \} ^{\T}\{  \hat{\bI}_{w 22 \cdot 1} ( \hat{\btheta}) \} ^{-1}
	\{ \bS_{w 2}^* ( \hat{\btheta} )  - \hat{
		\bB}_{21}   \bS_{w1}^* ( \hat{\btheta}) \}+ o_p ( 1),\end{eqnarray*}
where
$ \hat{\bB}_{21} = \hat{\bI}_{w21} ( \hat{\btheta} ) \{ \hat{\bI}_{w11} (\hat{\btheta}) \}^{-1} $
and
$ \hat{\bI}_{w 22 \cdot 1} (\hat{\btheta}) = \hat{\bI}_{w 22} ( \hat{\btheta} ) - \hat{\bI}_{w 21} ( \hat{\btheta} ) \{  \hat{\bI}_{w11} ( \hat{\btheta}) \}^{-1}
\hat{\bI}_{w12} ( \hat{\btheta}) . $
Therefore, by the limiting distribution associated with $\bS_{w 2}^* (\hat{\btheta} )$, using the same argument for proving  Theorem~\ref{theo: 1}, we can show that $W^*(\hat{\btheta}_2)$ converges in distribution to the weighted sum of $q$ independent $\chi^2(1)$ variables as $n\to\infty$,  where the weights are the eigenvalues of  $n \var  \{ \hat{\bS}_{w, 2 \cdot 1} ( \btheta) \}\bcalI_{2\cdot1}(\btheta_{1},\btheta_2^{(0)})^{-1}$,  $\hat{\bS}_{w,2\cdot1}(\btheta) = \hat{\bS}_{w2}(\btheta) - \hat{\bB}_{21}\hat{\bS}_{w1}(\btheta)$, and $\btheta$ is the true parameter in Theorem~\ref{theo: 3}. By a similar argument used in the proof of Lemma~1 and some basic algebra, we can show that the eigenvalues of  $n \var  \{ \hat{\bS}_{w, 2 \cdot 1} ( \btheta) \} \bcalI_{2\cdot1}(\btheta_{1},\btheta_2^{(0)})^{-1}$ are the same as those of $\bSigma_{\btheta,2}\bcalI_{2\cdot1}(\btheta_{1},\btheta_2^{(0)})$.

\section{Proof of Theorem~5}\label{ss: theorem 4}
Since $\hat{\btheta}$ solves $\hat{\bS}_w(\btheta) = \bzero$, we have
\begin{eqnarray}
	\bzero= \hat{\bS}_w(\hat\btheta)= \hat{\bS}_w(\btheta)- \hat{\bI}_w(\btheta)(\hat{\btheta} - \btheta) + o_p(n^{-1/2}),\label{eq: th4.1}
\end{eqnarray}
where the Taylor expansion holds {by (\ref{eq: consitency theta hat})}. Thus, by  (\ref{eq: th4.1}), we have
\begin{equation*}
	(\hat{\btheta} - \btheta) = \hat{\bI}_w(\btheta)^{-1}\hat{\bS}_w(\btheta)  + o_p(n^{-1/2}).
\end{equation*}
Since $\hat{\bS}_{w1} ( \hat{\btheta}^{(0)} ) =\bzero$, we have $\hat{\bS}_{w,2\cdot1}(\hat{\btheta}^{(0)}) = \hat{\bS}_{w2}(\hat{\btheta}^{(0)})$ and
\begin{equation}
	\hat{\btheta}_2 - \btheta_2^{(0)} = \{\hat{\bI}_{w,22\cdot1}(\hat{\btheta}^{(0)})\}^{-1}\hat{\bS}_{w2}(\hat{\btheta}^{(0)}) + o_p(n^{-1/2}).\label{eq: lab theta.hat 2}
\end{equation}

By (\ref{11}) and  (\ref{eq: lab theta.hat 2}), we have
\begin{eqnarray}
	X_{QS}^2(\btheta_2^{(0)}) &=& \hat{\bS}_{w2}(\hat{\btheta}^{(0)})^\T\{\hat{\bI}_{w,22\cdot1}(\hat{\btheta}^{(0)})\}^{-1}\hat{\bS}_{w2}(\hat{\btheta}^{(0)})\notag \\
	&=& (\hat{\btheta}_2 - \btheta_2^{(0)})^\T \hat{\bI}_{w,22\cdot1}(\hat{\btheta}^{(0)})(\hat{\btheta}_2 - \btheta_2^{(0)}) + o_p(n^{-1}). \label{eq: last}
\end{eqnarray}
By {Theorem~\ref{theo: CLT (2)},} Theorem~\ref{theo: 1} and (\ref{eq: last}), we have proved Theorem~\ref{cor: 2}.

\section{Proof of Theorem~6}\label{ss: theorem 5}
The proof of Theorem~\ref{cor: 3} is essentially similar to that of Theorem~\ref{cor: 2}. Instead of using the asymptotic normality result in (\ref{2}), we use the results in  (\ref{lemma1}) to establish Theorem~\ref{cor: 3}, and we omit the proof for simplicity.

\section{Proof of Theorem~7}\label{ss: theorem 6}

We can  express 
\begin{equation}
	X^{2*}(\hat{\bp}) = ( \hat{{\bp}}^* - \hat{\bp})^\T (\hat{\bP})^{-1} ( \hat{{\bp}}^* - \hat{\bp})
	\label{a2}
\end{equation}
where $\hat{\bP} = \mbox{diag} (\hat{\bp} ) - \hat{\bp}\hat{\bp}^\T$. {By Theorem~\ref{theo: (6)},} we can show that the proposed bootstrap method satisfies 
\begin{equation}
	\sqrt{n} ( \hat{{\bp}}^* - \hat{\bp} ) \mid A { \,\longrightarrow\,} N ( \bzero, \bSigma_p )
	\label{a3}
\end{equation}
in distribution as $n\to\infty$. 
It now follows from (\ref{a2}) and (\ref{a3}) that {(\ref{b3}) holds under $H_0$}.  Result (\ref{b4}) for $W^*(\hat{\bp})$ also holds noting that $X^{2*}(\hat{\bp})$ and $W^*(\hat{\bp})$ are asymptotically equivalent with respect to the bootstrap distribution.

\section{Proof of Theorem~8}\label{ss: theorem 7}

We present a brief justification of the proposed bootstrap method for testing independence in a two-way table of cell proportions or counts. Using the notation of \citet{Scott1981CHI}, let ${\bh} ( {\bp})$ be the $d=(R-1)(C-1)$ dimensional vector with elements $h_{ij} ( {\bp}) = p_{ij} -  p_{i+} p_{+j} $, $i=1, \ldots, R-1; j=1, \ldots, C-1$, where ${\bp} = (p_{11}, p_{12}, \ldots, p_{RC-1}){}^\T$. Then the chi-squared statistic $X_I^2$, under $H_0$, may be expressed in a matrix form as
$$
X_{I}^2 = n \{ {\bh} ( \hat{{\bp}} ) - {\bh} ( {\bp} )  \}^\T ( \hat{{\bP}}_{R+}^{-1} \otimes \hat{{\bP}}_{+C}^{-1} ) \{ {\bh} ( \hat{{\bp}} ) - {\bh} ( {\bp} )  \}, $$
where $\hat{{\bP}}_{R+}= \mbox{diag} ( \hat{{\bp}}_{R+} )- \hat{{\bp}}_{R+} \hat{{\bp}}_{R+} ^\T $, $\hat{{\bP}}_{+C}= \mbox{diag} ( \hat{{\bp}}_{+C})  - \hat{{\bp}}_{+C} \hat{{\bp}}_{+C}^\T$,
$\hat{{\bp}}_{R+} = ( \hat{p}_{1+}, \ldots, \hat{p}_{R-1,+}){}^\T$, $\hat{{\bp}}_{+C} = ( \hat{p}_{+1}, \ldots, \hat{p}_{+, C-1}){}^\T$, and $\otimes$ denotes the direct product. Now, noting that $\sqrt{n} ( \hat{{\bp}}- {\bp} )\to N ( \bzero, \bSigma_p ) $ in distribution as $n\to\infty$,  it follows that
$$ \sqrt{n} \{ {\bh} ( \hat{ {\bp}} ) -  {\bh} ( {\bp} ) \}   { \,\longrightarrow\,} N ( \bzero, {\bH} \bSigma_p {\bH}^\T )
$$
in distribution as $n\to\infty$,
where ${\bH} = \partial {\bh} ( {\bp}) / \partial {\bp}^\T$ is the $d \times (RC-1)$ matrix of partial derivatives of ${\bh} ( {\bp})$. Using the above result, we get (\ref{b6}) where the $\delta_l  \ (l=1, \ldots, d)$ are the eigenvalues of the design effect matrix ${\bD}_h = \left( {{\bP}}_{R+}^{-1} \otimes {{\bP}}_{+C}^{-1} \right) ( {\bH} \bSigma_p {\bH}^\T )$.

Turning to the proposed bootstrap method, we can express the bootstrap version of $X_{I}^2$ in a matrix form as
$$
X_{I}^{2*} = n \{ {\bh} ( \hat{{\bp}}^* ) - {\bh} (\hat{\bp} )  \}^\T ( \hat{{\bP}}_{R+}^{*-1} \otimes \hat{{\bP}}_{+C}^{*-1} ) \{ {\bh} ( \hat{{\bp}}^* ) - {\bh} (\hat{\bp} )  \}. $$
Similar to  the proof in Sections~\ref{ss: pois sampling}--\ref{sec: stscl}, we have 
$$ \sqrt{n} \left\{ {\bh} ( \hat{{\bp}}^* ) - {\bh} ( \hat{\bp}) \right\} \mid A{ \,\longrightarrow\,} N ( \bzero, {{\bH}} \bSigma_p {{\bH}}^\T )$$
in distribution as $n\to\infty$,
so the result (\ref{7}) for $X_{I}^{2*}$ holds.

Since $ X_{ I}^2$ and $W_I$ are asymptotically equivalent under $H_0$, we can show the results  for $W_I^*$ in a similar way by assuming a multinomial distribution for the super-population model.

\section{Additional Simulation Study: Test of Independence} \label{sec: SS test independence}

In this section, we consider test of independence in a $3\times3$ table of counts to check the performance of the proposed bootstrap testing method. For a finite population $U=\{\by_i:i=1,\ldots,N\}$, 
$
\by_i
$ is generated by  a multinomial distribution with one trial and a success probability vector $\bp$,
where  $\bp=(p_{11}, \ldots,p_{ij},\ldots, p_{33}){}^\T$, and $p_{ij}$ is the success probability for the cell in the $i$-th row and $j$-th column for $i,j=1,\ldots,3$. That is, $\by_i$ is a dummy variable consisting of eight 0's and a 1. For the success probability vector $\bp$, we consider three cases:
\begin{enumerate}
	\item [Case I]: $p_{11} = 1/4$, $p_{12}=p_{13} =  p_{21} =p_{31}=1/8$, $p_{22}=p_{23}=p_{32}=p_{33}=1/16$.
	\item [Case II]: $p_{11}=1/4$, $p_{12}=p_{13} = (1.4)/8$, $p_{21} =p_{31}=(0.6)/8$, $p_{22}=p_{33}=1/16$, $p_{23}=(1.4)/16$, $p_{32}=(0.6)/16$.
	\item [Case III]: $p_{11}=p_{2,3} = p_{3,2} = 1/6$, $p_{12}=p_{13} = p_{21} =p_{31}=p_{22}=p_{33}=1/12$.
\end{enumerate}
Case I satisfies independence for  the two-way table of counts, but  Cases II--III do not.  The level of non-independence can be expressed by
$ \gamma=  \sum_{i=1}^3 \sum_{j=1}^3{ ( p_{ij} - p_{i+} p_{+j} )^2}/( p_{i+} p_{+j} ) $, and $\gamma = 0$ corresponds to independence.
The values of $\gamma$ are $0$, $0.017$ and $0.125$ for Cases I--III, respectively.

For each $\by_i$, we generate an auxiliary variable $x_i= \bbeta{}^\T \by_i$, where $\bbeta = (\beta_1,\ldots,\beta_9){}^\T$, $\beta_j = 0.5+ e_j$ for $j=1,\ldots,9$, $e_j \sim \mbox{Ex}(1)$. Probability proportional to size sampling with replacement is used to generate a sample of size $n$ with selection probability proportional to $x_i$.  We consider two scenarios for the population and sample sizes: $(N,n) = (2\,000,75)$ and $(N,n) = (10\,000,150)$.

We are interested in testing independence   in the two-way table with $\alpha = 0.05$ nominal significance level. For each sample, we compare the following five test methods:
\begin{enumerate}
	\item Naive Pearson method based on  $X_{I}^2$ in (\ref{eq: 7chi}) with $\chi^2 (4)$ being the reference distribution.
	\item Naive likelihood-ratio method  using $ W_{I}$ in (\ref{eq: 7w}) with $ \chi^2 (4)$ being the reference distribution.
	\item The second-order  Rao-Scott correction method using the  Satterthwaite approximation \citep{ThomasRao1987}. This method is widely used in analyzing two-way contingency table and is implemented by the command  \verb/svychisq/ in the R package \verb|survey| \citep{surveyp}.
	\item Bootstrap Pearson method using $X_{I}^2$, and its distribution is approximated by that of $X_{I}^{2*}$ in (\ref{eq: 7chi*}).
	\item Bootstrap likelihood-ratio method based on  $ W_{I}$, and its distribution is approximated by that of $W_{I}^*$ in (\ref{eq: 7w*}).
\end{enumerate}

For each scenario, we generate $1\,000$ Monte Carlo samples, and {we consider $M=200$, $M=500$ and $M=1\,000$ iterations for both  bootstrap methods.} for  both bootstrap testing methods. Table \ref{tab: 4} summarizes the simulation results. For Case I when the null hypothesis is true, the type I error rates  of the two naive methods are much larger than the nominal significance level 0.05 for different sample sizes. The performance of the Rao-Scott method with second-order correction is similar to the two bootstrap testing methods in terms of type I error and power. 
The   power of the two bootstrap testing methods increases with the value of $\gamma$. {In addition, we get similar test results regardless the number of bootstrap repetitions.  }


\begin{table}
	\caption{Power of the test procedures for independence based on 1\,000 Monte Carlo simulation samples, Case I corresponds to the independence scenario, and the  nominal significance level is 0.05. {The numbers of bootstrap repetitions are set to be $M=200$, $M=500$ and $M=1\,000$. } }\label{tab: 4}
	\begin{center}
		\begin{tabular}{cccccccccccc}
			\hline\hline
			\multirow{2}{*}{$(N,n)$} & \multirow{2}{*}{Case} &&\multirow{2}{*}{NP}
			&\multirow{2}{*}{NLR}&\multirow{2}{*}{RS}&\multicolumn{2}{c}{$M=200$}
			& \multicolumn{2}{c}{$M=500$}&\multicolumn{2}{c}{$M=1\,000$}\\
			&&&&&&BP&BLR&BP&BLR&BP&BLR\\
			\hline
			\multirow{3}{*}{$(2\,000,75)$}&I &&0.16& 0.08 &0.06& 0.08& 0.05 &0.07& 0.04& 0.06& 0.04 \\
			&II&& 0.21 &0.13& 0.10& 0.12& 0.10& 0.12& 0.09& 0.11 &0.09 \\
			&III && 0.71& 0.67& 0.53 &0.56& 0.56 &0.56& 0.56 &0.55& 0.56  \\
			&&&&&&&&&&\\
			\multirow{3}{*}{$(10\,000,150)$}&I&&0.14 &0.13 &0.05& 0.06& 0.05 &0.07& 0.05& 0.06 &0.05 \\
			&II&&0.25& 0.24& 0.13& 0.15& 0.14& 0.14& 0.14& 0.14& 0.13  \\
			& III&&0.94& 0.93& 0.84& 0.85& 0.82 &0.84& 0.82 &0.84 &0.83 \\
			\hline
		\end{tabular}
	\end{center}
	\begin{tablenotes}
		\setlength\labelsep{0pt}
		\footnotesize
		\item NOTE: NP, naive Pearson method; NLR: naive likelihood-ratio method; RS, Rao-Scott method using second-order correction; BP, bootstrap Pearson method; BLR: bootstrap likelihood-ratio method.
	\end{tablenotes}
\end{table}

 \end{document}